
\input amstexl


\catcode`\@=11
\ifx\amstexloaded@\relax\else
 \errmessage{AmS-TeX must be loaded before LamS-TeX}\fi
\ifx\laxread@\undefined\else\catcode`\@=\active \fi
\def\err@#1{\errmessage{LamS-TeX error: #1}}
\def^^L{\par}
\let\+\tabalign
\def\newcount{\alloc@0\count\countdef\insc@unt}
\def\newdimen{\alloc@1\dimen\dimendef\insc@unt}
\def\newskip{\alloc@2\skip\skipdef\insc@unt}
\def\newmuskip{\alloc@3\muskip\muskipdef\@cclvi}
\def\newbox{\alloc@4\box\chardef\insc@unt}
\let\newtoks\relax
\def\newhelp#1#2{\newtoks#1#1\expandafter{\csname#2\endcsname}}
\def\newtoks{\alloc@5\toks\toksdef\@cclvi}
\def\newread{\alloc@6\read\chardef\sixt@@n}
\def\newwrite{\alloc@7\write\chardef\sixt@@n}
\def\newfam{\alloc@8\fam\chardef\sixt@@n}
\def\newlanguage{\alloc@9\language\chardef\@cclvi}
\def\newinsert#1{\global\advance\insc@unt by\m@ne
  \ch@ck0\insc@unt\count
  \ch@ck1\insc@unt\dimen
  \ch@ck2\insc@unt\skip
  \ch@ck4\insc@unt\box
  \allocationnumber=\insc@unt
  \global\chardef#1=\allocationnumber
  \wlog{\string#1=\string\insert\the\allocationnumber}}
\def\newif#1{\count@\escapechar \escapechar\m@ne
  \expandafter\expandafter\expandafter
   \edef\@if#1{true}{\let\noexpand#1=\noexpand\iftrue}%
  \expandafter\expandafter\expandafter
   \edef\@if#1{false}{\let\noexpand#1=\noexpand\iffalse}%
  \@if#1{false}\escapechar\count@}

\def\Err@#1{\errhelp\defaulthelp@\err@{#1}}
{\catcode`\@=\active
 \edef\next{\gdef\noexpand@{\futurelet\noexpand\next
  \csname at\string@\endcsname}}
 \next
}
\def\at@{\ifcat\noexpand\next a\let\next@\at@@\else
 \ifcat\noexpand\next0\let\next@\at@@\else
 \ifcat\noexpand\next\relax\let\next@\at@@\else
 \let\next@\at@@@\fi\fi\fi\next@}
\def\at@@@{\errhelp\athelp@\err@{Invalid use of @}}
\def\at@@#1{\expandafter
 \ifx\csname\string#1@at\endcsname\relax\let\next@\at@@@\else
 \DN@{\csname\string#1@at\endcsname}\fi\next@}
\def\atdef@#1{\expandafter\def\csname\string#1@at\endcsname}
\newif\iftest@
\def\tagin@#1{\tagin@false
 \DN@##1\tag##2##3\next@{\test@true\ifx\tagin@##2\test@false\fi}%
 \next@#1\tag\tagin@\next@\tagin@false\iftest@\tagin@true\fi}
\let\lkerns@\relax
\def\nolinebreak{\RIfM@\mathmodeerr@\nolinebreak\else
 \ifhmode\saveskip@\lastskip\unskip
 \nobreak\ifdim\saveskip@>\z@\hskip\saveskip@\fi\lkerns@
 \else\vmodeerr@\nolinebreak\fi\fi}
\def\allowlinebreak{\RIfM@\mathmodeerr@\allowlinebreak\else
 \ifhmode\saveskip@\lastskip\unskip
 \allowbreak\ifdim\saveskip@>\z@\hskip\saveskip@\fi\lkerns@
 \else\vmodeerr@\allowlinebreak\fi\fi}
\def\linebreak{\RIfM@\mathmodeerr@\linebreak\else
 \ifhmode\unskip\unkern\break\lkerns@
 \else\vmodeerr@\linebreak\fi\fi}
\let\nkerns@\relax
\def\newline{\RIfM@\mathmodeerr@\newline\else
 \ifhmode\unskip\unkern\null\hfill\break\nkerns@
 \else\vmodeerr@\newline\fi\fi}%
\def\newbox@{\alloc@@4\box\chardef\insc@unt}
\def\newcount@{\alloc@@0\count\countdef\insc@unt}
\def\accentedsymbol#1#2{\expandafter\newbox@\csname\exstring@#1@box\endcsname
 \setbox\csname\exstring@#1@box\endcsname\hbox{$\m@th#2$}%
 \define#1{\copy\csname\exstring@#1@box\endcsname{}}}
\def\rightadd@#1\to#2{\toks@{\\#1}\toks@@\expandafter{#2}\xdef#2{\the\toks@@
 \the\toks@}\toks@{}\toks@@{}}
\def\fontlist@{\\\tenrm\\\sevenrm\\\fiverm\\\teni\\\seveni\\\fivei
 \\\tensy\\\sevensy\\\fivesy\\\tenex\\\tenbf\\\sevenbf\\\fivebf
 \\\tensl\\\tenit}
\def\font@#1=#2 {\rightadd@#1\to\fontlist@\font#1=#2 }
\def\ismember@#1#2{\global\let\Next@ F\let\next@= #2%
 {\def\\##1{\let\nextii@##1\ifx\nextii@\next@\global\let\Next@ T\fi}#1}%
 \test@false\ifx\Next@ T\test@true\fi\let\next@\relax}
\def\FNSS@#1{\let\FNSS@@#1\FN@\FNSS@@@}
\def\FNSS@@@{\ifx\next\space@\def\FNSS@@@@. {\FN@\FNSS@@@}\else
 \def\FNSS@@@@.{\FNSS@@}\fi\FNSS@@@@.}
\atdef@"{\unskip
 \DN@{\ifx\next`\DN@`{\FN@\nextii@}%
  \else\ifx\next\lq\DN@\lq{\FN@\nextii@}%
  \else\DN@####1{\FN@\nextiii@}\fi\fi
  \next@}%
 \DNii@{\ifx\next`\DN@`{\sldl@``}%
  \else\ifx\next\lq\DN@\lq{\sldl@``}%
  \else\DN@{\dlsl@`}\fi\fi\next@}%
 \def\nextiii@{\ifx\next'\DN@'{\srdr@''}%
  \else\ifx\next\rq\DN@\rq{\srdr@''}%
  \else\DN@{\drsr@'}\fi\fi\next@}%
 \FNSS@\next@}
\def\root{%
  \DN@{\ifx\next\uproot\let\next@\nextii@\else
   \ifx\next\leftroot\let\next@\nextiii@\else
   \let\next@\plainroot@\fi\fi\next@}%
  \DNii@\uproot##1{\uproot@##1\relax\FNSS@\nextiv@}%
  \def\nextiv@{\ifx\next\leftroot\let\next@\nextv@\else
   \let\next@\plainroot@\fi\next@}%
  \def\nextv@\leftroot##1{\leftroot@##1\relax\plainroot@}%
  \def\nextiii@\leftroot##1{\leftroot@##1\relax\FNSS@\nextvi@}%
  \def\nextvi@{\ifx\next\uproot\let\next@\nextvii@\else
   \let\next@\plainroot@\fi\next@}%
  \def\nextvii@\uproot##1{\uproot@##1\relax\plainroot@}%
  \bgroup\uproot@\z@\leftroot@\z@
 \FNSS@\next@}
\def\loop#1\repeat{\def\iterate{#1\relax\expandafter\iterate\fi}%
 \iterate\let\iterate\relax}
\def\gloop@#1\repeat{\gdef\iterate@{#1\relax\expandafter\iterate@\fi}%
 \iterate@\global\let\iterate@\relax}
\def\printoptions{\W@{Do you want S(yntax check),
  G(alleys) or P(ages)?^^JType S, G or P, follow by <return>: }\loop
 \read\m@ne to\ans@
 \edef\next@{\def\noexpand\Ans@{\ans@}}\uppercase\expandafter{\next@}%
 \ifx\Ans@\S@\test@true\syntax\else
 \ifx\Ans@\G@\test@true\galleys\else
 \ifx\Ans@\P@\test@true\else
 \test@false\fi\fi\fi
 \iftest@\else\W@{Type S, G or P, follow by <return>: }%
 \repeat}
\expandafter\let\csname A@;\endcsname;
\expandafter\let\csname A@:\endcsname:
\expandafter\let\csname A@?\endcsname?
\expandafter\let\csname A@!\endcsname!
\def\APdef#1{\def\next@{\expandafter\let\csname A@\string#1\endcsname#1}%
 \afterassignment\next@\def#1}
\let\fextra@\,
\def\tdots@{\unskip
 \DN@{$\m@th\mathinner{\ldotp\ldotp\ldotp}\,
   \ifx\next,\,$\else\ifx\next.\,$\else
   \ifx\next;\,$\else
   \expandafter\ifx\csname A@\string;\endcsname\next\fextra@$\else
   \ifx\next:\,$\else
   \expandafter\ifx\csname A@\string:\endcsname\next\fextra@$\else
   \ifx\next?\,$\else
   \expandafter\ifx\csname A@\string?\endcsname\next\fextra@$\else
   \ifx\next!\,$\else
   \expandafter\ifx\csname A@\string!\endcsname\next\fextra@$\else
   $ \fi\fi\fi\fi\fi\fi\fi\fi\fi\fi}%
 \ \FN@\next@}
\def\extrap@#1{%
 \ifx\next,\DN@{#1\,}\else
 \ifx\next;\DN@{#1\,}\else
 \expandafter\ifx\csname A@\string;\endcsname\next\DN@{#1\fextra@}\else
 \ifx\next.\DN@{#1\,}\else\extra@
 \ifextra@\DN@{#1\,}\else
 \let\next@#1\fi\fi\fi\fi\fi\next@}
\def\dotsc{\DN@{\ifx\next;\plainldots@\,\else
 \expandafter\ifx\csname A@\string;\endcsname\next\plainldots@\fextra@\else
 \ifx\next.\plainldots@\,\else\extra@\plainldots@
 \ifextra@\,\fi\fi\fi\fi}%
 \FN@\next@}
\def\keybin@{\keybin@true
 \ifx\next+\else\ifx\next=\else\ifx\next<\else\ifx\next>\else\ifx\next-\else
 \ifx\next*\else\ifx\next:\else
 \expandafter\ifx\csname A@\string;\endcsname\next\else
 \keybin@false\fi\fi\fi\fi\fi\fi\fi\fi}
\def\boldkey#1{\ifcat\noexpand#1A%
  \ifcmmibloaded@{\fam\cmmibfam#1}\else
   \Err@{First bold symbol font not loaded}\fi
 \else
 \let\next=#1%
 \ifx#1!\mathchar"5\bffam@21 \else
 \expandafter\ifx\csname A@\string!\endcsname\next\mathchar"5\bffam@21 \else
 \ifx#1(\mathchar"4\bffam@28 \else\ifx#1)\mathchar"5\bffam@29 \else
 \ifx#1+\mathchar"2\bffam@2B \else\ifx#1:\mathchar"3\bffam@3A \else
 \expandafter\ifx\csname A@\string:\endcsname\next\mathchar"3\bffam@3A \else
 \ifx#1;\mathchar"6\bffam@3B \else
 \expandafter\ifx\csname A@\string;\endcsname\next\mathchar"6\bffam@3B \else
 \ifx#1=\mathchar"3\bffam@3D \else
 \ifx#1?\mathchar"5\bffam@3F \else
 \expandafter\ifx\csname A@\string?\endcsname\next\mathchar"5\bffam@3F \else
 \ifx#1[\mathchar"4\bffam@5B \else
 \ifx#1]\mathchar"5\bffam@5D \else
 \ifx#1,\mathchari@63B \else
 \ifx#1-\mathcharii@200 \else
 \ifx#1.\mathchari@03A \else
 \ifx#1/\mathchari@03D \else
 \ifx#1<\mathchari@33C \else
 \ifx#1>\mathchari@33E \else
 \ifx#1*\mathcharii@203 \else
 \ifx#1|\mathcharii@06A \else
 \ifx#10\bold0\else\ifx#11\bold1\else\ifx#12\bold2\else\ifx#13\bold3\else
 \ifx#14\bold4\else\ifx#15\bold5\else\ifx#16\bold6\else\ifx#17\bold7\else
 \ifx#18\bold8\else\ifx#19\bold9\else
  \Err@{\noexpand\boldkey can't be used with #1}%
 \fi\fi\fi\fi\fi\fi\fi\fi\fi\fi\fi\fi\fi\fi\fi
 \fi\fi\fi\fi\fi\fi\fi\fi\fi\fi\fi\fi\fi\fi\fi\fi\fi\fi}
\def\arabic#1{#1}
\def\alph#1{\count@#1\relax\advance\count@96 \ifnum\count@>122
 \Err@{\noexpand\alph invalid for numbers > 26}\else\char\count@\fi}
\def\Alph#1{\count@#1\relax\advance\count@64 \ifnum\count@>90
 \Err@{\noexpand\Alph invalid for numbers > 26}\else\char\count@\fi}

\def\Roman#1{\uppercase\expandafter{\romannumeral#1}}
\def\fnsymbol#1{\count@#1\relax
 \count@@\count@
 \advance\count@\m@ne\divide\count@7
 \count@@@\count@\advance\count@@@\@ne
 \multiply\count@7 \advance\count@@-\count@
 \count@\count@@@
 {\loop
  \ifcase\count@@\or*\or\dag\or\ddag\or\P\or\S\or\text{$\|$}\or\#\fi
  \advance\count@\m@ne\ifnum\count@>\z@\repeat}}
\def\cardnine@#1{\ifcase#1\or one\or two\or three\or four\or five\or
 six\or seven\or eight\or nine\fi}
\let\alloc@\alloc@@
\newcount\ten@
\ten@10
\def\cardinal#1{\count@#1\relax
 \ifnum\count@>99 \number\count@
 \else
  \ifnum\count@=\z@ zero%
  \else
   \ifnum\count@<\ten@\cardnine@\count@
   \else
    \ifnum\count@<20
     \advance\count@-\ten@
     \ifcase\count@ ten\or eleven\or twelve\or thirteen\or fourteen\or
      fifteen\or sixteen\or seventeen\or eighteen\or nineteen\fi
    \else
     \count@@\count@\count@@@\count@@
     \divide\count@\ten@\multiply\count@\ten@
     \advance\count@@@-\count@\divide\count@\ten@
     \ifcase\count@\or\or twenty\or thirty\or forty\or fifty\or sixty\or
      seventy\or eighty\or ninety\fi
     \ifnum\count@@@=\z@\else-\cardnine@\count@@@\fi
    \fi
   \fi
  \fi
 \fi}
\def\ordnine@#1{\ifcase#1\or first\or second\or third\or fourth\or fifth\or
 sixth\or seventh\or eighth\or ninth\fi}
\newcount\count@@@@
\def\ordsuffix@{\count@@@@\count@
 \divide\count@\ten@
 \count@@@\count@\count@@\count@
 \divide\count@@\ten@\multiply\count@@\ten@
 \advance\count@@@-\count@@
 \ifnum\count@@@=\@ne th%
 \else
  \count@@@\count@@@@
  \count@@\count@@@@
  \divide\count@@\ten@\multiply\count@@\ten@
  \advance\count@@@-\count@@
  \ifcase\count@@@ th\or st\or nd\or rd\else th\fi
 \fi}
\def\nordinal#1{\count@#1\relax\number\count@\ordsuffix@}
\def\spordinal#1{\count@#1\relax\number\count@$^{\text{\ordsuffix@}}$}
\def\ordinal#1{\count@#1\relax
 \ifnum\count@>99 \number\count@\ordsuffix@
 \else
   \ifnum\count@=\z@ zeroth%
  \else
    \ifnum\count@<\ten@\ordnine@\count@
    \else
     \ifnum\count@<20 \advance\count@-\ten@
      \ifcase\count@ tenth\or eleventh\or twelfth\or thirteenth\or
       fourteenth\or fifteenth\or sixteenth\or seventeenth\or eighteenth\or
       nineteenth\fi
     \else
      \count@@\count@
      \divide\count@\ten@\multiply\count@\ten@
      \count@@@\count@@\advance\count@@@-\count@
      \divide\count@\ten@
      \ifcase\count@\or\or twent\or thirt\or fort\or fift\or sixt\or sevent\or
       eight\or ninet\fi
      \ifnum\count@@@=\z@ ieth\else y-\ordnine@\count@@@\fi
     \fi
    \fi
  \fi
 \fi}
\font@\tensmc=cmcsc10
\textonlyfont@\smc\tensmc
\newtoks\noexpandtoks@
\noexpandtoks@{\let\arabic\relax\let\alph\relax\let\Alph\relax
 \let\Roman\relax\let\fnsymbol\relax\let\rm\relax
 \let\it\relax\let\bf\relax\let\sl\relax\let\smc\relax
 \let\/\relax\let\null\relax}
\def\noexpands@{\the\noexpandtoks@}
\def\Nonexpanding#1{\global\noexpandtoks@
 \expandafter{\the\noexpandtoks@\let#1\relax}}
\def\prevanish@{\saveskip@\z@\ifhmode\saveskip@\lastskip\unskip\fi}
\def\postvanish@{\ifdim\saveskip@>\z@\hskip\saveskip@\fi\FN@\postvanish@@}
\def\postvanish@@{\DN@.{}%
 \ifx\next\space@\ifdim\saveskip@>\z@\DN@. {}\fi\fi\next@.}
\def\invisible#1{\prevanish@\ignorespaces#1\unskip\postvanish@}
\def\vanishlist@{\\\invisible}
\let\noindent@\noindent
\def\noindent{\par\noindent@\FN@\pretendspace@}
\def\pretendspace@{\ismember@\vanishlist@\next
 \iftest@\nobreak\hskip-\p@\hskip\p@\fi}

\newtoks\everypartoks@
\def\noindent@@{\par\everypartoks@\expandafter{\the\everypar}\everypar{}%
 \noindent@\everypar\expandafter{\the\everypartoks@}}
\def\page{\Err@{\noexpand\page has no meaning by itself}}
\let\page@C\pageno
\let\page@P\empty
\let\page@Q\empty
\def\page@S#1{#1\/}
\def\page@F{\rm}
\def\page@N{\arabic}   
\newif\ifindexing@
\def\indexfile{\ifindexing@\else
 \alloc@@7\write\chardef\sixt@@n\ndx@
 \immediate\openout\ndx@=\jobname.ndx
 \global\indexing@true\fi}
\global\advance\insc@unt\m@ne
\ch@ck0\insc@unt\count
\ch@ck1\insc@unt\dimen
\ch@ck2\insc@unt\skip
\ch@ck4\insc@unt\box
\allocationnumber\insc@unt
\global\chardef\margin@\allocationnumber
\dimen\margin@\maxdimen
\count\margin@\z@
\skip\margin@\z@
\newif\ifindexproofing@
\def\indexproofing{\indexproofing@true}
\def\noindexproofing{\indexproofing@false}
\def\unmacro@#1:#2->#3\unmacro@{\def\macpar@{#2}\def\macdef@{#3}}
\def\starparts@#1{\def\stari@{#1}\def\starii@{#1}\let\stariii@\empty
 \test@false
 \DN@##1*##2##3\next@{\ifx\starparts@##2\test@false\else\test@true\fi}%
 \next@#1*\starparts@\next@
 \iftest@\DN@{\starparts@@#1\starparts@@}\else\let\next@\relax\fi\next@}
\def\starparts@@#1*#2\starparts@@{\def\starii@{#1}\def\stariii@{*#2}}
\def\windex@{\ifindexing@
 \expandafter\unmacro@\meaning\stari@\unmacro@
 \edef\macdef@{\string"\macdef@\string"}%
 \edef\next@{\write\ndx@{\macdef@}}\next@
 \write\ndx@{{\number\pageno}{\page@N}{\page@P}{\page@Q}}%
 \fi
 \ifindexproofing@
  \ifx\stariii@\empty\else
   \expandafter\unmacro@\meaning\stariii@\unmacro@\fi
  \insert\margin@{\hbox{\rm\vrule\height9\p@\depth2\p@\width\z@\starii@
  \ifx\stariii@\empty\else\tt\macdef@\fi}}\fi}
\catcode`\"=\active
\def"{\FN@\quote@}
\def\quote@{\ifx\next"\expandafter\quote@@\else\expandafter\quote@@@\fi}
\def\quote@@@#1"{\starparts@{#1}\starii@\windex@}
\def\quote@@"#1"{\prevanish@\starparts@{#1}\windex@\FN@\quote@@@@}
\def\quote@@@@{\ifx\next"\DN@"{\postvanish@}\else
 \let\next@\postvanish@\fi\next@}
\rightadd@"\to\vanishlist@
\def\idefine#1{\DN@{#1}\DNii@{\noexpand#1}%
 \afterassignment\idefine@\def\nextiii@}
\def\idefine@{\ifindexing@
 \expandafter\let\next@\nextiii@
 \expandafter\unmacro@\meaning\nextiii@\unmacro@
 \immediate\write\ndx@{\noexpand\define\nextii@\macpar@{\macdef@}}\fi}
\def\iabbrev*#1#2{\ifindexing@\toks@{#2}%
 \immediate\write\ndx@{\noexpand\abbrev*\noexpand#1{\the\toks@}}\fi}
\newread\laxread@
\newwrite\laxwrite@
\let\fnpages@\empty
\def\Finit@#1#2\Finit@{\let\nextii@#1\def\nextiii@{#2}}
\catcode`\~=11
\def\getparts@ @#1~#2~#3~#4~#5~#6{\def\nextiv@{#1}%
 \def\nextiii@{#2~#3~#4~#5~}\count@#6\relax}
\newif\ifdocument@
\def\document{\ifdocument@\else\global\document@true
 \let\fontlist@\empty
 \immediate\openin\laxread@=\jobname.lax\relax
 {\endlinechar\m@ne\noexpands@\catcode`\@=11 \catcode`\~=11
  \loop\ifeof\laxread@\else
   \read\laxread@ to\next@
   \ifx\next@\empty
   \else
    \expandafter\Finit@\next@\Finit@
    \if\nextii@ F%
     \expandafter\rightadd@\nextiii@\to\fnpages@
    \else
     \expandafter\getparts@\next@
     \edef\next@{\gdef\csname\nextiv@ @L\endcsname{\nextiii@\number\count@}}%
     \next@
    \fi
   \fi
  \repeat}%
 \immediate\closein\laxread@
 \immediate\openout\laxwrite@=\jobname.lax\relax\fi}
\let\thelabel@\relax
\def\thelabels@{\thelabel@ ~\thelabel@@ ~\thelabel@@@ ~\thelabel@@@@ ~}
\def\label#1{\prevanish@
 \ifx\thelabel@\relax
  \Err@{There's nothing here to be labelled}%
 \else
  {\noexpands@
  \expandafter\ifx\csname#1@L\endcsname\relax
   \expandafter\xdef\csname#1@L\endcsname{\thelabels@0}%
   \immediate\write\laxwrite@{@#1~\thelabels@1}%
  \else
   \edef\next@{@~\csname#1@L\endcsname}%
    \expandafter\getparts@\next@
    \ifodd\count@
    \expandafter\xdef\csname#1@L\endcsname{\thelabels@0}%
    \immediate\write\laxwrite@{@#1~\thelabels@1}%
   \else
    \Err@{Label #1 already used}%
   \fi
  \fi
  }%
 \fi
 \postvanish@}
\rightadd@\label\to\vanishlist@
\def\thepages@{\page@N{\number\page@C}~%
 \page@S{\page@P\page@N{\number\page@C}\page@Q}~%
 \number\page@C ~\page@P\page@N{\number\page@C}\page@Q ~}
\def\pagelabel#1{\prevanish@
 \expandafter\ifx\csname#1@L\endcsname\relax
  {\noexpands@
  \expandafter\xdef\csname#1@L\endcsname{\thepages@2}}%
  \write\laxwrite@{@#1~\thepages@3}%
 \else
  {\noexpands@
  \edef\next@{@~\csname#1@L\endcsname}%
  \expandafter\getparts@\next@
  \ifodd\count@
   \ifnum\count@=\@ne
    \expandafter\xdef\csname#1@L\endcsname{\thelabels@2}%
   \fi
   \write\laxwrite@{@#1~\thepages@3}%
  \else
   \Err@{Label #1 already used}%
  \fi
  }%
 \fi
 \postvanish@}
\rightadd@\pagelabel\to\vanishlist@
\newif\ifreferr@
\referr@true
\def\RefErrors{\global\referr@true}
\def\RefWarnings{\global\referr@false}
\setbox\z@\hbox{\global\count@=`^^30}
\ifnum\count@=48 \let\versionthree@\relax\fi
\def\nolabel@#1#2#3{\expandafter\ifx\csname#2@L\endcsname\relax
 \ifreferr@\Err@{No \noexpand\label found for #2}\else
 \W@{Warning: No \noexpand\label found for #2.}%
 \ifx\versionthree@\relax\W@{l.\number\inputlineno\space ... \string#1{#2}}\fi
 \fi#3\else}
\def\csL@#1{{\noexpands@\xdef\Next@{\csname#1@L\endcsname}}}
\def\ref#1{\nolabel@\ref{#1}\relax
 \DNii@##1~##2\nextii@{##1}%
 \csL@{#1}\expandafter\nextii@\Next@\nextii@\fi}
\def\Ref#1{\nolabel@\Ref{#1}\relax
 \DNii@##1~##2~##3\nextii@{##2}%
 \csL@{#1}\expandafter\nextii@\Next@\nextii@\fi}
\def\nref#1{\nolabel@\nref{#1}\relax
 \DNii@##1~##2~##3~##4\nextii@{##3}%
 \csL@{#1}\expandafter\nextii@\Next@\nextii@\fi}
\def\pref#1{\nolabel@\pref{#1}\relax
 \DNii@##1~##2~##3~##4~##5\nextii@{##4}%
 \csL@{#1}\expandafter\nextii@\Next@\nextii@\fi}
\let\pref@\pref
\def\Evaluatenref#1{\nolabel@\Evaluatenref{#1}{\gdef\Nref{-10000 }}%
 \DNii@##1~##2~##3~##4\nextii@{\DNii@{##3}}%
 \csL@{#1}\expandafter\nextii@\Next@\nextii@
 \xdef\Nref{\nextii@}\fi}
\def\Evaluatepref#1{\nolabel@\Evaluatepref{#1}{\global\let\Pref\empty}%
 \DNii@##1~##2~##3~##4~##5\nextii@{\DNii@{##4}}%
 \csL@{#1}\expandafter\nextii@\Next@\nextii@
 \xdef\Pref{\nextii@}\fi}
\def\readlax#1{\immediate\openin\laxread@=#1.lax\relax
 \ifeof\laxread@\W@{}\W@{File #1.lax not found.}\W@{}\fi
 {\endlinechar\m@ne\noexpands@\catcode`\@=11 \catcode`\~=11
  \loop\ifeof\laxread@\else
   \read\laxread@ to\nextv@
   \ifx\nextv@\empty
   \else
    \expandafter\Finit@\nextv@\Finit@
    \ifx\nextii@ F%
    \else
     \expandafter\getparts@\nextv@
     \expandafter\ifx\csname\nextiv@ @L\endcsname\relax
      \edef\next@{\gdef\csname\nextiv@ @L\endcsname
       {\nextiii@\ifnum\count@=\@ne0\else2\fi}}%
      \next@
     \else
      \Err@{Label \nextiv@\space in #1.lax already used}%
     \fi
    \fi
   \fi
  \repeat}%
 \immediate\closein\laxread@}
\catcode`\~=\active

\def\FNSSP@{\FNSS@\pretendspace@}
\everydisplay{\csname displaymath \endcsname}
\expandafter\def\csname displaymath \endcsname#1$${#1$$\FNSSP@}
\def\locallabel@{\let\thelabel@\Thelabel@\let\thelabel@@\Thelabel@@
 \let\thelabel@@@\Thelabel@@@\let\thelabel@@@@\Thelabel@@@@}
\newcount\tag@C
\tag@C\z@
\let\tag@P\empty
\let\tag@Q\empty
\def\tag@S#1{{\rm(}{#1\/}{\rm)}}
\let\tag@N\arabic
\def\tag@F{\rm}
\def\maketag@{\FN@\maketag@@}
\def\maketag@@{\ifx\next\relax\DN@\relax{\FN@\maketag@@}\else
 \ifx\next"\let\next@\maketag@@@\else
 \let\next@\maketag@@@@\fi\fi\next@}
\def\xdefThelabel@#1{\xdef\Thelabel@{#1{\Thelabel@@@}}}
\def\xdefThelabel@@#1{\xdef\Thelabel@@{#1{\Thelabel@@@@}}}
\def\maketag@@@@#1\maketag@{\global\advance\tag@C\@ne
 {\noexpands@
  \xdef\Thelabel@@@{\number\tag@C}%
  \xdefThelabel@\tag@N
  \xdef\Thelabel@@@@{\ifmathtags@$\tag@P\Thelabel@\tag@Q$\else
   \tag@P\Thelabel@\tag@Q\fi}%
  \xdefThelabel@@\tag@S
  }%
 \locallabel@
 \hbox{\tag@F\thelabel@@}%
 #1}
\def\Qlabel@#1{{\noexpands@\xdef\Thelabel@@{#1}%
 \let\style\empty\xdef\Thelabel@@@@{#1}%
 \let\pre\empty\let\post\empty\xdef\Thelabel@{#1}%
 \let\numstyle\empty\xdef\Thelabel@@@{#1}}}
\def\maketag@@@"#1"#2\maketag@{%
 {\let\pre\tag@P\let\post\tag@Q\let\style\tag@S\let\numstyle\tag@N
  \hbox{\tag@F#1}%
  \noexpands@
  \Qlabel@{#1}%
  }%
 \locallabel@
 #2}
\def\align@{\inalign@true\inany@true
 \vspace@\allowdisplaybreak@\displaybreak@\intertext@
 \def\tag{\global\tag@true\ifnum\and@=\z@
  \DN@{&\omit\global\rwidth@\z@&\relax}\else
  \DN@{&\relax}\fi\next@}%
 \iftagsleft@\DN@{\csname align \endcsname}\else
  \DN@{\csname align \space\endcsname}\fi\next@}
\def\noset@{\def\Offset##1##2{\prevanish@\postvanish@}%
 \def\Reset##1##2{\prevanish@\postvanish@}}
\def\measure@#1\endalign{\global\lwidth@\z@\global\rwidth@\z@
 \global\maxlwidth@\z@\global\maxrwidth@\z@
 \global\and@\z@
 \setbox\z@\vbox
  {\noset@\everycr{\noalign{\global\tag@false\global\and@\z@}}\Let@
  \halign{\setboxz@h{$\m@th\displaystyle{\@lign##}$}%
   \global\lwidth@\wdz@
   \ifdim\lwidth@>\maxlwidth@\global\maxlwidth@\lwidth@\fi
   \global\advance\and@\@ne
   &\setboxz@h{$\m@th\displaystyle{{}\@lign##}$}\global\rwidth@\wdz@
   \ifdim\rwidth@>\maxrwidth@\global\maxrwidth@\rwidth@\fi
   \global\advance\and@\@ne
   &\Tag@\eat@{##}\crcr#1\crcr}}%
 \totwidth@\maxlwidth@\advance\totwidth@\maxrwidth@}
\def\prepost@{\global\let\tag@P@\tag@P\global\let\tag@Q@\tag@Q}
\def\reprepost@{\let\tag@P\tag@P@\let\tag@Q\tag@Q@}
\expandafter\def\csname align \space\endcsname#1\endalign
 {\measure@#1\endalign\global\and@\z@
 \ifingather@\everycr{\noalign{\global\and@\z@}}\else\displ@y@\fi
 \Let@\tabskip\centering@
 \halign to\displaywidth
  {\hfil\strut@\setboxz@h{$\m@th\displaystyle{\@lign##\prepost@}$}%
  \boxz@\global\advance\and@\@ne
  \tabskip\z@skip
  &\setboxz@h{$\m@th\displaystyle{{}\@lign##\prepost@}$}%
  \global\rwidth@\wdz@\boxz@\hfil\global\advance\and@\@ne
  \tabskip\centering@
  &\setboxz@h{\@lign\strut@\reprepost@\maketag@##\maketag@}%
  \dimen@\displaywidth\advance\dimen@-\totwidth@
  \divide\dimen@\tw@\advance\dimen@\maxrwidth@\advance\dimen@-\rwidth@
  \ifdim\dimen@<\tw@\wdz@\llap{\vtop{\normalbaselines\null\boxz@}}%
  \else\llap{\boxz@}\fi
  \tabskip\z@skip
  \crcr#1\crcr
  \black@\totwidth@}}
\expandafter\def\csname align \endcsname#1\endalign{\measure@#1\endalign
 \global\and@\z@
 \ifdim\totwidth@>\displaywidth\let\displaywidth@\totwidth@\else
  \let\displaywidth@\displaywidth\fi
 \ifingather@\everycr{\noalign{\global\and@\z@}}\else\displ@y@\fi
 \Let@\tabskip\centering@\halign to\displaywidth
  {\hfil\strut@\setboxz@h{$\m@th\displaystyle{\@lign##\prepost@}$}%
  \global\lwidth@\wdz@\global\lineht@\ht\z@
  \boxz@\global\advance\and@\@ne
  \tabskip\z@skip&\setboxz@h{$\m@th\displaystyle{{}\@lign##\prepost@}$}%
  \ifdim\ht\z@>\lineht@\global\lineht@\ht\z@\fi
  \boxz@\hfil\global\advance\and@\@ne
  \tabskip\centering@&\kern-\displaywidth@
  \setboxz@h{\@lign\strut@\reprepost@\maketag@##\maketag@}%
  \dimen@\displaywidth\advance\dimen@-\totwidth@
  \divide\dimen@\tw@\advance\dimen@\maxlwidth@\advance\dimen@-\lwidth@
  \ifdim\dimen@<\tw@\wdz@
   \rlap{\vbox{\normalbaselines\boxz@\vbox to\lineht@{}}}\else
   \rlap{\boxz@}\fi
  \tabskip\displaywidth@\crcr#1\crcr\black@\totwidth@}}
\def\attag@#1{\let\Maketag@\maketag@\let\TAG@\Tag@
 \let\Prepost@\prepost@\let\Reprepost@\reprepost@
 \let\Tag@\relax\let\maketag@\relax
 \let\prepost@\relax\let\reprepost@\relax
 \ifmeasuring@
  \def\llap@##1{\setboxz@h{##1}\hbox to\tw@\wdz@{}}%
  \def\rlap@##1{\setboxz@h{##1}\hbox to\tw@\wdz@{}}%
 \else\let\llap@\llap\let\rlap@\rlap\fi
 \toks@{\hfil\strut@
  $\m@th\displaystyle{\@lign\the\hashtoks@\prepost@}$%
  \tabskip\z@skip\global\advance\and@\@ne&
  $\m@th\displaystyle{{}\@lign\the\hashtoks@\prepost@}$\hfil
  \ifxat@\tabskip\centering@\fi\global\advance\and@\@ne}%
 \iftagsleft@
  \toks@@{\tabskip\centering@&\Tag@\kern-\displaywidth
   \rlap@{\@lign\reprepost@\maketag@\the\hashtoks@\maketag@}%
   \global\advance\and@\@ne\tabskip\displaywidth}\else
  \toks@@{\tabskip\centering@&\Tag@\llap@{\@lign\reprepost@\maketag@
   \the\hashtoks@\maketag@}\global\advance\and@\@ne\tabskip\z@skip}\fi
 \atcount@#1\relax\advance\atcount@\m@ne
 \loop\ifnum\atcount@>\z@
  \toks@\expandafter{\the\toks@&\hfil$\m@th\displaystyle{\@lign
  \the\hashtoks@\prepost@}$\global\advance\and@\@ne
  \tabskip\z@skip
  &$\m@th\displaystyle{{}\@lign\the\hashtoks@\prepost@}$\hfil\ifxat@
  \tabskip\centering@\fi\global\advance\and@\@ne}\advance\atcount@\m@ne
 \repeat
 \edef\preamble@{\the\toks@\the\toks@@}%
 \edef\preamble@@{\preamble@}%
 \let\maketag@\Maketag@\let\Tag@\TAG@
 \let\prepost@\Prepost@\let\reprepost@\Reprepost@}
\def\unlabel@{\def\label##1{\prevanish@\postvanish@}%
 \def\pagelabel##1{\prevanish@\postvanish@}}
\newcount\tag@CC
\expandafter\def\csname alignat \endcsname#1#2\endalignat
 {\inany@true\xat@false
 \def\tag{\global\tag@true
  \count@#1\relax\multiply\count@\tw@\advance\count@\m@ne
  \gdef\tag@{&}%
  \loop\ifnum\count@>\and@\xdef\tag@{&\omit\tag@}%
  \advance\count@\m@ne\repeat
  \tag@\relax}%
 \vspace@\allowdisplaybreak@\displaybreak@\intertext@
 \displ@y@\measuring@true\tag@CC\tag@C
 \setbox\savealignat@\hbox{\noset@\unlabel@$\m@th\displaystyle\Let@
  \attag@{#1}\vbox{\halign{\span\preamble@@\crcr#2\crcr}}$}%
 \measuring@false
 \Let@\attag@{#1}\tag@C\tag@CC
 \tabskip\centering@\halign to\displaywidth
  {\span\preamble@@\crcr#2\crcr\black@{\wd\savealignat@}}}
\expandafter\def\csname xalignat \endcsname#1#2\endxalignat
 {\inany@true\xat@true
 \def\tag{\global\tag@true
  \count@#1\relax\multiply\count@\tw@\advance\count@\m@ne
  \gdef\tag@{&}%
  \loop\ifnum\count@>\and@\xdef\tag@{&\omit\tag@}%
  \advance\count@\m@ne\repeat
  \tag@\relax}%
 \vspace@\allowdisplaybreak@\displaybreak@\intertext@
 \displ@y@\measuring@true\tag@CC\tag@C
 \setbox\savealignat@\hbox{\noset@\unlabel@$\m@th\displaystyle\Let@
  \attag@{#1}\vbox{\halign{\span\preamble@@\crcr#2\crcr}}$}%
 \measuring@false\Let@\attag@{#1}\tag@C\tag@CC
 \tabskip\centering@\halign to\displaywidth
 {\span\preamble@@\crcr#2\crcr\black@{\wd\savealignat@}}}
\def\gather{\RIfMIfI@\DN@{\onlydmatherr@\gather}\else
 \ingather@true\inany@true\def\tag{&\relax}%
 \vspace@\allowdisplaybreak@\displaybreak@\intertext@
 \displ@y\Let@
 \iftagsleft@\DN@{\csname gather \endcsname}\else
  \DN@{\csname gather \space\endcsname}\fi\fi
 \else\DN@{\onlydmatherr@\gather}\fi\next@}
\def\exstring@{\expandafter\eat@\string}
\def\newcounter#1{\define#1{}%
 \edef\next@{\def\noexpand#1{\futurelet\noexpand\next
  \csname\exstring@#1@Z\endcsname}}\next@
 \edef\next@{\def\csname\exstring@#1@Z\endcsname
  {\global\advance\csname\exstring@#1@C\endcsname\@ne
  {\csname\exstring@#1@F\endcsname\csname\exstring@#1@S\endcsname
   {\csname\exstring@#1@P\endcsname\csname\exstring@#1@N\endcsname
   {\noexpand\number\csname\exstring@#1@C\endcsname}%
   \csname\exstring@#1@Q\endcsname}}%
  \noexpand\ifx\noexpand\next\noexpand\label
   \def\noexpand\next@\noexpand\label########1{{\noexpand\noexpands@
    \xdef\noexpand\Thelabel@{\csname\exstring@#1@N\endcsname
     {\noexpand\number\csname\exstring@#1@C\endcsname}}%
    \xdef\noexpand\Thelabel@@@{\noexpand\number
     \csname\exstring@#1@C\endcsname}%
    \xdef\noexpand\Thelabel@@{\csname\exstring@#1@S\endcsname
     {\csname\exstring@#1@P\endcsname
     \csname\exstring@#1@N\endcsname
     {\noexpand\number\csname\exstring@#1@C\endcsname}%
     \csname\exstring@#1@Q\endcsname}}%
    \xdef\noexpand\Thelabel@@@@{\csname\exstring@#1@P\endcsname
     \csname\exstring@#1@N\endcsname
     {\noexpand\number\csname\exstring@#1@C\endcsname}%
     \csname\exstring@#1@Q\endcsname}}%
    {\noexpand\locallabel@\noexpand\label{########1}}}%
   \noexpand\else\let\noexpand\next@\relax\noexpand\fi\noexpand\next@}}\next@
 \expandafter\newcount@\csname\exstring@#1@C\endcsname
 \expandafter\let\csname\exstring@#1@N\endcsname\arabic
 \expandafter\def\csname\exstring@#1@S\endcsname##1{##1\/}%
 \expandafter\let\csname\exstring@#1@P\endcsname\empty
 \expandafter\let\csname\exstring@#1@Q\endcsname\empty
 \expandafter\def\csname\exstring@#1@F\endcsname{\rm}%
 }
\def\HASH@#1#2{\ifnum#2=\z@\else
 \edef\next@{\toks@{\the\toks@\the\hashtoks@#2}%
 \toks@@{\the\toks@@{\the\hashtoks@#2}}}\next@\expandafter\HASH@\fi}
\def\HASH@@{\toks@{}\toks@@{}\expandafter\HASH@\macpar@00}
\def\usecounter#1#2{\expandafter\ifx\csname\exstring@#1@Z\endcsname
 \relax\Err@{\noexpand#1not created with \string\newcounter}\fi
 \expandafter\let\csname\exstring@#1@@Z\endcsname\relax
 \expandafter\let\csname\exstring@#1@@Z@\endcsname\relax
 \expandafter\let\csname\exstring@#1@@Z@@\endcsname\relax
 \edef\next@{\def\noexpand#2{\futurelet\noexpand\next
  \csname\exstring@#1@@Z\endcsname}}\next@
 \edef\next@{\def\csname\exstring@#1@@Z\endcsname{\noexpand\ifx
  \noexpand\next\noexpand\label\def\noexpand\next@\noexpand\label
   ########1{\csname\exstring@#1@@Z@\endcsname
   {\noexpand#1\noexpand\label{########1}}}%
   \noexpand\else\noexpand\ifx\noexpand\next
   \noexpand"\def\noexpand\next@\noexpand"########1\noexpand"%
   {\csname\exstring@#1@@Z@\endcsname{{\expandafter\noexpand
   \csname\exstring@#1@F\endcsname
   \let\noexpand\pre\expandafter\noexpand\csname\exstring@#1@P\endcsname
   \let\noexpand\post\expandafter\noexpand\csname\exstring@#1@Q\endcsname
   \let\noexpand\style\expandafter\noexpand\csname\exstring@#1@S\endcsname
   \let\noexpand\numstyle\expandafter\noexpand\csname\exstring@#1@N\endcsname
   ########1}}}\noexpand\else
   \def\noexpand\next@{\csname\exstring@#1@@Z@\endcsname{\noexpand#1}}%
   \noexpand\fi\noexpand\fi\noexpand\next@}}\next@
 \def\next@{\expandafter\expandafter\expandafter\unmacro@\expandafter
  \meaning\csname\exstring@#1@@Z@@\endcsname\unmacro@
  \HASH@@
  \edef\next@{\def\csname\exstring@#1@@Z@\endcsname\the\toks@{%
   \expandafter\noexpand\csname\exstring@#1@@Z@@\endcsname\the\toks@@
   \noexpand\FNSSP@}}\next@}%
 \afterassignment\next@
 \expandafter\def\csname\exstring@#1@@Z@@\endcsname}
\def\listbi@{\penalty50 \medskip}
\def\listbii@{\penalty100 \smallskip}
\let\listbiii@\relax
\let\listbiv@\relax
\let\listbv@\relax
\def\listmi@{\advance\leftskip30\p@\relax}
\let\listmii@\listmi@
\let\listmiii@\listmi@
\let\listmiv@\listmi@
\let\listmv@\listmi@
\def\itemi@#1{\noindent@@\llap{#1\hskip5\p@}}
\let\itemii@\itemi@
\let\itemiii@\itemi@
\let\itemiv@\itemi@
\let\itemv@\itemi@
\def\liste@{\penalty-50 \medskip}
\def\listei@{\penalty-100 \smallskip}
\let\listeii@\relax
\let\listeiii@\relax
\let\listeiv@\relax
\expandafter\newcount\csname list@C1\endcsname
\csname list@C1\endcsname\z@
\expandafter\newcount\csname list@C2\endcsname
\csname list@C2\endcsname\z@
\expandafter\newcount\csname list@C3\endcsname
\csname list@C3\endcsname\z@
\expandafter\newcount\csname list@C4\endcsname
\csname list@C4\endcsname\z@
\expandafter\newcount\csname list@C5\endcsname
\csname list@C5\endcsname\z@
\expandafter\let\csname list@P1\endcsname\empty
\expandafter\let\csname list@P2\endcsname\empty
\expandafter\let\csname list@P3\endcsname\empty
\expandafter\let\csname list@P4\endcsname\empty
\expandafter\let\csname list@P5\endcsname\empty
\expandafter\let\csname list@Q1\endcsname\empty
\expandafter\let\csname list@Q2\endcsname\empty
\expandafter\let\csname list@Q3\endcsname\empty
\expandafter\let\csname list@Q4\endcsname\empty
\expandafter\let\csname list@Q5\endcsname\empty
\expandafter\def\csname list@S1\endcsname#1{{\rm(}{#1\/}{\rm)}}
\expandafter\def\csname list@S2\endcsname#1{{\rm(}{#1\/}{\rm)}}
\expandafter\def\csname list@S3\endcsname#1{{\rm(}{#1\/}{\rm)}}
\expandafter\def\csname list@S4\endcsname#1{{\rm(}{#1\/}{\rm)}}
\expandafter\def\csname list@S5\endcsname#1{{\rm(}{#1\/}{\rm)}}
\expandafter\let\csname list@N1\endcsname\arabic
\expandafter\let\csname list@N2\endcsname\arabic
\expandafter\let\csname list@N3\endcsname\arabic
\expandafter\let\csname list@N4\endcsname\arabic
\expandafter\let\csname list@N5\endcsname\arabic
\expandafter\def\csname list@F1\endcsname{\rm}
\expandafter\def\csname list@F2\endcsname{\rm}
\expandafter\def\csname list@F3\endcsname{\rm}
\expandafter\def\csname list@F4\endcsname{\rm}
\expandafter\def\csname list@F5\endcsname{\rm}
\newcount\listlevel@
\listlevel@\z@
\def\list@@C{\csname list@C\number\listlevel@\endcsname}
\def\list@@P{\csname list@P\number\listlevel@\endcsname}
\def\list@@Q{\csname list@Q\number\listlevel@\endcsname}
\def\list@@S{\csname list@S\number\listlevel@\endcsname}
\def\list@@N{\csname list@N\number\listlevel@\endcsname}
\def\list@@F{\csname list@F\number\listlevel@\endcsname}
\newif\iffirstitemi@
\newif\iffirstitemii@
\newif\iffirstitemiii@
\newif\iffirstitemiv@
\newif\iffirstitemv@
\def\Firstitem@true{\csname firstitem\romannumeral\listlevel@
 @true\endcsname}
\def\Firstitem@false{\csname firstitem\romannumeral\listlevel@
 @false\endcsname}
\def\Listm@{\csname listm\romannumeral\listlevel@ @\endcsname}
\def\Item@{\csname item\romannumeral\listlevel@ @\endcsname}
\def\Liste@{\csname liste\romannumeral\listlevel@ @\endcsname}
\newif\iflistcontinue@
\def\keepitem{\listcontinue@true}
\newcount\list@C@
\def\list{%
 \iflistcontinue@\csname list@C1\endcsname\csname list@C@\endcsname\fi
 \global\csname list@C2\endcsname\z@
 \global\csname list@C3\endcsname\z@
 \global\csname list@C4\endcsname\z@
 \global\csname list@C5\endcsname\z@
 \begingroup
 \firstitemi@true
 \listlevel@\@ne
 \def\item{\FN@\item@}%
 \FN@\list@}
\Invalid@\runinitem
\def\list@{\ifx\next\par
 \DN@\par{\FN@\list@}\else
 \ifx\next\runinitem
  \DN@\runinitem{\FN@\runinitem@}\else
  \DN@{\par\dimen@\parskip\parskip\dimen@}\fi\fi\next@}
\newif\ifoutlevel@
\newif\ifrunin@
\def\item@{%
 \ifoutlevel@\Liste@\outlevel@false\fi
 \ifrunin@\runin@false\par
  \dimen@\parskip\parskip\dimen@
  \Listm@\fi
 \iffirstitemi@\listbi@\listmi@\firstitemi@false\else\par\fi
 \iffirstitemii@\listbii@\listmii@\firstitemii@false\else\par\fi
 \iffirstitemiii@\listbiii@\listmiii@\firstitemiii@false\else\par\fi
 \iffirstitemiv@\listbiv@\listmiv@\firstitemiv@false\else\par\fi
 \iffirstitemv@\listbv@\listmv@\firstitemv@false\else\par\fi
 \DN@"##1"{{\let\pre\list@@P\let\post\list@@Q
  \let\style\list@@S\let\numstyle\list@@N
  \vskip-\parskip
  \Item@{\list@@F##1}%
  \noexpands@
  \Qlabel@{##1}}%
  \locallabel@
  \FNSSP@}%
 \DNii@{\global\advance\list@@C\@ne
  {\noexpands@
   \xdef\Thelabel@@@{\number\list@@C}%
   \xdefThelabel@\list@@N
   \xdef\Thelabel@@@@{\list@@P\Thelabel@\list@@Q}%
   \xdefThelabel@@\list@@S
  }%
  \locallabel@
  \vskip-\parskip
  \Item@{\list@@F\thelabel@@}%
  \FN@\pretendspace@}%
 \ifx\next"\expandafter\next@\else\expandafter\nextii@\fi}
\def\runinitem@{%
  \runin@true
  \Firstitem@false
  \DN@"##1"{{\let\pre\list@@P\let\post\list@@Q
   \let\style\list@@S\let\numstyle\list@@N
   \unskip\space{\list@@F##1} %
   \noexpands@
   \Qlabel@{##1}}%
   \locallabel@
   \ignorespaces}%
  \DNii@{\global\advance\list@@C\@ne
   {\noexpands@
    \xdef\Thelabel@@@{\number\list@@C}%
    \xdefThelabel@\list@@N
    \xdef\Thelabel@@@@{\list@@P\Thelabel@\list@@Q}%
    \xdefThelabel@@\list@@S
   }%
   \locallabel@
   \unskip\space{\list@@F\thelabel@@} }%
  \ifx\next"\expandafter\next@\else\expandafter\nextii@\fi}
\def\inlevel{\ifnum\listlevel@=5
 \DN@{\Err@{Already 5 levels down}}\else
 \DN@{\begingroup\advance\listlevel@\@ne
 \Firstitem@true\FN@\inlevel@}\fi\next@}
\def\inlevel@{\ifx\next\par
 \DN@\par{\FN@\inlevel@}\else
 \ifx\next\runinitem
  \DN@\runinitem{\FN@\runinitem@}\else
  \let\next@\relax\fi\fi\next@}
\def\outlevel{\ifnum\listlevel@=\@ne
 \Err@{At top level}\else
 \par\global\list@@C\z@\endgroup\outlevel@true\fi}
\def\endlist{%
 \expandafter\global\csname list@C@\endcsname\csname list@C1\endcsname
 \par
 \global\toks\@ne{}\count@\listlevel@
 {\loop
  \ifnum\count@>\z@\global\toks\@ne\expandafter{\the\toks\@ne\endgroup}%
  \advance\count@\m@ne
  \repeat}%
 \the\toks\@ne
 \liste@
 \listcontinue@false\global\csname list@C1\endcsname\z@
 \vskip-\parskip
 \noindent@@
 \FN@\pretendspace@}
\newif\iffirstdescribe@
\def\describe{\par
 \begingroup\firstdescribe@true
 \def\item##1{%
  \iffirstdescribe@\penalty50 \medskip\vskip-\parskip
  \firstdescribe@false\else\par\fi
  \noindent@@\hangindent2pc\hangafter\@ne
  {\bf##1}\hskip.5em}}

\Invalid@\pullin
\Invalid@\pullinmore
\newif\iffirstpull@
\def\margins{\par\begingroup\firstpull@true
 \def\pullin##1##2{\par
  \iffirstpull@\firstpull@false\else\endgroup\fi
  \begingroup\DN@{##1}%
  \ifx\next@\empty\leftskip\z@\else\ifx\next@\space\leftskip\z@
  \else\leftskip##1\fi\fi
  \DN@{##2}\ifx\next@\empty\rightskip\z@\else\ifx\next@\space
  \rightskip\z@\else\rightskip##2\fi\fi\ignorespaces}%
 \def\pullinmore##1##2{\par
  \xdef\Next@{\leftskip\the\leftskip\relax\rightskip\the\rightskip\relax}%
  \iffirstpull@\firstpull@false\else\endgroup\fi
  \begingroup\Next@
  \DN@{##1}%
  \ifx\next@\empty\else\ifx\next@\space\else\advance\leftskip##1\fi\fi
  \DN@{##2}\ifx\next@\empty\else\ifx\next@\space\else
  \advance\rightskip##2\fi\fi\ignorespaces}}

\newif\ifnopunct@
\newif\ifnospace@
\newif\ifoverlong@
\let\nofrillslist@\empty
\let\overlonglist@\empty
\def\nopunct{\nopunct@true\FN@\nopunct@}
\def\nospace{\nospace@true\FN@\nospace@}
\def\overlong{\overlong@true\FN@\overlong@}
\def\nopunct@{\ifx\next\nospace
 \DN@\nospace{\nospace@true\FN@\nopnos@}\else\ifx\next\overlong
 \DN@\overlong{\overlong@true\FN@\nopol@}\else
 \let\next@\nopunct@@\fi\fi\next@}
\def\nopunct@@#1{\ismember@\nofrillslist@#1%
 \iftest@\let\next@#1\else
 \DN@{\nopunct@false\Err@{\noexpand\nopunct can't be used with
 \string#1}#1}\fi\next@}
\def\nospace@{\ifx\next\nopunct
 \DN@\nopunct{\nopunct@true\FN@\nopnos@}\else\ifx\next\overlong
 \DN@\overlong{\overlong@true\FN@\nosol@}\else
 \let\next@\nospace@@\fi\fi\next@}
\def\nospace@@#1{\ismember@\nofrillslist@#1%
 \iftest@\let\next@#1\else
 \DN@{\nospace@false\Err@{\noexpand\nospace can't be used with
 \string#1}#1}\fi\next@}
\def\overlong@{\ifx\next\nopunct
 \DN@\nopunct{\nopunct@true\FN@\nopol@}\else\ifx\next\nospace
 \DN@\nospace{\nospace@true\FN@\nosol@}\else
 \let\next@\overlong@@\fi\fi\next@}
\def\overlong@@#1{\ismember@\overlonglist@#1%
 \iftest@\let\next@#1\else
 \DN@{\overlong@false\Err@{\noexpand\overlong can't be used with
 \string#1}#1}\fi\next@}
\def\nopnos@{\ifx\next\overlong
 \DN@\overlong{\overlong@true\nopnosol@}\else
 \let\next@\nopnos@@\fi\next@}
\def\nopol@{\ifx\next\nospace
 \DN@\nospace{\nospace@true\nopnosol@}\else
 \let\next@\nopol@@\fi\next@}
\def\nosol@{\ifx\next\nopunct
 \DN@\nopunct{\nopunct@true\nopnosol@}\else
 \let\next@\nosol@@\fi\next@}
\def\nopnos@@#1{\ismember@\nofrillslist@#1%
 \iftest@\let\next@#1\else
 \DN@{\nopunct@false\nospace@false
  \Err@{\noexpand\nopunct\noexpand\nospace
   can't be used with \string#1}#1}\fi\next@}
\def\testii@#1{\ismember@\nofrillslist@#1%
 \iftest@\let\nextiii@ T\else\let\nextiii@ F\fi
 \ismember@\overlonglist@#1%
 \iftest@\let\nextiv@ T\else\let\nextiv@ F\fi
 \test@false\if\nextiii@ T\if\nextiv@ T\test@true\fi\fi}
\def\nopol@@#1{\testii@{#1}%
 \iftest@\let\next@#1%
 \else\DN@{\if\nextiii@ T\else\nopunct@false\fi
  \if\nextiv@ T\else\overlong@false\fi
  \Err@{\if\nextiii@ T\else\noexpand\nopunct\fi
  \if\nextiv@ T\else\noexpand\overlong\fi can't be used
  with \string#1}#1}\fi\next@}
\def\nosol@@#1{\testii@{#1}%
 \iftest@\let\next@#1%
 \else\DN@{\if\nextiii@ T\else\nospace@false\fi
  \if\nextiv@ T\else\overlong@false\fi
  \Err@{\if\nextiii@ T\else\noexpand\nospace\fi
  \if\nextiv@ T\else\noexpand\overlong\fi can't be used
  with \string#1}#1}\fi\next@}
\def\nopnosol@#1{\testii@{#1}%
 \iftest@\let\next@#1%
 \else\DN@{\if\nextiii@ T\else\nopunct@false\nospace@false\fi
  \if\nextiv@ T\else\overlong@false\fi
  \Err@{\if\nextiii@ T\else\noexpand\nopunct\noexpand\nospace\fi
  \if\nextiv@ T\else\noexpand\overlong\fi can't be used
  with \string#1}#1}\fi\next@}
\def\punct@#1{\ifnopunct@\else#1\fi}
\def\addspace@#1{\ifnospace@\else#1\fi}
\def\hss@{\ifoverlong@\z@ plus\@m\p@ minus\@m\p@
 \else \z@ plus\@m\p@\fi}
\rightadd@\demo\to\nofrillslist@
\newif\ifclaim@
\def\exxx@{\expandafter\expandafter\expandafter\eat@\expandafter\string}
\let\colon@:
\def\demo#1{\ifclaim@
 \Err@{Previous \expandafter\noexpand\claimtype@ has
  no matching \string\end\exxx@\claimtype@}%
 \let\next@\relax
 \else
  \par
  \ifdim\lastskip<\smallskipamount\removelastskip\smallskip\fi
  \begingroup
  \noindent@@{\smc\ignorespaces#1\unskip
   \punct@{\null\colon@}\addspace@\enspace}%
  \nopunct@false\nospace@false
  \rm
  \DN@{\FNSSP@}%
 \fi
 \next@}
\def\enddemo{\par\endgroup\nopunct@false\nospace@false\smallskip}
\rightadd@\claim\to\nofrillslist@
\def\claim@F{\smc}
\def\claim@@@F{\csname\exxx@\claimtype@ @F\endcsname}
\def\claimformat@#1#2#3{%
 \medbreak\noindent@@{\smc#1 {\claim@@@F#2} #3%
 \punct@{\null.}\addspace@\enspace}\sl}
\def\claimformat@@#1#2{\claimformat@{\ignorespaces#1\unskip}%
 {\ifx\thelabel@@\empty\unskip\else\thelabel@@\fi}%
 {\ignorespaces#2\unskip}%
 \let\Claimformat@@\claimformat@@\FNSSP@}
\let\Claimformat@@\claimformat@@
\def\claim@@@P{\csname\exxx@\claimtype@ @P\endcsname}
\def\claim@@@Q{\csname\exxx@\claimtype@ @Q\endcsname}
\def\claim@@@S{\csname\exxx@\claimtype@ @S\endcsname}
\def\claim@@@N{\csname\exxx@\claimtype@ @N\endcsname}
\def\claim@@@C{\csname claim@C\claimclass@\endcsname}
\newcount\claim@C
\claim@C\z@
\let\claim@P\empty
\let\claim@Q\empty
\def\claim@S#1{#1\/}
\let\claim@N\arabic
\def\claim{\claim@true\let\claimclass@\empty
 \def\claimtype@{\claim}\FN@\claim@}
\def\claim@{%
 \ifx\next\c
  \let\next@\claim@c
 \else
  \ifx\next"%
   \let\next@\claim@q
  \else
   \begingroup\global\advance\claim@C\@ne
   {\noexpands@
    \xdef\Thelabel@@@{\number\claim@C}%
    \xdefThelabel@\claim@N
    \xdef\Thelabel@@@@{\claim@P\Thelabel@\claim@Q}%
    \xdefThelabel@@\claim@S
   }%
   \locallabel@
   \let\next@\Claimformat@@
  \fi
 \fi
 \next@}
\def\claim@c\c#1{\claim@true\begingroup
 \expandafter
 \ifx\csname claim@C#1\endcsname\relax
  \expandafter\newcount@\csname claim@C#1\endcsname
  \global\csname claim@C#1\endcsname\@ne
 \else
  \global\advance\csname claim@C#1\endcsname\@ne
 \fi
 \def\claimclass@{#1}%
 {\noexpands@
  \xdef\Thelabel@@@{\number\claim@@@C}%
  \xdefThelabel@\claim@@@N
  \xdef\Thelabel@@@@{\claim@@@P\Thelabel@\claim@@@Q}%
  \xdefThelabel@@\claim@@@S
 }%
 \locallabel@
 \FNSS@\claim@c@}
\def\claim@q"#1"{\begingroup
 {\let\pre\claim@@@P\let\post\claim@@@Q
  \let\style\claim@@@S\let\numstyle\claim@@@N
  \noexpands@
  \Qlabel@{#1}}%
 \locallabel@
 \FNSS@\claim@q@}
\def\claim@c@{\ifx\next"%
 \global\advance\claim@@@C\m@ne\let\next@\claim@cq
 \else\let\next@\Claimformat@@\fi\next@}
\def\claim@cq"#1"{{\let\pre\claim@@@P\let\post\claim@@@Q
 \let\style\claim@@@S\let\numstyle\claim@@@N
 \noexpands@
 \Qlabel@{#1}}%
 \locallabel@
 \FNSS@\Claimformat@@}
\def\claim@q@{\ifx\next\c\expandafter\claim@qc
 \else\expandafter\Claimformat@@\fi}
\def\claim@qc\c#1{\expandafter\ifx\csname claim@C#1\endcsname\relax
 \expandafter\newcount@\csname claim@C#1\endcsname
 \global\csname claim@C#1\endcsname\z@\fi
 \FNSS@\Claimformat@@}
\def\endclaim{\endgroup\claim@false\nopunct@false\nospace@false
 \let\Claimformat@@\claimformat@@\medbreak}
\Invalid@\claimclause
\def\newclaim{\FN@\newclaim@}
\def\newclaim@{\ifx\next\claimclause
 \DN@\claimclause##1{\newclaim@@{##1}}\else
 \DN@{\newclaim@@\relax}\fi\next@}
\def\claimlist@{\\\claim}
\newtoks\claim@i
\newtoks\claim@v
\let\noclaimclause@=F
\def\newclaim@@#1#2#3\c#4#5{\define#2{}%
 \rightadd@#2\to\claimlist@\rightadd@#2\to\nofrillslist@%
 \expandafter\def\csname\exstring@#2@P\endcsname{\claim@P}%
 \expandafter\def\csname\exstring@#2@Q\endcsname{\claim@Q}%
 \expandafter\def\csname\exstring@#2@S\endcsname{\claim@S}%
 \expandafter\def\csname\exstring@#2@N\endcsname{\claim@N}%
 \expandafter\def\csname\exstring@#2@F\endcsname{\claim@F}%
 \expandafter\def\csname end\exstring@#2\endcsname{\endclaim}%
 \expandafter\ifx\csname claim@C#4\endcsname\relax
  \expandafter\newcount@\csname claim@C#4\endcsname
  \global\csname claim@C#4\endcsname\z@\fi
 \edef\next@{\let\csname\exstring@#2@C\endcsname
   \csname claim@C#4\endcsname}\next@
 \def#2{\ifx\noclaimclause@ T\else#1\fi
  \global\claim@i{#1}\gdef\claim@iv{#4}\global\claim@v{#5}%
  \def\claimtype@{#2}\def\Claimformat@@{\claimformat@@{#5}}\claim@c\c{#4}}}
\def\shortenclaim#1#2{\define#2{}%
 \ismember@\claimlist@#1%
 \iftest@
  \rightadd@#2\to\nofrillslist@%
  \expandafter\def\csname\exstring@#2@P\endcsname
   {\csname\exstring@#1@P\endcsname}%
  \expandafter\def\csname\exstring@#2@Q\endcsname
   {\csname\exstring@#1@Q\endcsname}%
  \expandafter\def\csname\exstring@#2@S\endcsname
   {\csname\exstring@#1@S\endcsname}%
  \expandafter\def\csname\exstring@#2@N\endcsname
   {\csname\exstring@#1@N\endcsname}%
  \expandafter\def\csname\exstring@#2@F\endcsname
   {\csname\exstring@#1@F\endcsname}%
  \expandafter\def\csname end\exstring@#2\endcsname{\endclaim}%
  \edef\next@{\let\csname\exstring@#2@C\endcsname
    \csname claim\exstring@#1C\endcsname}\next@
  \setbox\z@\vbox{\let\noclaimclause@ T#1""\relax\endgroup}%
  \edef#2{\the\claim@i
   \def\noexpand\claimtype@{\noexpand#2}%
   \def\noexpand\Claimformat@@{\noexpand\claimformat@@{\the\claim@v}\relax}%
   \noexpand\claim@c\noexpand\c{\claim@iv}}%
 \else
  \Err@{\noexpand#1not yet created by \string\newclaim}%
 \fi}
\def\classtest@#1{\DN@{#1}\ifx\next@\claimclass@
 \test@true\else\test@false\fi}
\def\typetest@#1{\DN@{#1}\ifx\next@\claimtype@\test@true\else
  \test@false\fi}
\newif\iftoc@
\def\tocfile{\iftoc@\else\alloc@@7\write\chardef\sixt@@n\toc@
 \immediate\openout\toc@=\jobname.toc
 \alloc@@7\write\chardef\sixt@@n\tic@
 \immediate\openout\tic@=\jobname.tic
 \global\toc@true\fi}
\rightadd@\hl\to\nofrillslist@
\rightadd@\HL\to\overlonglist@
\def\HL@@C{\csname HL@C\HLlevel@\endcsname}
\def\HL@@P{\csname HL@P\HLlevel@\endcsname}
\def\HL@@Q{\csname HL@Q\HLlevel@\endcsname}
\def\HL@@S{\csname HL@S\HLlevel@\endcsname}
\def\HL@@N{\csname HL@N\HLlevel@\endcsname}
\def\HL@@F{\csname HL@F\HLlevel@\endcsname}
\def\HL@@@C{\csname\exxx@\HLtype@ @C\endcsname}
\def\HL@@@P{\csname\exxx@\HLtype@ @P\endcsname}
\def\HL@@@Q{\csname\exxx@\HLtype@ @Q\endcsname}
\def\HL@@@S{\csname\exxx@\HLtype@ @S\endcsname}
\def\HL@@@N{\csname\exxx@\HLtype@ @N\endcsname}
\def\HL#1{\expandafter
 \ifx\csname HL@C#1\endcsname\relax
  \DN@{\Err@{\string\HL#1 not defined in this style}}%
 \else
  \DN@{\gdef\HLlevel@{#1}\def\HLname@{\HL{#1}}\let\HLtype@\relax\FNSS@\HL@}%
 \fi
 \next@}%
\newif\ifquoted@
\let\aftertoc@\relax
\def\HL@{%
 \DN@"##1"##2\endHL{\def\entry@{##2}\quoted@true
  {\noexpands@
  \ifx\HLtype@\relax
   \let\pre\HL@@P\let\post\HL@@Q\let\style\HL@@S\let\numstyle\HL@@N
  \else
   \let\pre\HL@@@P\let\post\HL@@@Q\let\style\HL@@@S\let\numstyle\HL@@@N
  \fi
  \Qlabel@{##1}\let\style\relax\xdef\Qlabel@@@@{##1}%
  \xdef\Thepref@{\Thelabel@@@@}}%
  \csname HL@\HLlevel@\endcsname##2\endHL
  \let\pref\Thepref@
  \csname HL@I\HLlevel@\endcsname
  \csname HL@J\HLlevel@\endcsname
  \let\pref\pref@
  \HLtoc@	
  \aftertoc@
  \let\aftertoc@\relax\overlong@false}%
 \DNii@##1\endHL{\def\entry@{##1}\quoted@false
  {\noexpands@
  \ifx\HLtype@\relax
   \global\advance\HL@@C\@ne
   \xdef\Thelabel@@@{\number\HL@@C}%
   \xdefThelabel@{\HL@@N}%
   \xdef\Thelabel@@@@{\HL@@P\Thelabel@\HL@@Q}%
   \xdefThelabel@@{\HL@@S}%
  \else
   \global\advance\HL@@@C\@ne
   \xdef\Thelabel@@@{\number\HL@@@C}%
   \xdefThelabel@{\HL@@@N}%
   \xdef\Thelabel@@@@{\HL@@@P\Thelabel@\HL@@@Q}%
   \xdefThelabel@@{\HL@@@S}%
  \fi
  \xdef\Thepref@{\Thelabel@@@@}}%
  \csname HL@\HLlevel@\endcsname##1\endHL
  \let\pref\Thepref@
  \csname HL@I\HLlevel@\endcsname
  \csname HL@J\HLlevel@\endcsname
  \let\pref\pref@
  \HLtoc@
  \aftertoc@
  \let\aftertoc@\relax\overlong@false}%
 \ifx\next"\expandafter\next@\else\expandafter\nextii@\fi}%
\Invalid@\endHL
\def\hl@@C{\csname hl@C\hllevel@\endcsname}
\def\hl@@P{\csname hl@P\hllevel@\endcsname}
\def\hl@@Q{\csname hl@Q\hllevel@\endcsname}
\def\hl@@S{\csname hl@S\hllevel@\endcsname}
\def\hl@@N{\csname hl@N\hllevel@\endcsname}
\def\hl@@F{\csname hl@F\hllevel@\endcsname}
\def\hl@@@C{\csname\exxx@\hltype@ @C\endcsname}
\def\hl@@@P{\csname\exxx@\hltype@ @P\endcsname}
\def\hl@@@Q{\csname\exxx@\hltype@ @Q\endcsname}
\def\hl@@@S{\csname\exxx@\hltype@ @S\endcsname}
\def\hl@@@N{\csname\exxx@\hltype@ @N\endcsname}
\def\hl#1{\expandafter
 \ifx\csname hl@C#1\endcsname\relax
  \DN@{\Err@{\string\hl#1 not defined in this style}}%
 \else
  \DN@{\gdef\hllevel@{#1}\def\hlname@{\hl{#1}}\let\hltype@\relax\FNSS@\hl@}%
 \fi
 \next@}
\def\hl@{%
 \DN@"##1"##2{\def\entry@{##2}\quoted@true
  {\noexpands@
  \ifx\hltype@\relax
   \let\pre\hl@@P\let\post\hl@@Q\let\style\hl@@S\let\numstyle\hl@@N
  \else
   \let\pre\hl@@@P\let\post\hl@@@Q\let\style\hl@@@S\let\numstyle\hl@@@N
  \fi
  \Qlabel@{##1}\let\style\relax\xdef\Qlabel@@@@{##1}%
  \xdef\Thepref@{\Thelabel@@@@}}%
  \csname hl@\hllevel@\endcsname{##2}%
  \let\pref\Thepref@
  \csname hl@I\hllevel@\endcsname
  \csname hl@J\hllevel@\endcsname
  \let\pref\pref@
  \hltoc@
  \aftertoc@
  \let\aftertoc@\relax\nopunct@false\nospace@false\FNSSP@}%
 \DNii@##1{\def\entry@{##1}\quoted@false
  {\noexpands@
  \ifx\hltype@\relax
   \global\advance\hl@@C\@ne
   \xdef\Thelabel@@@{\number\hl@@C}%
   \xdefThelabel@{\hl@@N}%
   \xdef\Thelabel@@@@{\hl@@P\Thelabel@\hl@@Q}%
   \xdefThelabel@@{\hl@@S}%
  \else
   \global\advance\hl@@@C\@ne
   \xdef\Thelabel@@@{\number\hl@@@C}%
   \xdefThelabel@{\hl@@@N}%
   \xdef\Thelabel@@@@{\hl@@@P\Thelabel@\hl@@@Q}%
   \xdefThelabel@@{\hl@@@S}%
  \fi
  \xdef\Thepref@{\Thelabel@@@@}}%
  \csname hl@\hllevel@\endcsname{##1}%
  \let\pref\Thepref@
  \csname hl@I\hllevel@\endcsname
  \csname hl@J\hllevel@\endcsname
  \let\pref\pref@
  \hltoc@
  \aftertoc@
  \let\aftertoc@\relax\nopunct@false\nospace@false\FNSSP@}%
 \ifx\next"\expandafter\next@\else\expandafter\nextii@\fi}%
\def\six@#1#2 #3 #4 #5 #6 #7 {\DN@{#2}\ifx\next@\empty
 \DN@##1\six@{}\else
 \write#1{ #2 #3 #4 #5 #6 #7}\DN@{\six@#1}\fi
 \next@}
\def\Sixtoc@{\ifx\macdef@\empty\else
 \DN@##1##2\next@{\def\macdef@{##1##2}}%
 \expandafter\next@\macdef@\next@
 \edef\next@
  {\noexpand\six@\toc@\macdef@
  \space\space\space\space\space\space\space\space\space\space\space\space
  \noexpand\six@}%
 \next@\let\macdef@\relax\fi}
\def\QorThelabel@@@@{\ifquoted@
 \noexpand\noexpand\noexpand"\Qlabel@@@@\noexpand\noexpand\noexpand"\else
 \Thelabel@@@@\fi}
\def\HLtoc@{%
 \iftoc@
 \expandafter\expandafter\expandafter\unmacro@
  \expandafter\meaning\csname HL@W\HLlevel@\endcsname\unmacro@
  {\noexpands@\let\style\relax
   \edef\next@{\write\toc@{\noexpand\noexpand\expandafter\noexpand\HLname@
   {\macdef@}{\QorThelabel@@@@}}}%
  \next@}%
  \expandafter\unmacro@\meaning\entry@\unmacro@
  \Sixtoc@
  \write\toc@{\noexpand\Page{\number\pageno}{\page@N}%
   {\page@P}{\page@Q}^^J}%
 \fi}
\def\hltoc@{%
 \iftoc@
 \expandafter\expandafter\expandafter\unmacro@
  \expandafter\meaning\csname hl@W\hllevel@\endcsname\unmacro@
  {\noexpands@\let\style\relax
  \edef\next@{\write\toc@{%
   \ifnopunct@\noexpand\noexpand\noexpand\nopunct\fi
   \ifnospace@\noexpand\noexpand\noexpand\nospace\fi
   \noexpand\noexpand\expandafter\noexpand\hlname@
   {\macdef@}{\QorThelabel@@@@}}}%
  \next@}%
  \expandafter\unmacro@\meaning\entry@\unmacro@
  \Sixtoc@
  \write\toc@{\noexpand\Page{\number\pageno}{\page@N}%
   {\page@P}{\page@Q}^^J}%
 \fi}
\def\mainfile#1{\def\mainfile@{#1}}
\def\checkmainfile@{\ifx\mainfile@\undefined
 \Err@{No \noexpand\mainfile specified}\fi}
\expandafter\newcount@\csname HL@C1\endcsname
\csname HL@C1\endcsname\z@
\expandafter\def\csname HL@S1\endcsname#1{#1\null.}
\expandafter\let\csname HL@N1\endcsname\arabic
\expandafter\let\csname HL@P1\endcsname\empty
\expandafter\let\csname HL@Q1\endcsname\empty
\expandafter\def\csname HL@F1\endcsname{\bf}
\expandafter\let\csname HL@W1\endcsname\empty
\expandafter\newcount@\csname hl@C1\endcsname
\csname hl@C1\endcsname\z@
\expandafter\def\csname hl@S1\endcsname#1{#1\/}
\expandafter\let\csname hl@N1\endcsname\arabic
\expandafter\let\csname hl@P1\endcsname\empty
\expandafter\let\csname hl@Q1\endcsname\empty
\expandafter\def\csname hl@F1\endcsname{\bf}
\expandafter\let\csname hl@W1\endcsname\empty
\expandafter\def\csname HL@1\endcsname#1\endHL{\bigbreak
 {\locallabel@
  \global\setbox\@ne\vbox{\Let@\tabskip\hss@
  \halign to\hsize{\bf\hfil\ignorespaces##\unskip\hfil\cr
  \expandafter\ifx\csname HL@W1\endcsname\empty\else
   \csname HL@W1\endcsname\space\fi
  {\HL@@F\ifx\thelabel@@\empty\else\thelabel@@\space\fi}%
  \ignorespaces#1\crcr}}%
  }%
 \unvbox\@ne\nobreak\medskip}
\expandafter\def\csname hl@1\endcsname#1{\medbreak\noindent@@
 {\locallabel@
 \bf{\hl@@F\ifx\thelabel@@\empty\else\thelabel@@\space\fi}%
 \ignorespaces#1\unskip\punct@{\null.}\addspace@\enspace}}
\expandafter\def\csname HL@I1\endcsname{\Reset\hl1{1}%
 \ifx\pref\empty\newpre\hl1{}\else\newpre\hl1{\pref.}\fi}
\def\NameHL#1#2{\define#2{}%
 \expandafter\ifx\csname HL@R#1\endcsname\relax
 \else
  \def\nextiv@{\let\nextiii@}%
  \expandafter\nextiv@\csname HL@R#1\endcsname
  \expandafter\let\nextiii@\undefined
  \expandafter\let\csname\exxx@\nextiii@ @C\endcsname\relax
  \expandafter\let\csname\exxx@\nextiii@ @P\endcsname\relax
  \expandafter\let\csname\exxx@\nextiii@ @Q\endcsname\relax
  \expandafter\let\csname\exxx@\nextiii@ @S\endcsname\relax
  \expandafter\let\csname\exxx@\nextiii@ @N\endcsname\relax
  \expandafter\let\csname\exxx@\nextiii@ @F\endcsname\relax
  \expandafter\let\csname\exxx@\nextiii@ @W\endcsname\relax
  \expandafter\let\csname end\exxx@\nextiii@\endcsname\undefined
 \fi
 \expandafter\gdef\csname HL@R#1\endcsname{#2}%
 \expandafter\gdef\csname\exstring@#2@R\endcsname{{HL}{#1}}%
 \iftoc@\write\toc@{\noexpand\NameHL#1\noexpand#2^^J}\fi
 \rightadd@#2\to\overlonglist@
 \edef\next@{\let\csname\exstring@#2@C\endcsname\expandafter\noexpand
  \csname HL@C#1\endcsname}\next@
 \edef\next@{\let\csname\exstring@#2@P\endcsname\expandafter\noexpand
  \csname HL@P#1\endcsname}\next@
 \edef\next@{\let\csname\exstring@#2@Q\endcsname\expandafter\noexpand
  \csname HL@Q#1\endcsname}\next@
 \edef\next@{\let\csname\exstring@#2@S\endcsname\expandafter\noexpand
  \csname HL@S#1\endcsname}\next@
 \edef\next@{\let\csname\exstring@#2@N\endcsname\expandafter\noexpand
  \csname HL@N#1\endcsname}\next@
 \edef\next@{\let\csname\exstring@#2@F\endcsname\expandafter\noexpand
  \csname HL@F#1\endcsname}\next@
 \edef\next@{\let\csname\exstring@#2@W\endcsname\expandafter\noexpand
  \csname HL@W#1\endcsname}\next@
 \edef\next@{\def\noexpand#2####1\expandafter\noexpand
  \csname end\exstring@#2\endcsname
  {\def\noexpand\HLtype@{\noexpand#2}%
   \def\noexpand\HLname@{\noexpand#2}%
   \gdef\noexpand\HLlevel@{#1}%
   \noexpand\FNSS@\noexpand\HL@####1\noexpand\endHL}}%
  \next@
 \edef\next@{\noexpand\Invalid@\expandafter\noexpand
  \csname end\exstring@#2\endcsname}%
 \next@}
\def\Namehl#1#2{\define#2{}%
 \expandafter\ifx\csname hl@R#1\endcsname\relax
 \else
  \def\nextiv@{\let\nextiii@}%
  \expandafter\nextiv@\csname hl@R#1\endcsname
  \expandafter\let\nextiii@\undefined
  \expandafter\let\csname\exxx@\nextiii@ @C\endcsname\relax
  \expandafter\let\csname\exxx@\nextiii@ @P\endcsname\relax
  \expandafter\let\csname\exxx@\nextiii@ @Q\endcsname\relax
  \expandafter\let\csname\exxx@\nextiii@ @S\endcsname\relax
  \expandafter\let\csname\exxx@\nextiii@ @N\endcsname\relax
  \expandafter\let\csname\exxx@\nextiii@ @F\endcsname\relax
  \expandafter\let\csname\exxx@\nextiii@ @W\endcsname\relax
 \fi
 \expandafter\gdef\csname hl@R#1\endcsname{#2}%
 \expandafter\gdef\csname\exstring@#2@R\endcsname{{hl}{#1}}%
 \iftoc@\write\toc@{\noexpand\Namehl#1\noexpand#2^^J}\fi
 \rightadd@#2\to\nofrillslist@%
 \edef\next@{\let\csname\exstring@#2@C\endcsname\expandafter\noexpand
  \csname hl@C#1\endcsname}\next@
 \edef\next@{\let\csname\exstring@#2@P\endcsname\expandafter\noexpand
  \csname hl@P#1\endcsname}\next@
 \edef\next@{\let\csname\exstring@#2@Q\endcsname\expandafter\noexpand
  \csname hl@Q#1\endcsname}\next@
 \edef\next@{\let\csname\exstring@#2@S\endcsname\expandafter\noexpand
  \csname hl@S#1\endcsname}\next@
 \edef\next@{\let\csname\exstring@#2@N\endcsname\expandafter\noexpand
  \csname hl@N#1\endcsname}\next@
 \edef\next@{\let\csname\exstring@#2@F\endcsname\expandafter\noexpand
  \csname hl@F#1\endcsname}\next@
 \edef\next@{\let\csname\exstring@#2@W\endcsname\expandafter\noexpand
  \csname hl@W#1\endcsname}\next@
 \edef\next@{\def\noexpand#2{%
  \def\noexpand\hltype@{\noexpand#2}%
  \def\noexpand\hlname@{\noexpand#2}%
  \gdef\noexpand\hllevel@{#1}%
  \noexpand\FNSS@\noexpand\hl@}}%
 \next@}%
\def\Initialize{\FN@\Init@}
\def\Init@{\ifx\next\HL\let\next@\InitH@\else\ifx\next\hl\let\next@\InitH@
  \else\let\next@\InitS@\fi\fi\next@}
\def\InitH@#1#2{\expandafter\ifx\csname\exstring@#1@C#2\endcsname\relax
 \DN@{\Err@{\noexpand#1level #2 not defined in this style}}\else
 \DN@{\expandafter\gdef\csname\exstring@#1@J#2\endcsname}\fi\next@}
\def\InitC@#1#2{\edef\nextii@{\expandafter\noexpand\csname#1\endcsname{#2}}}
\def\InitS@#1{\expandafter\ifx\csname\exstring@#1@R\endcsname\relax
 \Err@{\noexpand#1not defined in this style}\let\next@\relax\else
 \DN@{\let\next@}\expandafter\next@\csname\exstring@#1@R\endcsname
 \expandafter\InitC@\next@
 \DN@{\expandafter\InitH@\nextii@}\fi\next@}
\def\value#1{\expandafter
 \ifx\csname\exstring@#1@C\endcsname\relax
  \expandafter\ifx\csname\exstring@#1@C1\endcsname\relax
   \DN@{\Err@{\noexpand\value can't be used with \string#1}}%
  \else
   \DN@{\value@#1}%
  \fi
 \else
  \DN@{\number\csname\exstring@#1@C\endcsname\relax}%
 \fi
 \next@}
\def\value@#1#2{\expandafter
 \ifx\csname\exstring@#1@C#2\endcsname\relax
  \DN@{\Err@{\string\value\string#1 can't be followed by \string#2}}%
 \else
  \DN@{\number\csname\exstring@#1@C#2\endcsname\relax}%
 \fi
 \next@}
\newcount\Value
\def\Evaluate#1{\expandafter
 \ifx\csname\exstring@#1@C\endcsname\relax
  \expandafter\ifx\csname\exstring@#1@C1\endcsname\relax
   \DN@{\Err@{\noexpand\Evaluate can't be used with \string#1}}%
  \else
   \DN@{\Evaluate@#1}%
  \fi
 \else
  \DN@{\global\Value\csname\exstring@#1@C\endcsname}%
 \fi
 \next@}
\def\Evaluate@#1#2{\expandafter
 \ifx\csname\exstring@#1@C#2\endcsname\relax
  \DN@{\Err@{\string\Evaluate\string#1 can't be followed by \string#2}}%
 \else
  \DN@{\global\Value\csname\exstring@#1@C#2\endcsname}%
 \fi\next@}
\def\pre#1{\expandafter
 \ifx\csname\exstring@#1@P\endcsname\relax
  \expandafter\ifx\csname\exstring@#1@P1\endcsname\relax
   \DN@{\Err@{\noexpand\pre can't be used with \string#1}}%
  \else
   \DN@{\pre@#1}%
  \fi
 \else
  \DN@{{\csname\exstring@#1@P\endcsname}}%
 \fi
 \next@}
\def\pre@#1#2{\expandafter
 \ifx\csname\exstring@#1@P#2\endcsname\relax
  \DN@{\Err@{\string\pre\string#1 can't be followed by \string#2}}%
 \else
  \DN@{{\csname\exstring@#1@P#2\endcsname}}%
 \fi
 \next@}
\def\post#1{\expandafter
 \ifx\csname\exstring@#1@Q\endcsname\relax
  \expandafter\ifx\csname\exstring@#1@Q1\endcsname\relax
   \DN@{\Err@{\noexpand\post can't be used with \string#1}}%
  \else
   \DN@{\post@#1}%
  \fi
 \else
  \DN@{{\csname\exstring@#1@Q\endcsname}}%
 \fi
 \next@}
\def\post@#1#2{\expandafter
 \ifx\csname\exstring@#1@Q#2\endcsname\relax
  \DN@{\Err@{\string\post\string#1 can't be followed by \string#2}}%
 \else
  \DN@{{\csname\exstring@#1@Q#2\endcsname}}%
 \fi
 \next@}
\def\style#1{\expandafter
 \ifx\csname\exstring@#1@S\endcsname\relax
  \expandafter\ifx\csname\exstring@#1@S1\endcsname\relax
   \DN@{\Err@{\noexpand\style can't be used with \string#1}}%
  \else
   \DN@{\style@#1}%
  \fi
 \else
  \DN@{\csname\exstring@#1@S\endcsname}%
 \fi
 \next@}
\def\style@#1#2{\expandafter
 \ifx\csname\exstring@#1@S#2\endcsname\relax
  \DN@{\Err@{\string\style\string#1 can't be followed by \string#2}}%
 \else
  \DN@{\csname\exstring@#1@S#2\endcsname}%
 \fi
 \next@}
\def\fontstyle#1{\expandafter
 \ifx\csname\exstring@#1@F\endcsname\relax
  \expandafter\ifx\csname\exstring@#1@F1\endcsname\relax
   \DN@{\Err@{\noexpand\fontstyle can't be used with \string#1}}%
  \else
   \DN@{\fontstyle@#1}%
  \fi
 \else
  \DN@##1{{\csname\exstring@#1@F\endcsname##1}}%
 \fi
 \next@}
\def\fontstyle@#1#2{\expandafter
 \ifx\csname\exstring@#1@F#2\endcsname\relax
  \DN@{\Err@{\string\fontstyle\string#1 can't be followed by \string#2}}%
 \else
  \DN@##1{{\csname\exstring@#1@F#2\endcsname##1}}%
 \fi
 \next@}
\def\Reset#1{\expandafter
 \ifx\csname\exstring@#1@C\endcsname\relax
  \expandafter\ifx\csname\exstring@#1@C1\endcsname\relax
   \DN@{\Err@{\noexpand\Reset can't be used with \string#1}}%
  \else
   \DN@{\Reset@#1}%
  \fi
 \else
  \DN@##1{\count@##1\relax\ifx#1\page\else\advance\count@\m@ne\fi
   \global\csname\exstring@#1@C\endcsname\count@}%
 \fi
 \next@}
\def\Reset@#1#2{\expandafter
 \ifx\csname\exstring@#1@C#2\endcsname\relax
  \DN@{\Err@{\string\Reset\string#1 can't be followed by \string#2}}%
 \else
  \DN@##1{\count@##1\relax\advance\count@\m@ne
   \global\csname\exstring@#1@C#2\endcsname\count@}%
 \fi
 \next@}
\def\Offset#1{\expandafter
 \ifx\csname\exstring@#1@C\endcsname\relax
  \expandafter\ifx\csname\exstring@#1@C1\endcsname\relax
   \DN@{\Err@{\noexpand\Offset can't be used with \string#1}}%
  \else
   \DN@{\Offset@#1}%
  \fi
 \else
  \DN@##1{\count@##1\relax\advance\count@\m@ne\global\advance
   \csname\exstring@#1@C\endcsname\count@}%
 \fi
 \next@}
\def\Offset@#1#2{\expandafter
 \ifx\csname\exstring@#1@C#2\endcsname\relax
  \DN@{\Err@{\string\Offset\string#1 can't be followed by \string#2}}%
 \else
  \DN@##1{\count@##1\relax\advance\count@\m@ne
   \global\advance\csname\exstring@#1@C#2\endcsname\count@}%
 \fi
 \next@}
\def\getR@#1#2{\def\nextiv@{\let\nextiii@}\expandafter\nextiv@
 \csname\exstring@#1@R#2\endcsname}
\def\letR@#1#2#3{\expandafter\let\csname#1@#3#2\endcsname\Next@}
\def\letR@@#1#2{\expandafter\let\csname\exstring@#1@#2\endcsname\Next@}
\def\newpre#1{\expandafter
 \ifx\csname\exstring@#1@P\endcsname\relax
  \expandafter\ifx\csname\exstring@#1@P1\endcsname\relax
   \DN@{\Err@{\noexpand\newpre can't be used with \string#1}}%
  \else
   \DN@{\newpre@#1}%
  \fi
 \else
  \DN@{%
   \DNii@{%
    \endgroup
    \expandafter\let\csname\exstring@#1@P\endcsname\Next@
    \expandafter\ifx\csname\exstring@#1@R\endcsname\relax\else
    \getR@#1{}\expandafter\letR@\nextiii@ P\fi
    }%
   \begingroup\noexpands@\afterassignment\nextii@\xdef\Next@}%
 \fi
 \next@}
\def\newpre@#1#2{\expandafter
 \ifx\csname\exstring@#1@P#2\endcsname\relax
  \DN@{\Err@{\string\newpre\string#1 can't be followed by \string#2}}%
 \else
  \DN@{%
   \DNii@{%
    \endgroup
    \expandafter\let\csname\exstring@#1@P#2\endcsname\Next@
    \expandafter\ifx\csname\exstring@#1@R#2\endcsname\relax\else
    \getR@#1{#2}\expandafter\letR@@\nextiii@ P\fi
    }%
   \begingroup\noexpands@\afterassignment\nextii@\xdef\Next@}%
 \fi
 \next@}
\def\newpost#1{\expandafter
 \ifx\csname\exstring@#1@Q\endcsname\relax
  \expandafter\ifx\csname\exstring@#1@Q1\endcsname\relax
   \DN@{\Err@{\noexpand\newpost can't be used with \string#1}}%
  \else
   \DN@{\newpost@#1}%
  \fi
 \else
  \DN@{%
   \DNii@{%
    \endgroup
    \expandafter\let\csname\exstring@#1@Q\endcsname\Next@
    \expandafter\ifx\csname\exstring@#1@R\endcsname\relax\else
    \getR@#1{}\expandafter\letR@\nextiii@ Q\fi
    }%
   \begingroup\noexpands@\afterassignment\nextii@\xdef\Next@}%
 \fi
 \next@}
\def\newpost@#1#2{\expandafter
 \ifx\csname\exstring@#1@Q#2\endcsname\relax
  \DN@{\Err@{\string\newpost\string#1 can't be followed by \string#2}}%
 \else
  \DN@{%
   \DNii@{%
    \endgroup
    \expandafter\let\csname\exstring@#1@Q#2\endcsname\Next@
    \expandafter\ifx\csname\exstring@#1@R#2\endcsname\relax\else
    \getR@#1{#2}\expandafter\letR@@\nextiii@ Q\fi
    }%
   \begingroup\noexpands@\afterassignment\nextii@\xdef\Next@}%
 \fi
 \next@}
\def\newstyle#1{\expandafter
 \ifx\csname\exstring@#1@S\endcsname\relax
  \expandafter\ifx\csname\exstring@#1@S1\endcsname\relax
   \DN@{\Err@{\noexpand\newstyle can't be used
    with \string#1}}%
  \else
   \DN@{\newstyle@#1}%
  \fi
 \else
  \DN@{%
   \DNii@{%
    \expandafter\let\csname\exstring@#1@S\endcsname\Next@
    \expandafter\ifx\csname\exstring@#1@R\endcsname\relax\else
    \getR@#1{}\expandafter\letR@\nextiii@ S\fi
    }%
   \afterassignment\nextii@\gdef\Next@}%
 \fi
 \next@}
\def\newstyle@#1#2{\expandafter
 \ifx\csname\exstring@#1@S#2\endcsname\relax
  \DN@{\Err@{\string\newstyle\string#1 can't be followed by
   \string#2}}%
 \else
  \DN@{%
   \DNii@{%
    \expandafter\let\csname\exstring@#1@S#2\endcsname\Next@
    \expandafter\ifx\csname\exstring@#1@R#2\endcsname\relax\else
    \getR@#1{#2}\expandafter\letR@@\nextiii@ S\fi
    }%
   \afterassignment\nextii@\gdef\Next@}%
 \fi
 \next@}
\def\newnumstyle#1{\expandafter
 \ifx\csname\exstring@#1@N\endcsname\relax
  \expandafter\ifx\csname\exstring@#1@N1\endcsname\relax
   \DN@{\Err@{\noexpand\newnumstyle can't be used with
    \string#1}}%
  \else
   \DN@{\newnumstyle@#1}%
  \fi
 \else
  \DN@##1{%
   \gdef\Next@{##1}%
    \expandafter\let\csname\exstring@#1@N\endcsname\Next@
    \expandafter\ifx\csname\exstring@#1@R\endcsname\relax\else
    \getR@#1{}\expandafter\letR@\nextiii@ N\fi
    }%
 \fi
 \next@}
\def\newnumstyle@#1#2{\expandafter
 \ifx\csname\exstring@#1@N#2\endcsname\relax
  \DN@{\Err@{\string\newnumstyle\string#1 can't be followed by
   \string#2}}%
 \else
  \DN@##1{%
   \gdef\Next@{##1}%
    \expandafter\let\csname\exstring@#1@N#2\endcsname\Next@
    \expandafter\ifx\csname\exstring@#1@R#2\endcsname\relax\else
    \getR@#1{#2}\expandafter\letR@@\nextiii@ N\fi
    }%
  \fi
 \next@}
\def\newfontstyle#1{\expandafter
 \ifx\csname\exstring@#1@F\endcsname\relax
  \expandafter\ifx\csname\exstring@#1@F1\endcsname\relax
   \DN@{\Err@{\noexpand\newfontstyle can't be used with
    \string#1}}%
  \else
   \DN@{\newfontstyle@#1}%
  \fi
 \else
  \DN@##1{%
   \gdef\Next@{##1}%
    \expandafter\let\csname\exstring@#1@F\endcsname\Next@
    \expandafter\ifx\csname\exstring@#1@R\endcsname\relax\else
    \getR@#1{}\expandafter\letR@\nextiii@ F\fi
    }%
 \fi
 \next@}
\def\newfontstyle@#1#2{\expandafter
 \ifx\csname\exstring@#1@F#2\endcsname\relax
  \DN@{\Err@{\string\newfontstyle\string#1 can't be followed by
   \string#2}}%
 \else
  \DN@##1{%
   \gdef\Next@{##1}%
    \expandafter\let\csname\exstring@#1@F#2\endcsname\Next@
    \expandafter\ifx\csname\exstring@#1@R#2\endcsname\relax\else
    \getR@#1{#2}\expandafter\letR@@\nextiii@ F\fi
    }%
 \fi
 \next@}
\def\word#1{\expandafter
 \ifx\csname\exstring@#1@W\endcsname\relax
  \expandafter\ifx\csname\exstring@#1@W1\endcsname\relax
   \DN@{\Err@{\noexpand\word can't be used with \string#1}}%
  \else
   \DN@{\word@#1}%
  \fi
 \else
  \DN@{{\csname\exstring@#1@W\endcsname}}%
 \fi
 \next@}
\def\word@#1#2{\expandafter
 \ifx\csname\exstring@#1@W#2\endcsname\relax
  \DN@{\Err@{\string\word\noexpand#1can't be followed by \string#2}}%
 \else
  \DN@{{\csname\exstring@#1@W#2\endcsname}}%
 \fi
 \next@}
\def\newword#1{\expandafter
 \ifx\csname\exstring@#1@W\endcsname\relax
  \expandafter\ifx\csname\exstring@#1@W1\endcsname\relax
   \DN@{\Err@{\noexpand\newword can't be used  with \string#1}}%
  \else
   \DN@{\newword@#1}%
  \fi
 \else
  \DN@{%
   \DNii@{%
    \expandafter\let\csname\exstring@#1@W\endcsname\Next@
    \expandafter\ifx\csname\exstring@#1@R\endcsname\relax\else
     \getR@#1{}\expandafter\letR@\nextiii@ W\fi
    }%
   \afterassignment\nextii@\gdef\Next@}%
 \fi
 \next@}
\def\newword@#1#2{\expandafter
 \ifx\csname\exstring@#1@W#2\endcsname\relax
  \DN@{\Err@{\string\newword\noexpand#1can't be followed by \string#2}}%
 \else
  \DN@{%
   \DNii@{%
    \expandafter\let\csname\exstring@#1@W#2\endcsname\Next@
    \expandafter\ifx\csname\exstring@#1@R#2\endcsname\relax\else
     \getR@#1{#2}\expandafter\letR@@\nextiii@ W\fi
    }%
   \afterassignment\nextii@\gdef\Next@}%
 \fi
 \next@}
\newif\iffn@
\newcount\footmark@C
\footmark@C\z@
\def\footmark@S#1{$^{#1}$}
\let\footmark@N\arabic
\def\footmark@F{\rm}
\def\foottext@S#1{$^{#1}$}
\def\foottext@F{\rm}
\let\modifyfootnote@\relax
\def\modifyfootnote#1{\def\modifyfootnote@{#1}}
\def\vfootnote@#1{\insert\footins
 \bgroup
 \floatingpenalty\@MM\interlinepenalty\interfootnotelinepenalty
 \leftskip\z@\rightskip\z@\spaceskip\z@\xspaceskip\z@
 \rm\splittopskip\ht\strutbox\splitmaxdepth\dp\strutbox
 \locallabel@\noindent@@{\foottext@F#1}\modifyfootnote@
 \footstrut\FN@\fo@t}
\def\fo@t{\ifcat\bgroup\noexpand\next\expandafter\f@@t\else
 \expandafter\f@t\fi}
\def\f@t#1{#1\@foot}
\def\f@@t{\bgroup\aftergroup\@foot\afterassignment\FNSSP@\let\next@}
\def\@foot{\unskip\lower\dp\strutbox\vbox to\dp\strutbox{}\egroup
 \iffn@\expandafter\fn@false\else
 \expandafter\postvanish@\fi}
\newif\ifplainfn@
\plainfn@true
\def\fancyfootnotes{\plainfn@false}
\newcount\fancyfootmarkcount@
\fancyfootmarkcount@\z@
\newcount\lastfnpage@
\lastfnpage@-\@M
\let\justfootmarklist@\empty
\def\footmark{\let\@sf\empty
 \ifhmode\edef\@sf{\spacefactor\the\spacefactor}\/\fi
 \DN@{\ifx"\next\expandafter\nextii@\else\expandafter\footmark@\fi}%
 \DNii@"##1"{%
  \iffirstchoice@
   {\let\style\footmark@S\let\numstyle\footmark@N
   \footmark@F##1%
   \noexpands@
   \let\style\foottext@S
   \Qlabel@{##1}%
   }%
   \iffn@\else
    {\noexpands@
    \xdef\Next@{{\Thelabel@}{\Thelabel@@}{\Thelabel@@@}{\Thelabel@@@@}}%
    }%
    \expandafter\rightappend@\Next@\to\justfootmarklist@
   \fi
  \fi
  \@sf\relax}%
 \FN@\next@}
\def\footmark@{%
 \iffirstchoice@
  \global\advance\footmark@C\@ne
  \ifplainfn@
   \xdef\adjustedfootmark@{\number\footmark@C}%
  \else
   {\let\\\or\xdef\Next@{\ifcase\number\footmark@C\fnpages@\else
     -\@M\fi}}%
   \ifnum\Next@=-\@M
    \xdef\adjustedfootmark@{\number\footmark@C}%
   \else
    \ifnum\Next@=\lastfnpage@
     \global\advance\fancyfootmarkcount@\@ne
    \else
     \global\fancyfootmarkcount@\@ne
     \global\lastfnpage@\Next@
    \fi
    \xdef\adjustedfootmark@{\number\fancyfootmarkcount@}%
   \fi
  \fi
  {\noexpands@
  \xdef\Thelabel@@@{\adjustedfootmark@}%
  \xdefThelabel@\footmark@N
  \xdef\Thelabel@@@@{\Thelabel@}%
  \xdefThelabel@@\foottext@S
  }%
  \iffn@\else
   {\noexpands@
   \xdef\Next@{{\Thelabel@}{\Thelabel@@}{\Thelabel@@@}{\Thelabel@@@@}}%
   }%
   \expandafter\rightappend@\Next@\to\justfootmarklist@
  \fi
  \ifplainfn@
  \else
   \edef\next@{\write\laxwrite@{F\noexpand\the\pageno}}\next@
  \fi
 \fi
 \footmark@S{\footmark@N{\adjustedfootmark@}}%
 \@sf\relax}
\def\foottext{\prevanish@
 \ifx\justfootmarklist@\empty
  \Err@{There is no \noexpand\footmark for this \string\foottext}\fi
 \DN@\\##1##2\next@{\DN@{##1}\gdef\justfootmarklist@{##2}}%
 \expandafter\next@\justfootmarklist@\next@
 \expandafter\foottext@\next@}
\def\foottext@#1#2#3#4{{\noexpands@
  \xdef\Thelabel@{#1}\xdef\Thelabel@@{#2}%
  \xdef\Thelabel@@@{#3}\xdef\Thelabel@@@@{#4}}%
  \vfootnote@{\thelabel@@}}
\rightadd@\foottext\to\vanishlist@
\def\footnote{\fn@true
 \let\@sf\empty
 \ifhmode\edef\@sf{\spacefactor\the\spacefactor}\/\fi
 \DN@{\ifx"\next\expandafter\nextii@\else\expandafter\nextiii@\fi}%
 \DNii@"##1"{\footmark"##1"\vfootnote@{\let\style\foottext@S
  \let\numstyle\footmark@N##1}}%
 \def\nextiii@{\footmark\vfootnote@{\foottext@S{\footmark@N
  {\adjustedfootmark@}}}}%
 \FN@\next@}
\newdimen\litindent
\litindent20\p@
\newbox\litbox@
\newbox\Litbox@
\newcount\interlitpenalty@
\interlitpenalty@\@M
\newcount\litlines@
{\obeyspaces\gdef\defspace@{\def {\allowbreak\hskip.5emminus.15em}}}
{\obeylines\gdef\letM@{\let^^M\CtrlM@}}
\def\CtrlM@{\egroup
 \ifcase\litlines@\advance\litlines@\@ne\or
 \box\litbox@\advance\litlines@\@ne\else
 \penalty\interlitpenalty@\box\litbox@\fi
 \Lit@}
\def\Lit@{\setbox\litbox@\hbox\bgroup\litdefs@\hskip\litindent}
\newcount\littab@
\littab@8
\def\littab#1{\littab@#1\relax}
{\catcode`\^^I=\active\gdef\letTAB@{\let^^I\TAB@}}
\def\TAB@{\egroup
 \dimen@\wd\litbox@
 \advance\dimen@-\litindent
 \setboxz@h{\tt0}%
 \dimen@ii\littab@\wdz@
 \divide\dimen@\dimen@ii
 \multiply\dimen@\dimen@ii
 \advance\dimen@\littab@\wdz@
 \advance\dimen@\litindent
 \setbox\litbox@\hbox\bgroup\litdefs@\hbox to\dimen@{\unhbox\litbox@\hfil}}
{\catcode`\`=\active\gdef`{\relax\lq}}
\let\litbs@\relax
\let\litbs@@\relax
\def\litbackslash#1{%
 \edef\litbs@{\catcode`\string#1=\z@
 \def\noexpand\litbs@@{\def\expandafter\noexpand\csname\string#1\endcsname
  {\char`\string#1}}}}
\def\litcodes@{\catcode`\\=12
 \catcode`\{=12 \catcode`\}=12
 \catcode`\$=12 \catcode`\&=12
 \catcode`\#=12
 \catcode`\^=12 \catcode`\_=12
 \catcode`\@=12 \catcode`\~=12 \catcode`\"=12
 \catcode`\;=12 \catcode`\:=12 \catcode`\!=12 \catcode`\?=12
 \catcode`\%=12 \litbs@\catcode`\`=\active\obeyspaces\defspace@}
\def\activate@#1#2{{\lccode`\~=`#2%
 \lowercase{%
  \if0#1%
  \gdef\Next@{\def~{\egroup\endgroup\bigskip\vskip-\parskip
   \def\next@{\noindent@@\FN@\pretendspace@}\FNSS@\next@}}\else
  \gdef\Next@{\def~{\egroup\egroup\endgroup}}\fi
  }%
 }}
\def\litdefs@{\let\0\empty\let\1\litdelim@\def\ {\char32 }\litbs@@}%
\def\litdelimiter#1{%
 \edef\litdelim@{\char`#1}%
 \def\lit#1{\leavevmode\begingroup\litcodes@\litdefs@
  \tt\hyphenchar\tentt\m@ne\lit@}%
 \def\lit@##1#1{##1\endgroup\null}%
 \def\Lit#1{\ifhmode$$\abovedisplayskip\bigskipamount
  \abovedisplayshortskip\bigskipamount
  \belowdisplayskip\z@\belowdisplayshortskip\z@
  \postdisplaypenalty\@M
  $$\vskip-\baselineskip\else\bigskip\fi
  \begingroup\litlines@\z@
  \catcode`#1=\active\activate@0#1\Next@
  \def\displaybreak{\egroup\break\litlines@\z@\Lit@}%
  \def\allowdisplaybreak{\egroup\allowbreak\litlines@\z@\Lit@}%
  \def\allowdisplaybreaks{\egroup\allowbreak\interlitpenalty@\z@
   \litlines@\z@\Lit@}%
  \litcodes@\tt\catcode`\^^I=\active\letTAB@
  \obeylines\letM@\Lit@}%
 \def\Litbox##1=#1{\begingroup\ifodd##1\relax\aftergroup\global\fi
  \aftergroup\setbox\aftergroup##1\aftergroup\box\aftergroup\Litbox@
  \def\allowdisplaybreak{\egroup\allowbreak\litlines@\z@\Lit@}%
  \def\allowdisplaybreaks{\egroup\allowbreak\interlitpenalty@\z@
   \litlines@\z@\Lit@}%
  \catcode`#1=\active\activate@1#1\Next@
  \litcodes@\tt\catcode`\^^I=\active\letTAB@
  \obeylines\letM@\global\setbox\Litbox@\vbox\bgroup\litindent\z@%
  \litlines@\z@\Lit@}%
}
\newbox\titlebox@
\setbox\titlebox@\vbox{}
\rightadd@\title\to\overlonglist@
\def\title{\begingroup\Let@
 \global\setbox\titlebox@\vbox\bgroup\tabskip\hss@
 \halign to\hsize\bgroup\bf\hfil\ignorespaces##\unskip\hfil\cr}
\def\endtitle{\crcr\egroup\egroup\endgroup\overlong@false}
\newbox\authorbox@
\rightadd@\author\to\overlonglist@
\def\author{\begingroup\Let@
 \global\setbox\authorbox@\vbox\bgroup\tabskip\hss@
 \halign to\hsize\bgroup\rm\hfil\ignorespaces##\unskip\hfil\cr}
\def\endauthor{\crcr\egroup\egroup\endgroup\overlong@false}
\newbox\affilbox@
\def\affil{\begingroup\Let@
 \global\setbox\affilbox@\vbox\bgroup\tabskip\hss@
 \halign to\hsize\bgroup\rm\hfil\ignorespaces##\unskip\hfil\cr}%
\def\endaffil{\crcr\egroup\egroup\endgroup\overlong@false}
\let\date@\relax
\def\date#1{\gdef\date@{\ignorespaces#1\unskip}}
\def\today{\ifcase\month\or January\or February\or March\or April\or May\or
 June\or July\or August\or September\or October\or November\or December\fi
 \space\number\day, \number\year}
\def\maketitle{\hrule\height\z@\vskip-\topskip
 \vskip24\p@ plus12\p@ minus12\p@
 \unvbox\titlebox@
 \ifvoid\authorbox@\else\vskip12\p@ plus6\p@ minus3\p@\unvbox\authorbox@\fi
 \ifvoid\affilbox@\else\vskip10\p@ plus5\p@ minus2\p@\unvbox\affilbox@\fi
 \ifx\date@\relax\else\vskip6\p@ plus2\p@ minus\p@\centerline{\rm\date@}\fi
 \vskip18\p@ plus12\p@ minus6\p@}
\def\cite{%
 \DNii@(##1)##2{{\rm[}{##2}, {##1\/}{\rm]}}%
 \def\nextiii@##1{{\rm[}{##1\/}{\rm]}}%
 \DN@{\ifx\next(\expandafter\nextii@\else\expandafter\nextiii@\fi}%
 \FN@\next@}
\def\makebib@W{Bibliography}
\def\makebib{\begingroup\rm\bigbreak\centerline{\smc\makebib@W}%
 \nobreak\medskip
 \sfcode`\.=\@m\everypar{}\parindent\z@
 \def\nopunct{\nopunct@true}\def\nospace{\nospace@true}%
 \nopunct@false\nospace@false
 \def\lkerns@{\null\kern\m@ne sp\kern\@ne sp}%
 \def\nkerns@{\null\kern-\tw@ sp\kern\tw@ sp}%
}

\newif\ifnoprepunct@
\newif\ifnoprespace@
\newif\ifnoquotes@
\def\noprepunct{\noprepunct@true}
\def\noprespace{\noprespace@true}
\def\noquotes{\noquotes@true}
\newbox\nobox@
\newbox\keybox@
\newbox\bybox@
\newbox\paperbox@
\newbox\paperinfobox@
\newbox\jourbox@
\newbox\volbox@
\newbox\issuebox@
\newbox\yrbox@
\newbox\pgbox@
\newbox\ppbox@
\newbox\bookbox@
\newbox\inbookbox@
\newbox\bookinfobox@
\newbox\publbox@
\newbox\publaddrbox@
\newbox\edbox@
\newbox\edsbox@
\newbox\langbox@
\newbox\translbox@
\newbox\finalinfobox@
\def\setbibinfo@#1{\edef\next@{\ifnopunct@1\else0\fi
 \ifnospace@1\else0\fi\ifnoprepunct@1\else0\fi\ifnoprespace@1\else0\fi
 \ifnoquotes@1\else0\fi}%
 \DNii@{00000}%
 \ifx\next@\nextii@\else\xdef\bibinfo@{\bibinfo@\the#1,\next@}%
 \fi}
\def\getbibinfo@#1{\ifx\bibinfo@\empty
 \let\next@0\let\nextii@0\let\nextiii@0\let\nextiv@0\let\nextv@0\else
 \edef\next@{\def
  \noexpand\next@####1\the#1,####2####3####4####5####6####7\noexpand\next@
  {\let\noexpand\next@####2\let\noexpand\nextii@####3%
  \let\noexpand\nextiii@####4\let\noexpand\nextiv@####5%
  \let\noexpand\nextv@####6}%
  \noexpand\next@\bibinfo@\the#1,00000\noexpand\next@}\next@
 \fi}
\newif\ifbookinquotes@
\def\bookinquotes{\bookinquotes@true}
\newif\ifpaperinquotes@
\def\paperinquotes{\paperinquotes@true}
\newif\ifininbook@
\def\ininbook{\ininbook@true}
\newif\ifopenquotes@
\def\closequotes@{\ifopenquotes@''\openquotes@false\fi}
\newif\ifbeginbib@
\newif\ifendbib@
\newif\ifprevjour@
\newif\ifprevbook@
\newdimen\bibindent@
\bibindent@20\p@
\def\bib{\global\let\bibinfo@\empty\global\let\translinfo@\relax\beginbib@true
 \begingroup\noindent@
 \hangindent\bibindent@\hangafter\@ne\bib@}
\def\v@id#1{\setbox#1\box\voidb@x}
\def\bib@{\v@id\nobox@\v@id\keybox@\v@id\bybox@\v@id\paperbox@
 \v@id\paperinfobox@\v@id\jourbox@\v@id\volbox@\v@id\issuebox@
 \v@id\yrbox@\v@id\pgbox@\v@id\ppbox@\v@id\bookbox@\v@id\inbookbox@
 \v@id\bookinfobox@\v@id\publbox@\v@id\publaddrbox@\v@id\edbox@
 \v@id\edsbox@\v@id\langbox@\v@id\translbox@\v@id\finalinfobox@
 \bgroup}
\def\Setnonemptybox@#1#2{\unskip\setbibinfo@#1\egroup#2%
 \def\aftergroup@{\ifdim\wd#1=\z@\setbox#1\box\voidb@x\fi}%
 \setbox#1\vbox\bgroup\aftergroup\aftergroup@\hsize\maxdimen\leftskip\z@
 \rightskip\z@\hbadness\@M\hfuzz\maxdimen\noindent}
\def\setnonemptybox@#1{\Setnonemptybox@#1\relax}
\def\no{\setnonemptybox@\nobox@}
\def\key{\setnonemptybox@\keybox@\bf}
\def\by{\setnonemptybox@\bybox@}
\def\bysame{\setnonemptybox@\bybox@\leaders\hrule\hskip3em\null}
\def\paper{\setnonemptybox@\paperbox@
 \ifpaperinquotes@\getbibinfo@\paperbox@
 \if\nextv@1\else``\fi\else\it\fi}
\def\paperinfo{\setnonemptybox@\paperinfobox@}
\def\jour{\Setnonemptybox@\jourbox@\prevjour@true}
\def\vol{\setnonemptybox@\volbox@\bf}
\def\issue{\setnonemptybox@\issuebox@}
\def\yr{\setnonemptybox@\yrbox@}

\def\pg{\setnonemptybox@\pgbox@}
\def\pp{\setnonemptybox@\ppbox@}
\def\book{\Setnonemptybox@\bookbox@\prevbook@true
 \ifbookinquotes@\getbibinfo@\bookbox@
 \if\nextv@1\else``\fi\else\it\fi}
\def\inbook{\Setnonemptybox@\inbookbox@\prevbook@true
 \ifininbook@ in \fi\ifbookinquotes@\getbibinfo@\inbookbox@
 \if\nextv@1\else``\fi\fi}
\def\bookinfo{\setnonemptybox@\bookinfobox@}
\def\publ{\setnonemptybox@\publbox@}
\def\publaddr{\setnonemptybox@\publaddrbox@}
\def\ed{\setnonemptybox@\edbox@}
\def\eds{\setnonemptybox@\edsbox@}
\def\lang{\setnonemptybox@\langbox@}
\def\finalinfo{\setnonemptybox@\finalinfobox@}
\def\setboxzl@{\setbox\z@\lastbox}
\def\getbox@#1{\setbox\z@\vbox{\vskip-\@M\p@
 \unvbox#1%
 \setboxzl@
 \global\setbox\@ne\hbox{\unhbox\z@\unskip\unskip\unpenalty}%
 \ifdim\lastskip=-\@M\p@\else
 \loop\ifdim\lastskip=-\@M\p@
 \else\unskip\unpenalty\setboxzl@
 \global\setbox\@ne\hbox{\unhbox\z@\unhbox\@ne}%
 \repeat\fi}%
 \unhbox\@ne}
\def\adjustpunct@#1{\count@\lastkern
 \ifnum\count@=\z@#1\closequotes@\else
 \ifnum\count@>\tw@#1\closequotes@\else
 \ifnum\count@<-\tw@#1\closequotes@\else
  \unkern\unkern\setboxzl@
  \skip@\lastskip\unskip
  \count@@\lastpenalty\unpenalty
  \ifnum\count@=\tw@\unskip\setboxzl@\fi
  \ifdim\skip@=\z@\else\hskip\skip@\fi
  #1\closequotes@
  \ifnum\count@=\tw@\null\hfill\fi
  \penalty\count@@
 \fi\fi\fi}
\def\prepunct@#1#2{\getbibinfo@#2%
 \ifnopunct@
 \else
  \if\nextiii@0\adjustpunct@#1\fi
 \fi
 \closequotes@
 \ifnospace@
 \else
  \if\nextiv@0\space\else\fi
 \fi
 \nopunct@false\nospace@false
 \if\next@1\nopunct@true\fi
 \if\nextii@1\nospace@true\fi}
\def\ppunbox@#1#2{\prepunct@{#1}#2%
 \getbox@#2}
\let\semicolon@;
\def\endbib@{%
 \ifbeginbib@
  \ifvoid\nobox@
   \ifvoid\keybox@\else\hbox to\bibindent@{[\getbox@\keybox@]\hss}\fi
  \else\hbox to\bibindent@{\hss\getbox@\nobox@. }\fi
  \ifvoid\bybox@\else\getbox@\bybox@\fi
 \else
  \nopunct@true
  \ifvoid\bybox@\else\ppunbox@\relax\bybox@\fi
 \fi
 \ifvoid\translbox@\else\ppunbox@,\translbox@\fi
 \ifvoid\paperbox@\else\ppunbox@,\paperbox@\ifpaperinquotes@
  \if\nextv@1\else\openquotes@true\fi\fi
 \fi
 \ifvoid\paperinfobox@\else\ppunbox@,\paperinfobox@\fi
 \test@false
 \ifvoid\jourbox@\else\test@true\ppunbox@,\jourbox@\fi
 \ifprevjour@\test@true\fi
 \iftest@
  \ifvoid\volbox@\else\ppunbox@\relax\volbox@\fi
  \ifvoid\issuebox@
   \else\prepunct@\relax\issuebox@ no.~\getbox@\issuebox@\fi
  \ifvoid\yrbox@\else\prepunct@\relax\yrbox@(\getbox@\yrbox@)\fi
  \ifvoid\ppbox@\else\ppunbox@,\ppbox@\fi
  \ifvoid\pgbox@\else\prepunct@,\pgbox@ p.~\getbox@\pgbox@\fi
 \fi
 \test@false
 \ifvoid\bookbox@\else\test@true\ppunbox@,\bookbox@\ifbookinquotes@
  \if\nextv@1\else\openquotes@true\fi\fi\fi
 \ifvoid\inbookbox@\else\test@true\ppunbox@,\inbookbox@\ifbookinquotes@
  \if\nextv@1\else\openquotes@true\fi\fi\fi
 \ifprevbook@\test@true\fi
 \iftest@
  \ifvoid\edbox@\else\prepunct@\relax\edbox@(\getbox@\edbox@, ed.)\fi
  \ifvoid\edsbox@\else\prepunct@\relax\edsbox@(\getbox@\edsbox@, eds.)\fi
  \ifvoid\bookinfobox@\else\ppunbox@,\bookinfobox@\fi
  \ifvoid\publbox@\else\ppunbox@,\publbox@\fi
  \ifvoid\publaddrbox@\else\ppunbox@,\publaddrbox@\fi
  \ifvoid\yrbox@\else\ppunbox@,\yrbox@\fi
  \ifvoid\ppbox@\else\prepunct@,\ppbox@ pp.~\getbox@\ppbox@\fi
  \ifvoid\pgbox@\else\prepunct@,\pgbox@ p.~\getbox@\pgbox@\fi
 \fi
 \ifvoid\finalinfobox@
  \ifendbib@
   \ifnopunct@\else.\closequotes@\fi
  \else
  \ifvoid\langbox@\else\space(\getbox@\langbox@)\fi
   \/\semicolon@\closequotes@
  \fi
 \else
  \ifendbib@
   \ppunbox@{.\spacefactor3000\relax}\finalinfobox@
    \ifnopunct@\else.\fi
  \else
   \ppunbox@,\finalinfobox@\/\semicolon@\fi
 \fi
 \ifvoid\langbox@\else\space(\getbox@\langbox@)\fi
}
\def\endbib{\unskip\egroup\endbib@true\endbib@\par\endgroup}
\def\morebib{\unskip\egroup
 \endbib@false\endbib@
 \global\let\bibinfo@\empty\beginbib@false
 \bib@}
\def\anotherbib{\unskip\egroup
 \endbib@false\endbib@
 \global\let\bibinfo@\empty\beginbib@false
 \prevjour@false\prevbook@false\bib@}
\def\transl{\unskip
 \xdef\translinfo@{\the\translbox@,\ifnopunct@1\else0\fi
 \ifnospace@1\else0\fi\ifnoprepunct@1\else0\fi\ifnoprespace@1\else0\fi0}%
 \egroup\endbib@false\endbib@
 \global\let\bibinfo@\translinfo@\beginbib@false
 \bib@
 \egroup
 \def\aftergroup@{\ifdim\wd\translbox@=\z@\setbox\translbox@\box\voidb@x\fi}%
 \setbox\translbox@\vbox\bgroup\aftergroup\aftergroup@
 \hsize\maxdimen\leftskip\z@\rightskip\z@\hbadness\@M\hfuzz\maxdimen
 \noindent}
\newwrite\auxwrite@
\newread\bbl@
\def\UseBibTeX{\immediate\openout\auxwrite@=\jobname.aux
 \let\cite\BTcite@
 \def\nocite##1{\immediate\write\auxwrite@{\string\citation{##1}}}%
 \def\bibliographystyle##1{\immediate\write\auxwrite@{\string
  \bibstyle{##1}}}%
 \def\bibliography@W{Bibliography}%
 \def\bibliography##1{\immediate\write\auxwrite@{\string\bibdata{##1}}%
  \immediate\openin\bbl@=\jobname.bbl
  \ifeof\bbl@
   \W@{No .bbl file}%
  \else
   \immediate\closein\bbl@
   \begingroup\input bibtex \input\jobname.bbl \endgroup
  \fi}%
 }
\def\BTcite@{%
 \DNii@(##1)##2{{\rm[}\BTcite@@##2,\BTcite@@{\rm, }{##1\/}{\rm]}%
  \immediate\write\auxwrite@{\string\citation{##2}}}%
 \def\nextiii@##1{{\rm[}\BTcite@@##1,\BTcite@@\/{\rm]}%
  \immediate\write\auxwrite@{\string\citation{##1}}}%
 \DN@{\ifx\next(\expandafter\nextii@\else\expandafter\nextiii@\fi}%
 \FN@\next@}%
\def\BTcite@@#1,{\BTcite@@@{#1}\FN@\BTcite@@@@}
\def\BTcite@@@@{\ifx\next\BTcite@@
 \expandafter\eat@\else{\rm, }\expandafter\BTcite@@\fi}
\catcode`\~=11
\def\BTcite@@@#1{\nolabel@\cite{#1}\relax
 \DNii@##1~##2\nextii@{##1}%
 \csL@{#1}\expandafter\nextii@\Next@\nextii@\fi}
\catcode`\~=\active

\def\beginthebibliography@#1{\rm\setboxz@h{#1\ }\bibindent@\wdz@
 \bigbreak\centerline{\smc\bibliography@W}\nobreak\medskip
 \sfcode`\.=\@m\everypar{}\parindent\z@}

\newif\iffigproofing@
\def\Figureproofing{\figproofing@true}
\def\noFigureproofing{\figproofing@false}
\newif\ifHby@
\def\Hbyw#1{\global\Hby@true\hbyw\vsize{#1}}
\def\hbyw#1#2{%
 \hbox{%
  \ifHby@
  \else
   \iffigproofing@
    \setbox\z@\vbox{\hrule\width5\p@}\ht\z@\z@
    \vbox to#1{\hrule\height5\p@\width.4\p@\vfil\hrule\height5\p@\width.4\p@}%
    \kern-.4\p@\rlap{\copy\z@}\raise#1\hbox{\rlap{\copy\z@}}%
   \fi
  \fi
  \vbox to#1{\hbox to#2{}\vfil}%
  \ifHby@
  \else
   \iffigproofing@
    \vbox to#1{\hrule\height5\p@\width.4\p@\vfil\hrule\height5\p@\width.4\p@}%
    \kern-.4\p@\llap{\copy\z@}\raise#1\hbox{\llap{\boxz@}}%
   \fi
  \fi}}
\newcount\island@C
\let\island@P\empty
\let\island@Q\empty
\def\island@S#1{#1\null.}
\let\island@N\arabic
\def\island@F{\rm}
\def\island@@@P{\csname\exxx@\islandtype@ @P\endcsname}
\def\island@@@Q{\csname\exxx@\islandtype@ @Q\endcsname}
\def\island@@@S{\csname\exxx@\islandtype@ @S\endcsname}
\def\island@@@N{\csname\exxx@\islandtype@ @N\endcsname}
\def\island@@@F{\csname\exxx@\islandtype@ @F\endcsname}
\def\island@@@C{\csname island@C\islandclass@\endcsname}
\newif\ifplace@
\newif\ifisland@
\def\island{%
 \ifplace@
  \DN@{\let\islandclass@\empty\def\islandtype@{\island}\FN@\island@}%
 \else
  \long\DN@##1\endisland{\Err@{\noexpand\island must be used after some
   type of \string\...place}}%
 \fi
 \next@}
\def\island@{\ifx\next\c\let\next@\island@c\else
 \DN@{\FN@\island@@}\fi\next@}
\def\island@@{\ifcat\bgroup\noexpand\next\let\next@\island@@@\else
 \DN@{\Err@{\noexpand\island must be followed by a {prefix} for
 \string\caption's}}\fi\next@}
\newbox\islandbox@
\newcount\captioncount@
\def\island@@@#1{\def\captionprefix@{#1}\captioncount@\z@
 \global\setbox\islandbox@\vbox\bgroup}
\def\island@c\c#1{%
 \ifplace@
 \DN@{\def\islandclass@{#1}%
  \expandafter\ifx\csname island@C#1\endcsname\relax
  \expandafter\newcount@\csname island@C#1\endcsname
   \global\csname island@C#1\endcsname\z@\fi
  \FNSS@\island@c@}%
 \else
 \DN@{\edef\next@{\long\def\noexpand\next@########1\expandafter\noexpand
  \csname end\exxx@\islandtype@\endcsname{\noexpand\Err@{\noexpand\noexpand
  \expandafter\noexpand
  \islandtype@ must be used after some type of \noexpand\string
   \noexpand\...place}}}\next@\next@}%
 \fi
 \next@}
\def\island@c@{%
 \ifcat\bgroup\noexpand\next
  \let\next@\island@c@@
 \else
  \DN@{\Err@{\noexpand\island\string\c{\expandafter\string\islandclass@} must
   be followed by a {prefix} for \string\caption's}}%
 \fi\next@}
\def\island@c@@#1{\def\captionprefix@{#1}%
 \captioncount@\z@\global\setbox\islandbox@\vbox\bgroup}
\rightadd@\caption\to\nofrillslist@
\newbox\captionbox@
\newbox\Captionbox@
\def\caption{%
 \ifnum\captioncount@=\z@
  \ifnopunct@
   \DN@{\egroup\nopunct@true}%
  \else
   \let\next@\egroup
  \fi
 \else
  \let\next@\relax
 \fi
 \next@
 \advance\captioncount@\@ne
 \FN@\caption@}
\def\caption@{\ifx\next"\expandafter\caption@q\else\expandafter\caption@@\fi}
\def\caption@q"#1"{\quoted@true
 {\noexpands@
 \let\pre\island@@@P\let\post\island@@@Q
 \let\style\island@@@S\let\numstyle\island@@@N
 \Qlabel@{#1}\let\style\relax\xdef\Qlabel@@@@{#1}}%
 \finishcaption@}
\def\caption@@{\quoted@false
 \global\advance\island@@@C\@ne
 {\noexpands@
 \xdef\Thelabel@@@{\number\island@@@C}%
 \xdefThelabel@\island@@@N
 \xdef\Thelabel@@@@{\island@@@P\Thelabel@\island@@@Q}%
 \xdefThelabel@@\island@@@S
 \xdef\Thepref@{\Thelabel@@@@}}%
 \finishcaption@}
\long\def\captionformat@#1#2#3{\rm\strut#1 {\island@@@F#2} #3%
 \punct@.\strut}
\long\def\widerthanisland@#1#2#3{\test@true\setbox\z@\vbox{\hsize\maxdimen
 \noindent@@\captionformat@{#1}{#2}{#3}\par\setboxzl@}%
 \ifdim\wdz@=\z@
  \global\setbox\captionbox@\hbox{\noset@\unlabel@
   \captionformat@{#1}{#2}{#3}}%
  \ifdim\wd\captionbox@>\wd\islandbox@\else\test@false\fi
 \fi}
\long\def\captionformat@@#1#2#3{\widerthanisland@{#1}{#2}{#3}%
 \iftest@
  \global\setbox\captionbox@\vbox{\hsize\wd\islandbox@
   \vskip-\parskip\noindent@@\noset@\unlabel@
   \captionformat@{#1}{#2}{#3}\par}%
 \else
  \global\setbox\captionbox@
   \hbox to\wd\islandbox@{\hfil\box\captionbox@\hfil}%
 \fi}
\long\def\finishcaption@#1{\def\entry@{#1}%
 {\locallabel@
 \captionformat@@
  {\expandafter\ignorespaces\captionprefix@\unskip}%
  {\ifx\thelabel@@\empty\unskip\else\thelabel@@\fi}%
  {\ignorespaces#1\unskip}%
 \ifnum\captioncount@=\@ne
  \global\setbox\islandbox@\vbox{\ticwrite@\vbox{\box\islandbox@}}%
  \global\setbox\Captionbox@\vbox{\box\captionbox@}%
 \else
  \global\setbox\islandbox@\vbox{\unvbox\islandbox@\setboxzl@
   \ticwrite@\boxz@}%
  \global\setbox\Captionbox@\vbox{\unvbox\Captionbox@
   \smallskip\box\captionbox@}%
 \fi}%
 \nopunct@false\nospace@false\ignorespaces}
\def\Sixtic@{\ifx\macdef@\empty\else
 \DN@##1##2\next@{\def\macdef@{##1##2}}%
 \expandafter\next@\macdef@\next@
 \edef\next@
  {\noexpand\six@\tic@\macdef@
  \space\space\space\space\space\space\space\space\space\space\space\space
  \noexpand\six@}%
 \next@\let\macdef@\relax\fi}
\def\ticwrite@{%
 \iftoc@
  {\noexpands@\let\style\relax
  \DN@{\island}%
  \edef\next@{\write\tic@{%
   \ifnopunct@\noexpand\noexpand\noexpand\nopunct\fi
   \ifx\islandtype@\next@\noexpand\noexpand\noexpand\island
    \noexpand\string\noexpand\c{\islandclass@}{\captionprefix@}%
     {\QorThelabel@@@@}\else\noexpand\noexpand\expandafter\noexpand
     \islandtype@{\QorThelabel@@@@}}\fi}%
  \next@}%
  \expandafter\unmacro@\meaning\entry@\unmacro@
  \Sixtic@
  \write\tic@{\noexpand\Page{\number\pageno}{\page@N}{\page@P}{\page@Q}^^J}%
 \fi}
\def\Htrim@#1{%
 \ifHby@
  \dimen@\vsize
  \ifnum\captioncount@=\z@
  \else
   \advance\dimen@-\ht\Captionbox@
   \advance\dimen@-#1%
  \fi
  \global\Hby@false
  \dimen@ii\wd\islandbox@
  \global\setbox\islandbox@\vbox
   {\unvbox\islandbox@\setboxzl@
   \vbox to\z@{\vss\boxz@}\nointerlineskip\hbyw\dimen@\dimen@ii}%
  \global\Hby@true
 \fi}
\newif\ifdata@
\def\iclasstest@#1{\DN@{#1}\ifx\next@\islandclass@
 \test@true\else\test@false\fi}
\skipdef\skipi@=1
\def\endisland{\ifnum\captioncount@=\z@\expandafter\egroup\fi
 \ifdata@
 \else
  \iclasstest@{T}%
  \iftest@
   {\rm\global\skipi@-\dp\strutbox}\global\advance\skipi@\bigskipamount
   \Htrim@\skipi@
   \global\setbox\islandbox@\vbox
    {\ifnum\captioncount@=\z@\else
     \box\Captionbox@
     \nointerlineskip
     \vskip\skipi@\fi
     \box\islandbox@}%
  \else
   {\rm\global\skipi@\dp\strutbox}\global\advance\skipi@\medskipamount
   \Htrim@\skipi@
   \global\setbox\islandbox@\vbox
    {\box\islandbox@
     \ifnum\captioncount@=\z@\else
     \nointerlineskip
     \vskip\skipi@
     \box\Captionbox@
     \fi}%
  \fi
  \ifHby@
  \else
   \dimen@\ht\islandbox@\advance\dimen@\dp\islandbox@
   \ifdim\dimen@>\vsize
    \DN@{\island}%
    \Err@{%
     \ifx\islandtype@\next@\noexpand\island\else
      \expandafter\noexpand\islandtype@\fi
     \ifnum\captioncount@=\z@\else
       with \noexpand\caption\fi
      is larger than page}%
     \ht\islandbox@=\vsize
   \fi
  \fi
 \fi
 \global\Hby@false\island@true}
\def\newisland#1\c#2#3{\define#1{}%
 \iftoc@\immediate\write\tic@{\noexpand\newisland\noexpand#1%
  \string\c{#2}{#3}^^J}\fi
 \expandafter\def\csname\exstring@#1@S\endcsname{\island@S}%
 \expandafter\def\csname\exstring@#1@N\endcsname{\island@N}%
 \expandafter\def\csname\exstring@#1@P\endcsname{\island@P}%
 \expandafter\def\csname\exstring@#1@Q\endcsname{\island@Q}%
 \expandafter\def\csname\exstring@#1@F\endcsname{\island@F}%
 \expandafter\def\csname end\exstring@#1\endcsname{\endisland}%
 \expandafter
 \ifx\csname island@C#2\endcsname\relax
  \expandafter\newcount@\csname island@C#2\endcsname
  \global\csname island@C#2\endcsname\z@
 \fi
 \edef\next@{\noexpand\expandafter\noexpand\let\noexpand
  \csname\exstring@#1@C\noexpand\endcsname
  \csname island@C#2\endcsname}%
 \next@
 \def#1{\def\islandtype@{#1}\island@c\c{#2}{#3}}}
\newisland\Figure\c{F}{Figure}
\newisland\Table\c{T}{Table}
\newbox\islandboxi
\newbox\islandboxii
\newbox\islandboxiii
\newbox\captionboxi
\newbox\captionboxii
\newbox\captionboxiii
\long\def\islandpairdata#1#2{{\data@true
 \place@true
 #1%
 \global\setbox\islandboxi\box\islandbox@
 \global\setbox\captionboxi\box\Captionbox@
 #2%
 \global\setbox\islandboxii\box\islandbox@
 \global\setbox\captionboxii\box\Captionbox@
 }}
\long\def\islandpairbox#1#2{\islandpairdata{#1}{#2}%
 \dimen@\ht\captionboxi
 \ifdim\ht\captionboxii>\dimen@\dimen@\ht\captionboxii\fi
 \ifdim\dimen@>\z@
  \ifdim\ht\captionboxi<\dimen@
   \global\setbox\captionboxi\vbox to\dimen@{\unvbox\captionboxi\vfil}\fi
  \ifdim\ht\captionboxii<\dimen@
   \global\setbox\captionboxii\vbox to\dimen@{\unvbox\captionboxii\vfil}\fi
 \fi
 \global\setbox\islandbox@\vbox
 {\hbox to\hsize{\hfil\box\islandboxi\hfil\box\islandboxii\hfil}%
 \ifdim\dimen@>\z@\nointerlineskip
 {\rm\global\skipi@\dp\strutbox}\global\advance\skipi@\medskipamount
  \vskip\skipi@
  \hbox to\hsize{\hfil\box\captionboxi\hfil\box\captionboxii\hfil}\fi}}	
\long\def\islandpairboxa#1#2{\islandpairdata{#1}{#2}%
 \dimen@\ht\captionboxi
 \ifdim\ht\captionboxii>\dimen@\dimen@\ht\captionboxii\fi
 \ifdim\dimen@>\z@
  \ifdim\ht\captionboxi<\dimen@
   \global\setbox\captionboxi\vbox to\dimen@{\vfil\unvbox\captionboxi}\fi
  \ifdim\ht\captionboxii<\dimen@
   \global\setbox\captionboxii\vbox to\dimen@{\vfil\unvbox\captionboxii}\fi
 \fi
 \dimen@ii\ht\islandboxi
 \ifdim\ht\islandboxii>\dimen@ii \dimen@ii\ht\islandboxii\fi
 \ifdim\dimen@ii>\z@
  \ifdim\ht\islandboxi<\dimen@ii
   \global\setbox\islandboxi\vbox to\dimen@ii{\box\islandboxi\vfil}\fi
  \ifdim\ht\islandboxii<\dimen@ii
   \global\setbox\islandboxii\vbox to\dimen@ii{\box\islandboxii\vfil}\fi
 \fi
 \global\setbox\islandbox@\vbox{\ifdim\dimen@>\z@
  \hbox to\hsize{\hfil\box\captionboxi\hfil\box\captionboxii\hfil}%
  \nointerlineskip{\rm\global\skipi@-\dp\strutbox}%
  \global\advance\skipi@\bigskipamount\vskip\skipi@\fi
  \hbox to\hsize{\hfil\box\islandboxi\hfil\box\islandboxii\hfil}}}
\long\def\islandtripledata#1#2#3{{\data@true\place@true
 #1%
 \global\setbox\islandboxi\box\islandbox@
 \global\setbox\captionboxi\box\Captionbox@
 #2%
 \global\setbox\islandboxii\box\islandbox@
 \global\setbox\captionboxii\box\Captionbox@
 #3%
 \global\setbox\islandboxiii\box\islandbox@
 \global\setbox\captionboxiii\box\Captionbox@
 }}
\long\def\islandtriplebox#1#2#3{\islandtripledata{#1}{#2}{#3}%
 \dimen@\ht\captionboxi
 \ifdim\ht\captionboxii>\dimen@ \dimen@\ht\captionboxii\fi
 \ifdim\ht\captionboxiii>\dimen@ \dimen@\ht\captionboxiii\fi
 \ifdim\dimen@>\z@
  \ifdim\ht\captionboxi<\dimen@
   \global\setbox\captionboxi\vbox to\dimen@{\unvbox\captionboxi\vfil}\fi
  \ifdim\ht\captionboxii<\dimen@
   \global\setbox\captionboxii\vbox to\dimen@{\unvbox\captionboxii\vfil}\fi
  \ifdim\ht\captionboxiii<\dimen@
   \global\setbox\captionboxiii\vbox to\dimen@{\unvbox\captionboxiii\vfil}\fi
 \fi
 \global\setbox\islandbox@\vbox
  {\hbox to\hsize{\hfil\box\islandboxi\hfil\box\islandboxii\hfil
   \box\islandboxiii\hfil}%
 \ifdim\dimen@>\z@\nointerlineskip
  {\rm\global\skipi@\dp\strutbox}\global\advance\skipi@\medskipamount
  \vskip\skipi@
  \hbox to\hsize{\hfil\box\captionboxi\hfil\box\captionboxii\hfil
   \box\captionboxiii\hfil}\fi}}
\def\islandtripleboxa#1#2#3{\islandtripledata{#1}{#2}{#3}%
 \dimen@\ht\captionboxi
 \ifdim\ht\captionboxii>\dimen@ \dimen@\ht\captionboxii\fi
 \ifdim\ht\captionboxiii>\dimen@ \dimen@\ht\captionboxiii\fi
 \ifdim\dimen@>\z@
  \ifdim\ht\captionboxi<\dimen@
   \global\setbox\captionboxi\vbox to\dimen@{\vfil\unvbox\captionboxi}\fi
  \ifdim\ht\captionboxii<\dimen@
   \global\setbox\captionboxii\vbox to\dimen@{\vfil\unvbox\captionboxii}\fi
  \ifdim\ht\captionboxiii<\dimen@
   \global\setbox\captionboxiii\vbox to\dimen@{\vfil\unvbox\captionboxiii}\fi
 \fi
 \dimen@ii\ht\islandboxi
 \ifdim\ht\islandboxii>\dimen@ii \dimen@ii\ht\islandboxii\fi
 \ifdim\ht\islandboxiii>\dimen@ii \dimen@ii\ht\islandboxiii\fi
 \ifdim\dimen@ii>\z@
  \ifdim\ht\islandboxi<\dimen@ii
   \global\setbox\islandboxi\vbox to\dimen@ii{\box\islandboxi\vfil}\fi
  \ifdim\ht\islandboxii<\dimen@ii
   \global\setbox\islandboxii\vbox to\dimen@ii{\box\islandboxii\vfil}\fi
  \ifdim\ht\islandboxiii<\dimen@ii
   \global\setbox\islandboxiii\vbox to\dimen@ii{\box\islandboxiii\vfil}\fi
 \fi
 \global\setbox\islandbox@\vbox
  {\ifdim\dimen@>\z@
  \hbox to\hsize{\hfil\box\captionboxi\hfil\box\captionboxii\hfil
   \box\captionboxiii\hfil}%
  \nointerlineskip{\rm\global\skipi@-\dp\strutbox}%
  \global\advance\skipi@\bigskipamount\vskip\skipi@\fi
  \hbox to\hsize{\hfil\box\islandboxi\hfil\box\islandboxii\hfil
   \box\islandboxiii\hfil}}}
\def\Figurepair#1\and#2\endFigurepair{\island@true
 \islandpairbox{\Figure#1\endFigure}{\Figure#2\endFigure}}
\def\Figuretriple#1\and#2\and#3\endFiguretriple{\island@true
 \islandtriplebox{\Figure#1\endFigure}{\Figure#2\endFigure}%
  {\Figure#3\endFigure}}
\def\Tablepair#1\and#2\endTablepair{\island@true
 \islandpairboxa{\Table#1\endTable}{\Table#2\endTable}}
\def\Tabletriple#1\and#2\and#3\endTabletriple{\island@true
 \islandtripleboxa{\Table#1\endTable}{\Table#2\endTable}%
 {\Table#3\endTable}}
\def\place#1{\place@true\island@false
 #1%
 \ifisland@
  \box\islandbox@
 \else
  \Err@{Whoa ... there's no \string\Figure, \string\Table,
   etc., here}%
 \fi
 \place@false}
\newskip\belowtopfigskip
\belowtopfigskip 15\p@ plus 5\p@ minus5\p@
\newskip\abovebotfigskip
\abovebotfigskip 18\p@ plus 6\p@ minus6\p@
\newdimen\minpagesize
\minpagesize 5pc
\dimen@\belowtopfigskip
\advance\dimen@-\abovebotfigskip
\skip\topins\dimen@
\dimen\topins\z@
\newcount\topinscount@
\newbox\topinsdims@
\def\storedim@{\global\setbox\topinsdims@
 \vbox{\hbox to\dimen@{}\unvbox\topinsdims@}}
\def\advancedimtopins@{%
 \ifnum\pageno=\@ne
 \else
   \advance\dimen@\dimen\topins
   \global\dimen\topins\dimen@
 \fi}
\newcount\flipcount@
\def\fliptopins@{%
 \global\flipcount@\z@
 \ifvoid\topins\else
 \setbox\z@\vbox
  {\vskip\p@
   \unvbox\topins
   \global\setbox\topins\vbox{}%
   \loop
    \test@false
    \ifdim\lastskip=\z@\unskip
     \ifdim\lastskip=\z@
      \test@true\fi\fi
    \iftest@
    \global\advance\flipcount@\@ne
    \setboxzl@
    \global\setbox\topins\vbox{\unvbox\topins\boxz@}%
    \unpenalty
   \repeat}\fi}
\newif\ifPar@
\newcount\Parcount@
\newbox\Parbox@
\expandafter\newbox\csname Parfigbox1\endcsname
\expandafter\newbox\csname Parfigbox2\endcsname
\expandafter\newbox\csname Parfigbox3\endcsname
\expandafter\newbox\csname Parfigbox4\endcsname
\expandafter\newbox\csname Parfigbox5\endcsname
\expandafter\newdimen\csname Parprev1\endcsname
\expandafter\newdimen\csname Parprev2\endcsname
\expandafter\newdimen\csname Parprev3\endcsname
\expandafter\newdimen\csname Parprev4\endcsname
\expandafter\newdimen\csname Parprev5\endcsname
\expandafter\newdimen\csname Parprev6\endcsname
\def\Par{\par\global\csname Parprev1\endcsname\prevdepth
 \global\Parcount@\@ne
 \global\Par@true\global\let\Parlist@\empty
 \global\setbox\Parbox@\vbox\bgroup\break}
\def\place@#1#2{%
 \ifisland@
  \ifhmode
   \ifPar@
    \ifnum\Parcount@>5
     \Err@{Only 5 \string\place's allowed per
      \string\Par...\noexpand\endPar paragraph}%
    \else
     \expandafter\expandafter\expandafter
      \global\expandafter\setbox
       \csname Parfigbox\number\Parcount@\endcsname\box\islandbox@
     \global\advance\Parcount@\@ne
     \xdef\Parlist@{\Parlist@#1}%
    \fi
   \else
    \vadjust{#2}%
   \fi
  \else
   #2%
  \fi
 \else
  \Err@{Whoa ... there's no \string\Figure,
   \string\Table, etc., here}%
 \fi
 \place@false}
\long\def\Aplace#1{\prevanish@
 \place@true\island@false
 #1%
 \place@ a\Aplace@
 \postvanish@}
\long\def\AAplace#1{\prevanish@\place@true\island@false
 #1%
 \place@ A\AAplace@
 \postvanish@}
\newif\ifAA@
\def\AAplace@{\AA@true\Aplace@\AA@false}
\let\AAlist@\empty
\def\Aplace@{\allowbreak
 \dimen@=\ht\islandbox@
 \advance\dimen@\abovebotfigskip
 \ht\islandbox@\dimen@
 \advance\dimen@\dp\islandbox@
 \storedim@
 \ifAA@
  \xdef\AAlist@{\AAlist@1}%
  \advancedimtopins@
 \else
  \xdef\AAlist@{\AAlist@0}%
  \ifnum\topinscount@>\@ne\else\advancedimtopins@\fi
 \fi
 \insert\topins{\penalty\z@\splittopskip\z@\floatingpenalty\z@
  \box\islandbox@}%
 \global\advance\topinscount@\@ne}
\long\def\Bplace#1{\prevanish@\place@true\island@false
 #1%
 \place@ b\Bplace@
 \postvanish@}
\def\Bplace@{\allowbreak
 \ifnum\topinscount@=\z@
  \setbox\z@\vbox{\vbox to-\belowtopfigskip{}}%
  \dimen@-\skip\topins
  \ht\z@\dimen@
  \storedim@
  \advancedimtopins@
  \insert\topins{\boxz@}%
  \global\advance\topinscount@\@ne
  \xdef\AAlist@{\AAlist@0}%
 \fi
 \dimen@\ht\islandbox@
 \advance\dimen@\abovebotfigskip
 \ht\islandbox@\dimen@
 \advance\dimen@\dp\islandbox@
 \storedim@
 \xdef\AAlist@{\AAlist@0}%
 \ifnum\topinscount@>\@ne\else\advancedimtopins@\fi
 \insert\topins{\penalty\z@\splittopskip\z@
  \floatingpenalty\z@
  \box\islandbox@}%
 \global\advance\topinscount@\@ne}
\def\breakisland@{\global\setbox\@ne\lastbox\global\skipi@\lastskip\unskip
 \global\setbox\thr@@\lastbox}%
\def\printisland@{\centerline{\box\thr@@}\nobreak\nointerlineskip
 \vskip\skipi@
 \ifdim\ht\@ne<\z@\box\@ne\else\centerline{\box\@ne}\fi}
\def\bottomfigs@{%
 \count@\@ne
 \loop
  \ifnum\count@<\flipcount@
  \nointerlineskip
  \vskip\abovebotfigskip
  \global\setbox\topins\vbox{\unvbox\topins\setboxzl@
   \unvbox\z@
   \breakisland@}%
  \printisland@
  \advance\count@\@ne
  \repeat}
\def\resetdimtopins@{%
 \global\advance\topinscount@-\flipcount@
 \global\setbox\topinsdims@\vbox
  {\unvbox\topinsdims@
   \count@\z@
   \DN@##1##2\next@{\gdef\AAlist@{##2}}%
   \loop
    \ifnum\count@<\flipcount@\setboxzl@
    \expandafter\next@\AAlist@\next@
    \advance\count@\@ne
    \repeat
   \dimen@\z@
   \count@\z@
   \setbox\tw@\vbox{}%
   \edef\nextiii@{\AAlist@}%
   \DN@##1##2\next@{\DNii@{##1}\def\nextiii@{##2}}%
   \loop
    \test@false
    \ifnum\count@<\topinscount@
    \expandafter\next@\nextiii@\next@
     \ifnum\count@<\tw@
      \test@true
     \else
      \if\nextii@ 1\test@true\fi
     \fi
    \fi
    \iftest@
     \setboxzl@
     \advance\dimen@\wdz@
     \setbox\tw@\vbox{\boxz@\unvbox\tw@}%
     \advance\count@\@ne
    \repeat
    \unvbox\tw@
    \global\dimen\topins\dimen@}}
\def\Place@#1#2{%
 \ifisland@
  \ifhmode
   \ifPar@
    \ifnum\Parcount@>5
     \Err@{Only 5 \string\place's allowed per
       \string\Par...\noexpand\endPar paragraph}%
    \else
     \expandafter\expandafter\expandafter\global\expandafter\setbox
      \csname Parfigbox\number\Parcount@\endcsname\box\islandbox@
     \global\advance\Parcount@\@ne
     \xdef\Parlist@{\Parlist@#1}%
     \vadjust{\break}%
    \fi
   \else
    \Err@{\noexpand#2allowed only in a \string\Par...\noexpand\endPar
     paragraph}%
   \fi
  \else
   #2%
  \fi
 \else
  \Err@{Who ... there's no \string\Figure, \string\Table,
   etc., here}%
 \fi
 \place@false}
\newif\ifC@
\newdimen\Cdim@
\long\def\Cplace#1{\prevanish@\place@true\island@false
 #1%
 \Place@ c\Cplace@
 \postvanish@}
\def\Cplace@{\allowbreak
 \ifnum\topinscount@>\z@\else
  \global\C@true\global\Cdim@\pagetotal\fi
 \Aplace@}
\long\def\Mplace#1{\prevanish@\place@true\island@false
 #1%
 \Place@ m\Mplace@
 \postvanish@}
\long\def\MXplace#1{\prevanish@\place@true\island@false
 #1%
 \Place@ M\MXplace@
 \postvanish@}
\newif\ifMX@
\def\MXplace@{\MX@true\Mplace@\MX@false}
\def\Mplace@{\allowbreak
 \dimen@\ht\islandbox@\advance\dimen@\dp\islandbox@
 \ifdim\pagetotal=\z@\else
  \ifdim\lastskip<\abovebotfigskip\advance\dimen@\abovebotfigskip
  \advance\dimen@-\lastskip\fi
 \fi
 \advance\dimen@\pagetotal
 \ifdim\dimen@>\pagegoal
  \Aplace@
 \else
  \nointerlineskip
  \ifdim\lastskip<\abovebotfigskip\removelastskip\vskip\abovebotfigskip\fi
  \setbox\z@\vbox{\unvbox\islandbox@
   \breakisland@}%
  \printisland@
  \ifnum\topinscount@=\z@
   \setbox\z@\vbox{\vbox to-\belowtopfigskip{}}%
   \dimen@-\skip\topins
   \ht\z@\dimen@
   \storedim@
   \advancedimtopins@
   \insert\topins{\boxz@}%
   \global\advance\topinscount@\@ne
   \xdef\AAlist@{\AAlist@0}%
  \fi
  \ifMX@
   \ifnum\topinscount@=\@ne
    \setbox\z@\vbox{\vbox to-\abovebotfigskip{}}%
    \ht\z@\z@
    \dimen@\z@
    \storedim@
    \advancedimtopins@
    \insert\topins{\boxz@}%
    \global\advance\topinscount@\@ne
    \xdef\AAlist@{\AAlist@0}%
   \fi
  \fi
  \nointerlineskip
  \vskip\belowtopfigskip
 \fi}
\expandafter\newbox\csname Parbox1\endcsname
\expandafter\newbox\csname Parbox2\endcsname
\expandafter\newbox\csname Parbox3\endcsname
\expandafter\newbox\csname Parbox4\endcsname
\expandafter\newbox\csname Parbox5\endcsname
\def\endPar{\egroup
 \count@\@ne
 {\vbadness\@M\vfuzz\maxdimen\splitmaxdepth\maxdimen\splittopskip\ht\strutbox
 \setbox\z@\vsplit\Parbox@ to\ht\Parbox@
 \loop
  \ifnum\count@<\Parcount@
  \expandafter\expandafter\expandafter\global\expandafter\setbox
   \csname Parbox\number\count@\endcsname\vsplit\Parbox@ to\ht\Parbox@
  \count@@\count@\advance\count@@\@ne
  \global\csname Parprev\number\count@@\endcsname
   \dp\csname Parbox\number\count@\endcsname
  \advance\count@\@ne
  \repeat}%
 \vskip\parskip
 \count@\@ne
 \def\nextv@##1##2\nextv@{\DN@{##1}\gdef\Parlist@{##2}}%
 \loop
  \ifnum\count@<\Parcount@
   \dimen@\csname Parprev\number\count@\endcsname
   \advance\dimen@\ht\strutbox
   \ifdim\dimen@<\baselineskip
    \advance\dimen@-\baselineskip\vskip-\dimen@
   \else
    \vskip\lineskip
   \fi
   \unvbox\csname Parbox\number\count@\endcsname
   \global\setbox\islandbox@\box\csname Parfigbox\number\count@\endcsname
   \expandafter\nextv@\Parlist@\nextv@
   \if a\next@\Aplace@\else
   \if A\next@\AAplace@\else
   \if b\next@\Bplace@\else
   \if c\next@\Cplace@\else
   \if m\next@\Mplace@\else
   \if M\next@\MXplace@\fi\fi\fi\fi\fi\fi
  \advance\count@\@ne
  \repeat
 \global\Par@false
 \ifvoid\Parbox@
  \prevdepth\csname Parprev\number\count@\endcsname
 \else
  \dimen@\csname Parprev\number\count@\endcsname\advance\dimen@\ht\strutbox
  \ifdim\dimen@<\baselineskip
    \advance\dimen@-\baselineskip\vskip-\dimen@
  \else
    \vskip\lineskip
  \fi
  \dimen@\dp\Parbox@
  \unvbox\Parbox@
  \prevdepth\dimen@
 \fi}
\def\folio{{\page@F\page@S{\page@P\page@N{\number\page@C}\page@Q}}}
\def\advancepageno{\global\advance\pageno\@ne}
\newif\ifspecialsplit@
\newbox\outbox@
\let\shipout@\shipout
\def\plainoutput{\specialsplit@false\ifvoid\topins\else\ifdim\ht\topins=\z@
 \specialsplit@true\advance\minpagesize-\skip\topins\fi\fi
 \fliptopins@
 \setbox\outbox@\vbox{\makeheadline\pagebody\makefootline}%
 {\noexpands@\let\style\relax
 \shipout@\box\outbox@}%
 \advancepageno
 \resetdimtopins@
 \ifvoid\@cclv\else\unvbox\@cclv\penalty\outputpenalty\fi
 \ifnum\outputpenalty>-\@MM\else\dosupereject\fi}
\def\pagebody{\vbox to\vsize{\boxmaxdepth\maxdepth
 \ifvoid\margin@\else
 \rlap{\kern\hsize\vbox to\z@{\kern4\p@\box\margin@\vss}}\fi
 \pagecontents}}
\newif\ifonlytop@
\def\pagecontents{%
 \onlytop@false
 \ifdim\ht\@cclv<\minpagesize\ifnum\flipcount@<\tw@\ifvoid\footins
  \onlytop@true\fi\fi\fi
 \test@false
 \ifC@
  \ifnum\flipcount@=\@ne
   \global\multiply\Cdim@\tw@
   \ifdim\Cdim@>\ht\@cclv
    \test@true
   \fi
  \fi
 \fi
 \global\C@false
 \iftest@
  \dimen@\ht\@cclv
  \advance\dimen@\skip\topins
  {\vfuzz\maxdimen\vbadness\@M
  \splitmaxdepth\maxdepth\splittopskip\topskip
  \setbox\z@\vsplit\@cclv to\dimen@
  \unvbox\z@}%
  \global\setbox\topins\vbox{\unvbox\topins
   \global\setbox\@ne\lastbox}%
  \setbox\z@\vbox{\unvbox\@ne
   \breakisland@}%
  \nointerlineskip
  \vskip\abovebotfigskip
  \printisland@
 \else
  \ifnum\flipcount@>\z@
   \global\setbox\topins\vbox{\unvbox\topins\global\setbox\@ne\lastbox}%
   \setbox\z@\vbox{\unvbox\@ne
    \breakisland@}%
   \printisland@
   \ifonlytop@\kern-\prevdepth\vfill\else\vskip\belowtopfigskip\fi
  \fi
 \fi
 \ifdim\ht\@cclv<\minpagesize
  \ifonlytop@\else\vfill\fi
 \else
  \ifspecialsplit@
   {\vfuzz\maxdimen\vbadness\@M
   \splitmaxdepth\maxdepth\splittopskip\topskip
   \dimen@ii\ht\@cclv \advance\dimen@ii\skip\topins
   \setbox\z@\vsplit\@cclv to\dimen@ii
   \unvbox\z@}%
  \else
   \unvbox\@cclv
  \fi
 \fi
 \bottomfigs@
 \ifvoid\footins\else\vskip\skip\footins\footnoterule\unvbox\footins\fi}
\newread\readdata@
\def\readthedata@#1{\expandafter
 \ifx\csname#1@D\endcsname\relax
  \immediate\openin\readdata@=#1.dat
  \ifeof\readdata@
   \Err@{No file #1.dat}%
  \else
   {\endlinechar\m@ne\gdef\Next@{}%
   \DNii@##1 ##2 ##3pt{\global\data@ht##1\global\data@dp##2%
    \global\data@wd##3pt}%
   \loop
    \ifeof\readdata@
    \else
    \read\readdata@ to\next@
    \ifx\next@\empty\else
     \edef\next@{\expandafter\nextii@\next@}%
     \expandafter\rightadd@\next@\to\Next@
    \fi
    \repeat}%
   \immediate\closein\readdata@
   \expandafter\expandafter\expandafter\global\expandafter
    \let\csname#1@D\endcsname\Next@\global\let\Next@\relax
  \fi
 \fi}
\newdimen\data@ht
\newdimen\data@dp
\newdimen\data@wd
\newif\ifgetdata@
\def\getdata@#1#2{\global\getdata@true\count@#2\relax
 {\let\\\or\xdef\Next@{\ifcase\number\count@#1\else
 \global\noexpand\getdata@false\fi}}\Next@}
\def\paste#1#2{\readthedata@{#1}%
 \getdata@{\csname#1@D\endcsname}{#2}%
 \ifgetdata@
 \dimen@\data@ht \advance\dimen@\data@dp
  \hbox{\special{dvipaste: #1 #2}%
   \lower\data@dp\vbox to\dimen@{\hbox to\data@wd{}\vfil}}%
 \else
  {\lccode`\Z=`\#\lccode`\N=`\N\lccode`\F=`\F%
   \lowercase{\Err@{No data for File [#1], Z#2}}}%
 \fi}
\newdimen\httable
\newdimen\dptable
\newdimen\wdtable
\def\measuretable#1#2{\readthedata@{#1}%
 \getdata@{\csname#1@D\endcsname}{#2}%
 \ifgetdata@
  \httable\data@ht \dptable\data@dp \wdtable\data@wd
 \else
  {\lccode`\Z=`\#\lccode`\N=`\N\lccode`\F=`\F%
  \lowercase{\Err@{No data for File [#1], Z#2}}}%
 \fi}
\def\East#1#2{\setboxz@h{$\m@th\ssize\;{#1}\;\;$}%
 \setbox\tw@\hbox{$\m@th\ssize\;{#2}\;\;$}\setbox4=\hbox{$\m@th#2$}%
 \dimen@\minaw@
 \ifdim\wdz@>\dimen@\dimen@\wdz@\fi\ifdim\wd\tw@>\dimen@\dimen@\wd\tw@\fi
 \ifdim\wd4 >\z@
  \mathrel{\mathop{\hbox to\dimen@{\rightarrowfill}}\limits^{#1}_{#2}}%
 \else
  \mathrel{\mathop{\hbox to\dimen@{\rightarrowfill}}\limits^{#1}}%
 \fi}
\def\West#1#2{\setboxz@h{$\m@th\ssize\;\;{#1}\;$}%
 \setbox\tw@\hbox{$\m@th\ssize\;\;{#2}\;$}\setbox4=\hbox{$\m@th#2$}%
 \dimen@\minaw@
 \ifdim\wdz@>\dimen@\dimen@\wdz@\fi\ifdim\wd\tw@>\dimen@\dimen@\wd\tw@\fi
 \ifdim\wd4 >\z@
  \mathrel{\mathop{\hbox to\dimen@{\leftarrowfill}}\limits^{#1}_{#2}}%
 \else
  \mathrel{\mathop{\hbox to\dimen@{\leftarrowfill}}\limits^{#1}}%
 \fi}
\font\arrow@i=lams1
\font\arrow@ii=lams2
\font\arrow@iii=lams3
\font\arrow@iv=lams4
\font\arrow@v=lams5
\newdimen\standardcgap
\standardcgap40\p@
\newdimen\hunit
\hunit\tw@\p@
\newdimen\standardrgap
\standardrgap32\p@
\newdimen\vunit
\vunit1.6\p@
\def\Cgaps#1{\RIfM@
 \standardcgap#1\standardcgap\relax\hunit#1\hunit\relax
 \else\nonmatherr@\Cgaps\fi}
\def\Rgaps#1{\RIfM@
 \standardrgap#1\standardrgap\relax\vunit#1\vunit\relax
 \else\nonmatherr@\Rgaps\fi}
\newdimen\getdim@
\def\getcgap@#1{\ifcase#1\or\getdim@\z@\else\getdim@\standardcgap\fi}
\def\getrgap@#1{\ifcase#1\getdim@\z@\else\getdim@\standardrgap\fi}
\def\cgaps{\RIfM@\expandafter\cgaps@\else\expandafter\nonmatherr@
 \expandafter\cgaps\fi}
\def\cgaps@{\ifnum\catcode`\;=\active\expandafter\cgapsA@\else
 \expandafter\cgapsO@\fi}
\def\cgapsO@#1{\toks@{\ifcase\i@\or\getdim@=\z@}%
 \gaps@@\standardcgap#1;\gaps@@\gaps@@
 \edef\next@{\the\toks@\noexpand\else\noexpand\getdim@\noexpand\standardcgap
  \noexpand\fi}%
 \toks@=\expandafter{\next@}%
 \edef\getcgap@##1{\i@##1\relax\the\toks@}\toks@{}}
{\catcode`\;=\active
 \gdef\cgapsA@#1{\toks@{\ifcase\i@\or\getdim@=\z@}%
 \gaps@@\standardcgap#1;\gaps@@\gaps@@
 \edef\next@{\the\toks@\noexpand\else\noexpand\getdim@\noexpand\standardcgap
  \noexpand\fi}%
 \toks@=\expandafter{\next@}%
 \edef\getcgap@##1{\i@##1\relax\the\toks@}\toks@{}}
}
\def\Gaps@@{\gaps@@}
\def\gaps@@#1#2;#3{\mgaps@#1#2\mgaps@
 \edef\next@{\the\toks@\noexpand\or\noexpand\getdim@
  \noexpand#1\the\mgapstoks@@}%
 \toks@\expandafter{\next@}%
 \DN@{#3}%
 \ifx\next@\Gaps@@\def\next@##1\gaps@@{}\else
  \def\next@{\gaps@@#1#3}\fi\next@}
{\catcode`\;=\active
 \gdef\rgaps#1{\RIfM@{\ifnum\catcode`\;=\active\def;{\string;}\fi
   \xdef\Next@{\noexpand\rgaps@{#1}}}%
  \Next@\edef\getrgap@##1{\i@##1\relax\the\toks@}\toks@{}\else
  \nonmatherr@\rgaps\fi}
}
\def\rgaps@#1{\toks@{\ifcase\i@\getdim@=\z@}%
 \gaps@@\standardrgap#1;\gaps@@\gaps@@
 \edef\next@{\the\toks@\noexpand\else\noexpand\getdim@\noexpand\standardrgap
  \noexpand\fi}%
 \toks@=\expandafter{\next@}}
\newbox\ZER@
\def\mgaps@#1{\let\mgapsnext@#1\FNSS@\mgaps@@}
\def\mgaps@@{\ifx\next\w\expandafter\mgaps@@@\else
 \expandafter\mgaps@@@@\fi}
\newtoks\mgapstoks@@
\def\mgaps@@@@#1\mgaps@{\getdim@\mgapsnext@\getdim@#1\getdim@
 \edef\next@{\noexpand\getdim@\the\getdim@}%
 \mgapstoks@@\expandafter{\next@}}
\def\mgaps@@@\w#1#2\mgaps@{\mgaps@@@@#2\mgaps@
 \setbox\ZER@\hbox{$\m@th\hskip15\p@\tsize@#1$}%
 \dimen@\wd\ZER@
 \ifdim\dimen@>\getdim@\getdim@\dimen@\fi
 \edef\next@{\noexpand\getdim@\the\getdim@}%
 \mgapstoks@@\expandafter{\next@}}
\def\changewidth#1#2{\setbox\ZER@{$\m@th#2}%
 \hbox to\wd\ZER@{\hss$\m@th#1$\hss}}
\atdef@({\FN@\ARROW@}
\def\ARROW@{\ifx\next)\let\next@\OPTIONS@\else
 \DN@{\csname\string @(\endcsname}\fi\next@}
\newif\ifoptions@
\def\OPTIONS@){\ifoptions@\let\next@\relax\else
 \DN@{\global\options@true\begingroup\optioncodes@}\fi\next@}
\newif\ifN@
\newif\ifE@
\newif\ifNESW@
\newif\ifH@
\newif\ifV@
\newif\ifHshort@
\expandafter\def\csname\string @(\endcsname #1,#2){%
 \ifoptions@\expandafter\endgroup\fi
 \N@false\E@false\H@false\V@false\Hshort@false
 \ifnum#1>\z@\E@true\fi
 \ifnum#1=\z@\V@true\global\tX@false\global\tY@false\global\a@false\fi
 \ifnum#2>\z@\N@true\fi
 \ifnum#2=\z@\H@true\global\tX@false\global\tY@false\global\a@false
  \ifshort@\Hshort@true\fi\fi
 \NESW@false
 \ifN@\ifE@\NESW@true\fi\else\ifE@\else\NESW@true\fi\fi
 \arrow@{#1}{#2}%
 \global\options@false
 \global\scount@\z@\global\tcount@\z@\global\arrcount@\z@
 \global\s@false\global\sxdimen@\z@\global\sydimen@\z@
 \global\tX@false\global\tXdimen@i\z@\global\tXdimen@ii\z@
 \global\tY@false\global\tYdimen@i\z@\global\tYdimen@ii\z@
 \global\a@false\global\exacount@\z@
 \global\x@false\global\xdimen@\z@
 \global\X@false\global\Xdimen@\z@
 \global\y@false\global\ydimen@\z@
 \global\Y@false\global\Ydimen@\z@
 \global\p@false\global\pdimen@\z@
 \global\label@ifalse\global\label@iifalse
 \global\dl@ifalse\global\ldimen@i\z@
 \global\dl@iifalse\global\ldimen@ii\z@
 \global\short@false\global\unshort@false}
\newif\iflabel@i
\newif\iflabel@ii
\newcount\scount@
\newcount\tcount@
\newcount\arrcount@
\newif\ifs@
\newdimen\sxdimen@
\newdimen\sydimen@
\newif\iftX@
\newdimen\tXdimen@i
\newdimen\tXdimen@ii
\newif\iftY@
\newdimen\tYdimen@i
\newdimen\tYdimen@ii
\newif\ifa@
\newcount\exacount@
\newif\ifx@
\newdimen\xdimen@
\newif\ifX@
\newdimen\Xdimen@
\newif\ify@
\newdimen\ydimen@
\newif\ifY@
\newdimen\Ydimen@
\newif\ifp@
\newdimen\pdimen@
\newif\ifdl@i
\newif\ifdl@ii
\newdimen\ldimen@i
\newdimen\ldimen@ii
\newif\ifshort@
\newif\ifunshort@
\def\zero@#1{\ifnum\scount@=\z@
 \if#1e\global\scount@\m@ne\else
 \if#1t\global\scount@\tw@\else
 \if#1h\global\scount@\thr@@\else
 \if#1'\global\scount@6 \else
 \if#1`\global\scount@7 \else
 \if#1(\global\scount@8 \else
 \if#1)\global\scount@9 \else
 \if#1s\global\scount@12 \else
 \if#1H\global\scount@13 \else
 \Err@{\Invalid@@ option \string\0}\fi\fi\fi\fi\fi\fi\fi\fi\fi
 \fi}
\def\one@#1{\ifnum\tcount@=\z@
 \if#1e\global\tcount@\m@ne\else
 \if#1h\global\tcount@\tw@\else
 \if#1t\global\tcount@\thr@@\else
 \if#1'\global\tcount@4 \else
 \if#1`\global\tcount@5 \else
 \if#1(\global\tcount@\ten@ \else
 \if#1)\global\tcount@11 \else
 \if#1s\global\tcount@12 \else
 \if#1H\global\tcount@13 \else
 \Err@{\Invalid@@ option \string\1}\fi\fi\fi\fi\fi\fi\fi\fi\fi
 \fi}
\def\a@#1{\ifnum\arrcount@=\z@
 \if#10\global\arrcount@\m@ne\else
 \if#1+\global\arrcount@\@ne\else
 \if#1-\global\arrcount@\tw@\else
 \if#1=\global\arrcount@\thr@@\else
 \Err@{\Invalid@@ option \string\a}\fi\fi\fi\fi
 \fi}
\def\ds@{\ifnum\catcode`\;=\active\expandafter\dsA@\else
 \expandafter\dsO@\fi}
\def\dsO@(#1;#2){\ds@@{#1}{#2}}
\def\ds@@#1#2{\ifs@\else
 \global\s@true
 \global\sxdimen@\hunit\global\sxdimen@#1\sxdimen@\relax
 \global\sydimen@\vunit\global\sydimen@#2\sydimen@\relax
 \fi}
\def\dtX@{\ifnum\catcode`\;=\active\expandafter\dtXA@\else
 \expandafter\dtXO@\fi}
\def\dtXO@(#1;#2){\dtX@@{#1}{#2}}
\def\dtX@@#1#2{\iftX@\else
 \global\tX@true
 \global\tXdimen@i\hunit\global\tXdimen@i#1\tXdimen@i\relax
 \global\tXdimen@ii\vunit\global\tXdimen@ii#2\tXdimen@ii\relax
 \fi}
\def\dtY@{\ifnum\catcode`\;=\active\expandafter\dtYA@\else
 \expandafter\dtYO@\fi}
\def\dtYO@(#1;#2){\dtY@@{#1}{#2}}
\def\dtY@@#1#2{\iftY@\else
 \global\tY@true
 \global\tYdimen@i\hunit\global\tYdimen@i#1\tYdimen@i\relax
 \global\tYdimen@ii\vunit\global\tYdimen@ii#2\tYdimen@ii\relax
 \fi}
{\catcode`\;=\active
 \gdef\dsA@(#1;#2){\ds@@{#1}{#2}}
 \gdef\dtXA@(#1;#2){\dtX@@{#1}{#2}}
 \gdef\dtYA@(#1;#2){\dtY@@{#1}{#2}}
}
\def\da@#1{\ifa@\else\global\a@true\global\exacount@#1\relax\fi}
\def\dx@#1{\ifx@\else
 \global\x@true
 \global\xdimen@\hunit\global\xdimen@#1\xdimen@\relax
 \fi}
\def\dX@#1{\ifX@\else
 \global\X@true
 \global\Xdimen@\hunit\global\Xdimen@#1\Xdimen@\relax
 \fi}
\def\dy@#1{\ify@\else
 \global\y@true
 \global\ydimen@\vunit\global\ydimen@#1\ydimen@\relax
 \fi}
\def\dY@#1{\ifY@\else
 \global\Y@true
 \global\Ydimen@\vunit\global\Ydimen@#1\Ydimen@\relax
 \fi}
\def\p@@#1{\ifp@\else
 \global\p@true
 \global\pdimen@\hunit\global\divide\pdimen@\tw@
 \global\pdimen@#1\pdimen@\relax
 \fi}
\def\L@#1{\iflabel@i\else
 \global\label@itrue\gdef\label@i{#1}%
 \fi}
\def\l@#1{\iflabel@ii\else
 \global\label@iitrue\gdef\label@ii{#1}%
 \fi}
\def\dL@#1{\ifdl@i\else
 \global\dl@itrue\global\ldimen@i\hunit\global\ldimen@i#1\ldimen@i\relax
 \fi}
\def\dl@#1{\ifdl@ii\else
 \global\dl@iitrue\global\ldimen@ii\hunit\global\ldimen@ii#1\ldimen@ii\relax
 \fi}
\def\s@{\ifunshort@\else\global\short@true\fi}
\def\uns@{\ifshort@\else\global\unshort@true\global\short@false\fi}
\def\optioncodes@{\let\0\zero@\let\1\one@\let\a\a@\let\ds\ds@\let\dtX\dtX@
 \let\dtY\dtY@\let\da\da@\let\dx\dx@\let\dX\dX@\let\dY\dY@\let\dy\dy@
 \let\p\p@@\let\L\L@\let\l\l@\let\dL\dL@\let\dl\dl@\let\s\s@\let\uns\uns@}
\def\slopes@{\\161\\152\\143\\134\\255\\126\\357\\238\\349\\45{10}\\56{11}%
 \\11{12}\\65{13}\\54{14}\\43{15}\\32{16}\\53{17}\\21{18}\\52{19}\\31{20}%
 \\41{21}\\51{22}\\61{23}}
\newcount\tan@i
\newcount\tan@ip
\newcount\tan@ii
\newcount\tan@iip
\newdimen\slope@i
\newdimen\slope@ip
\newdimen\slope@ii
\newdimen\slope@iip
\newcount\angcount@
\newcount\extracount@
\def\slope@{{\slope@i\secondy@\advance\slope@i-\firsty@
 \ifN@\else\multiply\slope@i\m@ne\fi
 \slope@ii\secondx@\advance\slope@ii-\firstx@
 \ifE@\else\multiply\slope@ii\m@ne\fi
 \ifdim\slope@ii<\z@
  \global\tan@i6 \global\tan@ii\@ne\global\angcount@23
 \else
  \dimen@\slope@i\multiply\dimen@6
  \ifdim\dimen@<\slope@ii
   \global\tan@i\@ne\global\tan@ii6 \global\angcount@\@ne
  \else
   \dimen@\slope@ii\multiply\dimen@6
   \ifdim\dimen@<\slope@i
    \global\tan@i6 \global\tan@ii\@ne\global\angcount@23
   \else
    \global\tan@ip\z@\global\tan@iip\@ne
    \def\\##1##2##3{\global\angcount@##3\relax
     \slope@ip\slope@i\slope@iip\slope@ii
     \multiply\slope@iip##1\relax\multiply\slope@ip##2\relax
     \ifdim\slope@iip<\slope@ip
      \global\tan@ip##1\relax\global\tan@iip##2\relax
     \else
      \global\tan@i##1\relax\global\tan@ii##2\relax
      \def\\####1####2####3{}%
     \fi}%
    \slopes@
    \slope@i\secondy@\advance\slope@i-\firsty@
    \ifN@\else\multiply\slope@i\m@ne\fi
    \multiply\slope@i\tan@ii\multiply\slope@i\tan@iip\multiply\slope@i\tw@
    \count@\tan@i\multiply\count@\tan@iip
    \extracount@\tan@ip\multiply\extracount@\tan@ii
    \advance\count@\extracount@
    \slope@ii\secondx@\advance\slope@ii-\firstx@
    \ifE@\else\multiply\slope@ii\m@ne\fi
    \multiply\slope@ii\count@
    \ifdim\slope@i<\slope@ii
     \global\tan@i\tan@ip\global\tan@ii\tan@iip
     \global\advance\angcount@\m@ne
    \fi
   \fi
  \fi
 \fi}%
}
\def\slope@a#1{{\def\\##1##2##3{\ifnum##3=#1\global\tan@i##1\relax
 \global\tan@ii##2\relax\fi}\slopes@}}
\newcount\i@
\newcount\j@
\newcount\colcount@
\newcount\Colcount@
\newcount\tcolcount@
\newdimen\rowht@
\newdimen\rowdp@
\newcount\rowcount@
\newcount\Rowcount@
\newcount\maxcolrow@
\newtoks\colwidthtoks@
\newtoks\Rowheighttoks@
\newtoks\Rowdepthtoks@
\newtoks\widthtoks@
\newtoks\Widthtoks@
\newtoks\heighttoks@
\newtoks\Heighttoks@
\newtoks\depthtoks@
\newtoks\Depthtoks@
\newif\iffirstCDcr@
\def\dotoks@i{%
 \global\widthtoks@\expandafter{\the\widthtoks@\else\getdim@\z@\fi}%
 \global\heighttoks@\expandafter{\the\heighttoks@\else\getdim@\z@\fi}%
 \global\depthtoks@\expandafter{\the\depthtoks@\else\getdim@\z@\fi}}
\def\dotoks@ii{%
 \global\widthtoks@{\ifcase\j@}%
 \global\heighttoks@{\ifcase\j@}%
 \global\depthtoks@{\ifcase\j@}}
\def\preCD@#1\endCD{\setbox\ZER@
 \vbox{%
  \def\arrow@##1##2{{}}%
  \global\rowcount@\m@ne\global\colcount@\z@\global\Colcount@\z@
  \global\firstCDcr@true\toks@{}%
  \global\widthtoks@{\ifcase\j@}%
  \global\Widthtoks@{\ifcase\i@}%
  \global\heighttoks@{\ifcase\j@}%
  \global\Heighttoks@{\ifcase\i@}%
  \global\depthtoks@{\ifcase\j@}%
  \global\Depthtoks@{\ifcase\i@}%
  \global\Rowheighttoks@{\ifcase\i@}%
  \global\Rowdepthtoks@{\ifcase\i@}%
  \Let@
  \everycr{%
   \noalign{%
    \global\advance\rowcount@\@ne
    \ifnum\colcount@<\Colcount@
    \else
     \global\Colcount@\colcount@\global\maxcolrow@\rowcount@
    \fi
    \global\colcount@\z@
    \iffirstCDcr@
     \global\firstCDcr@false
    \else
     \edef\next@{\the\Rowheighttoks@\noexpand\or\noexpand\getdim@\the\rowht@}%
      \global\Rowheighttoks@\expandafter{\next@}%
     \edef\next@{\the\Rowdepthtoks@\noexpand\or\noexpand\getdim@\the\rowdp@}%
      \global\Rowdepthtoks@\expandafter{\next@}%
     \global\rowht@\z@\global\rowdp@\z@
     \dotoks@i
     \edef\next@{\the\Widthtoks@\noexpand\or\the\widthtoks@}%
      \global\Widthtoks@\expandafter{\next@}%
     \edef\next@{\the\Heighttoks@\noexpand\or\the\heighttoks@}%
      \global\Heighttoks@\expandafter{\next@}%
     \edef\next@{\the\Depthtoks@\noexpand\or\the\depthtoks@}%
      \global\Depthtoks@\expandafter{\next@}%
     \dotoks@ii
    \fi}}%
  \tabskip\z@
  \halign{&\setbox\ZER@\hbox{\vrule\height\ten@\p@\width\z@\depth\z@     
   $\m@th\displaystyle{##}$}\copy\ZER@
   \ifdim\ht\ZER@>\rowht@\global\rowht@\ht\ZER@\fi
   \ifdim\dp\ZER@>\rowdp@\global\rowdp@\dp\ZER@\fi
   \global\advance\colcount@\@ne
   \edef\next@{\the\widthtoks@\noexpand\or\noexpand\getdim@\the\wd\ZER@}%
    \global\widthtoks@\expandafter{\next@}%
   \edef\next@{\the\heighttoks@\noexpand\or\noexpand\getdim@\the\ht\ZER@}%
    \global\heighttoks@\expandafter{\next@}%
   \edef\next@{\the\depthtoks@\noexpand\or\noexpand\getdim@\the\dp\ZER@}%
    \global\depthtoks@\expandafter{\next@}%
   \cr#1\crcr}}%
 \Rowcount@\rowcount@
 \global\Widthtoks@\expandafter{\the\Widthtoks@\fi\relax}%
 \edef\Width@##1##2{\i@##1\relax\j@##2\relax\the\Widthtoks@}%
 \global\Heighttoks@\expandafter{\the\Heighttoks@\fi\relax}%
 \edef\Height@##1##2{\i@##1\relax\j@##2\relax\the\Heighttoks@}%
 \global\Depthtoks@\expandafter{\the\Depthtoks@\fi\relax}%
 \edef\Depth@##1##2{\i@##1\relax\j@##2\relax\the\Depthtoks@}%
 \edef\next@{\the\Rowheighttoks@\noexpand\fi\relax}%
 \global\Rowheighttoks@\expandafter{\next@}%
 \edef\Rowheight@##1{\i@##1\relax\the\Rowheighttoks@}%
 \edef\next@{\the\Rowdepthtoks@\noexpand\fi\relax}%
 \global\Rowdepthtoks@\expandafter{\next@}%
 \edef\Rowdepth@##1{\i@##1\relax\the\Rowdepthtoks@}%
 \global\colwidthtoks@{\fi}%
 \setbox\ZER@\vbox{%
  \unvbox\ZER@
  \count@\rowcount@
  \loop
   \unskip\unpenalty
   \setbox\ZER@\lastbox
   \ifnum\count@>\maxcolrow@\advance\count@\m@ne
   \repeat
  \hbox{%
   \unhbox\ZER@
   \count@\z@
   \loop
    \unskip
    \setbox\ZER@\lastbox
    \edef\next@{\noexpand\or\noexpand\getdim@\the\wd\ZER@\the\colwidthtoks@}%
     \global\colwidthtoks@\expandafter{\next@}%
    \advance\count@\@ne
    \ifnum\count@<\Colcount@
    \repeat}}%
 \edef\next@{\noexpand\ifcase\noexpand\i@\the\colwidthtoks@}%
  \global\colwidthtoks@\expandafter{\next@}%
 \edef\Colwidth@##1{\i@##1\relax\the\colwidthtoks@}%
 \global\colwidthtoks@{}\global\Rowheighttoks@{}\global\Rowdepthtoks@{}%
 \global\widthtoks@{}\global\Widthtoks@{}\global\heighttoks@{}%
 \global\Heighttoks@{}\global\depthtoks@{}\global\Depthtoks@{}%
}
\newcount\xoff@
\newcount\yoff@
\newcount\endcount@
\newcount\rcount@
\newdimen\firstx@
\newdimen\firsty@
\newdimen\secondx@
\newdimen\secondy@
\newdimen\tocenter@
\newdimen\charht@
\newdimen\charwd@
\def\outside@{\Err@{This arrow points outside the \string\CD}}
\newif\ifsvertex@
\newif\iftvertex@
\def\arrow@#1#2{\global\xoff@#1\relax\global\yoff@#2\relax
 \count@\rowcount@\advance\count@-\yoff@
 \ifnum\count@<\@ne\outside@\else\ifnum\count@>\Rowcount@\outside@\fi\fi
 \count@\colcount@\advance\count@\xoff@
 \ifnum\count@<\@ne\outside@\else\ifnum\count@>\Colcount@\outside@\fi\fi
 \tcolcount@\colcount@\advance\tcolcount@\xoff@
 \Width@\rowcount@\colcount@\divide\getdim@\tw@\tocenter@-\getdim@
 \ifdim\getdim@=\z@
  \firstx@\z@\firsty@\mathaxis@\svertex@true
 \else
  \svertex@false
  \ifHshort@
   \Colwidth@\colcount@\divide\getdim@\tw@
   \ifE@ \firstx@\getdim@ \else \firstx@-\getdim@ \fi
  \else
   \ifE@ \firstx@\getdim@ \else \firstx@-\getdim@ \fi
  \fi
  \ifE@
   \ifH@ \advance\firstx@\thr@@\p@ \else \advance\firstx@-\thr@@\p@ \fi  
  \else
   \ifH@ \advance\firstx@-\thr@@\p@ \else \advance\firstx@\thr@@\p@ \fi  
  \fi
  \ifN@
   \Height@\rowcount@\colcount@ \firsty@\getdim@                         
   \ifV@ \advance\firsty@\thr@@\p@ \fi                                   
  \else
   \ifV@
    \Depth@\rowcount@\colcount@ \firsty@-\getdim@                        
    \advance\firsty@-\thr@@\p@                                           
   \else
    \firsty@\z@                                                          
   \fi
  \fi
 \fi
 \ifV@
 \else
  \Colwidth@\colcount@\divide\getdim@\tw@
  \ifE@\secondx@\getdim@\else\secondx@-\getdim@\fi
  \ifE@\else\getcgap@\colcount@\advance\secondx@-\getdim@\fi
  \endcount@\colcount@\advance\endcount@\xoff@
  \count@\colcount@
  \ifE@
   \advance\count@\@ne
   \loop
    \ifnum\count@<\endcount@
    \Colwidth@\count@\advance\secondx@\getdim@
    \getcgap@\count@\advance\secondx@\getdim@
    \advance\count@\@ne
    \repeat
  \else
   \advance\count@\m@ne
   \loop
    \ifnum\count@>\endcount@
    \Colwidth@\count@\advance\secondx@-\getdim@
    \getcgap@\count@\advance\secondx@-\getdim@
    \advance\count@\m@ne
    \repeat
  \fi
  \Colwidth@\count@\divide\getdim@\tw@
  \ifHshort@
  \else
   \ifE@\advance\secondx@\getdim@\else\advance\secondx@-\getdim@\fi
  \fi
  \ifE@\getcgap@\count@\advance\secondx@\getdim@\fi
  \rcount@\rowcount@\advance\rcount@-\yoff@
  \Width@\rcount@\count@\divide\getdim@\tw@
  \tvertex@false
  \ifH@\ifdim\getdim@=\z@\tvertex@true\Hshort@false\fi\fi
  \ifHshort@
  \else
   \ifE@\advance\secondx@-\getdim@\else\advance\secondx@\getdim@\fi
  \fi
  \iftvertex@
   \advance\secondx@.4\p@
  \else
   \ifE@\advance\secondx@-\thr@@\p@\else\advance\secondx@\thr@@\p@\fi    
  \fi
 \fi
 \ifH@
 \else
  \ifN@
   \Rowheight@\rowcount@\secondy@\getdim@
  \else
   \Rowdepth@\rowcount@\secondy@-\getdim@
   \getrgap@\rowcount@\advance\secondy@-\getdim@
  \fi
  \endcount@\rowcount@\advance\endcount@-\yoff@
  \count@\rowcount@
  \ifN@
   \advance\count@\m@ne
   \loop
    \ifnum\count@>\endcount@
    \Rowheight@\count@\advance\secondy@\getdim@
    \Rowdepth@\count@\advance\secondy@\getdim@
    \getrgap@\count@\advance\secondy@\getdim@
    \advance\count@\m@ne
    \repeat
  \else
   \advance\count@\@ne
   \loop
    \ifnum\count@<\endcount@
    \Rowheight@\count@\advance\secondy@-\getdim@
    \Rowdepth@\count@\advance\secondy@-\getdim@
    \getrgap@\count@\advance\secondy@-\getdim@
    \advance\count@\@ne
    \repeat
  \fi
  \tvertex@false
  \ifV@\Width@\count@\colcount@\ifdim\getdim@=\z@\tvertex@true\fi\fi
  \ifN@
   \getrgap@\count@\advance\secondy@\getdim@
   \Rowdepth@\count@\advance\secondy@\getdim@
   \iftvertex@
    \advance\secondy@\mathaxis@
   \else
    \Depth@\count@\tcolcount@\advance\secondy@-\getdim@
    \advance\secondy@-\thr@@\p@                                          
   \fi
  \else
   \Rowheight@\count@\advance\secondy@-\getdim@
   \iftvertex@
    \advance\secondy@\mathaxis@
   \else
    \Height@\count@\tcolcount@\advance\secondy@\getdim@
    \advance\secondy@\thr@@\p@                                           
   \fi
  \fi
 \fi
 \ifV@\else\advance\firstx@\sxdimen@\fi
 \ifH@\else\advance\firsty@\sydimen@\fi
 \iftX@
  \advance\secondy@\tXdimen@ii
  \advance\secondx@\tXdimen@i
  \slope@
 \else
  \iftY@
   \advance\secondy@\tYdimen@ii
   \advance\secondx@\tYdimen@i
   \slope@
   \secondy@\secondx@\advance\secondy@-\firstx@
   \ifNESW@\else\multiply\secondy@\m@ne\fi
   \multiply\secondy@\tan@i\divide\secondy@\tan@ii\advance\secondy@\firsty@
  \else
   \ifa@
    \slope@
    \ifNESW@\global\advance\angcount@\exacount@\else
     \global\advance\angcount@-\exacount@\fi
    \ifnum\angcount@>23 \global\angcount@23 \fi
    \ifnum\angcount@<\@ne\global\angcount@\@ne\fi
    \slope@a\angcount@
    \ifY@
     \advance\secondy@\Ydimen@
    \else
     \ifX@
      \advance\secondx@\Xdimen@
      \dimen@\secondx@\advance\dimen@-\firstx@
      \ifNESW@\else\multiply\dimen@\m@ne\fi
      \multiply\dimen@\tan@i\divide\dimen@\tan@ii
      \advance\dimen@\firsty@\secondy@\dimen@
     \fi
    \fi
   \else
    \ifH@\else\ifV@\else\slope@\fi\fi
   \fi
  \fi
 \fi
 \ifH@\else\ifV@\else\ifsvertex@\else
  \dimen@6\p@\multiply\dimen@\tan@ii
  \count@\tan@i\advance\count@\tan@ii\divide\dimen@\count@
  \ifE@\advance\firstx@\dimen@\else\advance\firstx@-\dimen@\fi
  \multiply\dimen@\tan@i\divide\dimen@\tan@ii
  \ifN@\advance\firsty@\dimen@\else\advance\firsty@-\dimen@\fi
 \fi\fi\fi
 \ifp@
  \ifH@\else\ifV@\else
   \getcos@\pdimen@\advance\firsty@\dimen@\advance\secondy@\dimen@
   \ifNESW@\advance\firstx@-\dimen@ii\else\advance\firstx@\dimen@ii\fi
  \fi\fi
 \fi
 \ifH@\else\ifV@\else
  \ifnum\tan@i>\tan@ii
   \charht@\ten@\p@\charwd@\ten@\p@
   \multiply\charwd@\tan@ii\divide\charwd@\tan@i
  \else
   \charwd@\ten@\p@\charht@\ten@\p@
   \divide\charht@\tan@ii\multiply\charht@\tan@i
  \fi
  \ifnum\tcount@=\thr@@
   \ifN@\advance\secondy@-.3\charht@\else\advance\secondy@.3\charht@\fi
  \fi
  \ifnum\scount@=\tw@
   \ifE@\advance\firstx@.3\charht@\else\advance\firstx@-.3\charht@\fi
  \fi
  \ifnum\tcount@=12
   \ifN@\advance\secondy@-\charht@\else\advance\secondy@\charht@\fi
  \fi
  \iftY@
  \else
   \ifa@
    \ifX@
    \else
     \secondx@\secondy@\advance\secondx@-\firsty@
     \ifNESW@\else\multiply\secondx@\m@ne\fi
     \multiply\secondx@\tan@ii\divide\secondx@\tan@i
     \advance\secondx@\firstx@
    \fi
   \fi
  \fi
 \fi\fi
 \ifH@\harrow@\else\ifV@\varrow@\else\arrow@@\fi\fi}
\newdimen\mathaxis@
\mathaxis@90\p@\divide\mathaxis@36
\def\harrow@b{\ifE@\hskip\tocenter@\hskip\firstx@\fi}
\def\harrow@bb{\ifE@\hskip\xdimen@\else\hskip\Xdimen@\fi}
\def\harrow@e{\ifE@\else\hskip-\firstx@\hskip-\tocenter@\fi}
\def\harrow@ee{\ifE@\hskip-\Xdimen@\else\hskip-\xdimen@\fi}
\def\harrow@{\dimen@\secondx@\advance\dimen@-\firstx@
 \ifE@\let\next@\rlap\else\multiply\dimen@\m@ne\let\next@\llap\fi
 \next@{%
  \harrow@b
  \smash{\raise\pdimen@\hbox to\dimen@
   {\harrow@bb\arrow@ii
    \ifnum\arrcount@=\m@ne\else\ifnum\arrcount@=\thr@@\else
     \ifE@
      \ifnum\scount@=\m@ne
      \else
       \ifcase\scount@\or\or\char118 \or\char117 \or\or\or\char119 \or
       \char120 \or\char121 \or\char122 \or\or\or\arrow@i\char125 \or
       \char117 \hskip\thr@@\p@\char117 \hskip-\thr@@\p@\fi
      \fi
     \else
      \ifnum\tcount@=\m@ne
      \else
       \ifcase\tcount@\char117 \or\or\char117 \or\char118 \or\char119 \or
       \char120 \or\or\or\or\or\char121 \or\char122 \or\arrow@i\char125
       \or\char117 \hskip\thr@@\p@\char117 \hskip-\thr@@\p@\fi
      \fi
     \fi
    \fi\fi
    \dimen@\mathaxis@\advance\dimen@.2\p@
    \dimen@ii\mathaxis@\advance\dimen@ii-.2\p@
    \ifnum\arrcount@=\m@ne
     \let\leads@\null
    \else
     \ifcase\arrcount@
      \def\leads@{\hrule\height\dimen@\depth-\dimen@ii}\or
      \def\leads@{\hrule\height\dimen@\depth-\dimen@ii}\or
      \def\leads@{\hbox to\ten@\p@{%
       \leaders\hrule\height\dimen@\depth-\dimen@ii\hfil
       \hfil
      \leaders\hrule\height\dimen@\depth-\dimen@ii\hskip\z@ plus2fil\relax
       \hfil
       \leaders\hrule\height\dimen@\depth-\dimen@ii\hfil}}\or
     \def\leads@{\hbox{\hbox to\ten@\p@{\dimen@\mathaxis@\advance\dimen@1.2\p@
       \dimen@ii\dimen@\advance\dimen@ii-.4\p@
       \leaders\hrule\height\dimen@\depth-\dimen@ii\hfil}%
       \kern-\ten@\p@
       \hbox to\ten@\p@{\dimen@\mathaxis@\advance\dimen@-1.2\p@
       \dimen@ii\dimen@\advance\dimen@ii-.4\p@
       \leaders\hrule\height\dimen@\depth-\dimen@ii\hfil}}}\fi
    \fi
    \cleaders\leads@\hfil
    \ifnum\arrcount@=\m@ne\else\ifnum\arrcount@=\thr@@\else
     \arrow@i
     \ifE@
      \ifnum\tcount@=\m@ne
      \else
       \ifcase\tcount@\char119 \or\or\char119 \or\char120 \or\char121 \or
       \char122 \or \or\or\or\or\char123 \or\char124 \or
       \char125 \or\char119 \hskip-\thr@@\p@\char119 \hskip\thr@@\p@\fi
      \fi
     \else
      \ifcase\scount@\or\or\char120 \or\char119 \or\or\or\char121 \or\char122
      \or\char123 \or\char124 \or\or\or\char125 \or
      \char119 \hskip-\thr@@\p@\char119 \hskip\thr@@\p@\fi
     \fi
    \fi\fi
    \harrow@ee}}%
  \harrow@e}%
 \iflabel@i
  \dimen@ii\z@\setbox\ZER@\hbox{$\m@th\tsize@@\label@i$}%
  \ifnum\arrcount@=\m@ne
  \else
   \advance\dimen@ii\mathaxis@
   \advance\dimen@ii\dp\ZER@\advance\dimen@ii\tw@\p@
   \ifnum\arrcount@=\thr@@\advance\dimen@ii\tw@\p@\fi
  \fi
  \advance\dimen@ii\pdimen@
  \next@{\harrow@b\smash{\raise\dimen@ii\hbox to\dimen@
   {\harrow@bb\hskip\tw@\ldimen@i\hfil\box\ZER@\hfil\harrow@ee}}\harrow@e}%
 \fi
 \iflabel@ii
  \ifnum\arrcount@=\m@ne
  \else
   \setbox\ZER@\hbox{$\m@th\tsize@\label@ii$}%
   \dimen@ii-\ht\ZER@\advance\dimen@ii-\tw@\p@
   \ifnum\arrcount@=\thr@@\advance\dimen@ii-\tw@\p@\fi
   \advance\dimen@ii\mathaxis@\advance\dimen@ii\pdimen@
   \next@{\harrow@b\smash{\raise\dimen@ii\hbox to\dimen@
    {\harrow@bb\hskip\tw@\ldimen@ii\hfil\box\ZER@\hfil\harrow@ee}}\harrow@e}%
  \fi
 \fi}
\let\tsize@\tsize
\def\tsizeCDlabels{\let\tsize@\tsize}
\def\ssizeCDlabels{\let\tsize@\ssize}
\def\tsize@@{\ifnum\arrcount@=\m@ne\else\tsize@\fi}
\def\varrow@{\dimen@\secondy@\advance\dimen@-\firsty@
 \ifN@\else\multiply\dimen@\m@ne\fi
 \setbox\ZER@\vbox to\dimen@
  {\ifN@\vskip-\Ydimen@\else\vskip\ydimen@\fi
   \ifnum\arrcount@=\m@ne\else\ifnum\arrcount@=\thr@@\else
    \hbox{\arrow@iii
     \ifN@
      \ifnum\tcount@=\m@ne
      \else
       \ifcase\tcount@\char117 \or\or\char117 \or\char118 \or\char119 \or
       \char120 \or\or\or\or\or\char121 \or\char122 \or\char123 \or
       \vbox{\hbox{\char117}\nointerlineskip\vskip\thr@@\p@
       \hbox{\char117}\vskip-\thr@@\p@}\fi
      \fi
     \else
      \ifcase\scount@\or\or\char118 \or\char117 \or\or\or\char119 \or
      \char120 \or\char121 \or\char122 \or\or\or\char123 \or
      \vbox{\hbox{\char117}\nointerlineskip\vskip\thr@@\p@
      \hbox{\char117}\vskip-\thr@@\p@}\fi
     \fi}%
    \nointerlineskip
   \fi\fi
   \ifnum\arrcount@=\m@ne
    \let\leads@\null
   \else
    \ifcase\arrcount@\let\leads@\vrule\or\let\leads@\vrule\or
    \def\leads@{\vbox to\ten@\p@{%
     \hrule\height1.67\p@\depth\z@\width.4\p@
     \vfil
     \hrule\height3.33\p@\depth\z@\width.4\p@
     \vfil
     \hrule\height1.67\p@\depth\z@\width.4\p@}}\or
    \def\leads@{\hbox{\vrule\height\p@\hskip\tw@\p@\vrule}}\fi
   \fi
  \cleaders\leads@\vfill\nointerlineskip
   \ifnum\arrcount@=\m@ne\else\ifnum\arrcount@=\thr@@\else
    \hbox{\arrow@iv
     \ifN@
      \ifcase\scount@\or\or\char118 \or\char117 \or\or\or\char119 \or
      \char120 \or\char121 \or\char122 \or\or\or\arrow@iii\char123 \or
      \vbox{\hbox{\char117}\nointerlineskip\vskip-\thr@@\p@
      \hbox{\char117}\vskip\thr@@\p@}\fi
     \else
      \ifnum\tcount@=\m@ne
      \else
       \ifcase\tcount@\char117 \or\or\char117 \or\char118 \or\char119 \or
       \char120 \or\or\or\or\or\char121 \or\char122 \or\arrow@iii\char123 \or
       \vbox{\hbox{\char117}\nointerlineskip\vskip-\thr@@\p@
       \hbox{\char117}\vskip\thr@@\p@}\fi
      \fi
     \fi}%
   \fi\fi
   \ifN@\vskip\ydimen@\else\vskip-\Ydimen@\fi}%
 \ifN@
  \dimen@ii\firsty@
 \else
  \dimen@ii-\firsty@\advance\dimen@ii\ht\ZER@\multiply\dimen@ii\m@ne
 \fi
 \rlap{\smash{\hskip\tocenter@\hskip\pdimen@\raise\dimen@ii\box\ZER@}}%
 \iflabel@i
  \setbox\ZER@\vbox to\dimen@{\vfil
   \hbox{$\m@th\tsize@@\label@i$}\vskip\tw@\ldimen@i\vfil}%
  \rlap{\smash{\hskip\tocenter@\hskip\pdimen@
  \ifnum\arrcount@=\m@ne\let\next@\relax\else\let\next@\llap\fi
  \next@{\raise\dimen@ii\hbox{\ifnum\arrcount@=\m@ne\hskip-.5\wd\ZER@\fi
   \box\ZER@\ifnum\arrcount@=\m@ne\else\hskip\tw@\p@\fi}}}}%
 \fi
 \iflabel@ii
  \ifnum\arrcount@=\m@ne
  \else
   \setbox\ZER@\vbox to\dimen@{\vfil
    \hbox{$\m@th\tsize@\label@ii$}\vskip\tw@\ldimen@ii\vfil}%
   \rlap{\smash{\hskip\tocenter@\hskip\pdimen@
   \rlap{\raise\dimen@ii\hbox{\ifnum\arrcount@=\thr@@\hskip4.5\p@\else
    \hskip2.5\p@\fi\box\ZER@}}}}%
  \fi
 \fi
}
\newdimen\goal@
\newdimen\shifted@
\newcount\Tcount@
\newcount\Scount@
\newbox\shaft@
\newcount\slcount@
\def\getcos@#1{%
 \ifnum\tan@i<\tan@ii
  \dimen@#1%
  \ifnum\slcount@<8 \count@9 \else \ifnum\slcount@<12 \count@8 \else
   \count@7 \fi\fi
  \multiply\dimen@\count@\divide\dimen@\ten@
  \dimen@ii\dimen@\multiply\dimen@ii\tan@i\divide\dimen@ii\tan@ii
 \else
  \dimen@ii#1%
  \count@-\slcount@\advance\count@24
  \ifnum\count@<8 \count@9 \else \ifnum\count@<12 \count@8
   \else\count@7 \fi\fi
  \multiply\dimen@ii\count@\divide\dimen@ii\ten@
  \dimen@\dimen@ii\multiply\dimen@\tan@ii\divide\dimen@\tan@i
 \fi}
\newdimen\adjust@
\def\Nnext@{\ifN@\let\next@\raise\else\let\next@\lower\fi}
\def\arrow@@{\slcount@\angcount@
 \ifNESW@
  \ifnum\angcount@<\ten@
   \let\arrowfont@\arrow@i\global\advance\angcount@\m@ne
   \global\multiply\angcount@13
  \else
   \ifnum\angcount@<19
    \let\arrowfont@\arrow@ii\global\advance\angcount@-\ten@
    \global\multiply\angcount@13
   \else
    \let\arrowfont@\arrow@iii\global\advance\angcount@-19
    \global\multiply\angcount@13
  \fi\fi
  \Tcount@\angcount@
 \else
  \ifnum\angcount@<5
   \let\arrowfont@\arrow@iii\global\advance\angcount@\m@ne
   \global\multiply\angcount@13 \global\advance\angcount@65
  \else
   \ifnum\angcount@<14
    \let\arrowfont@\arrow@iv\global\advance\angcount@-5
    \global\multiply\angcount@13
   \else
    \ifnum\angcount@<23
     \let\arrowfont@\arrow@v\global\advance\angcount@-14
     \global\multiply\angcount@13
    \else
     \let\arrowfont@\arrow@i\global\angcount@117
  \fi\fi\fi
  \ifnum\angcount@=117 \Tcount@115 \else\Tcount@\angcount@\fi
 \fi
 \Scount@\Tcount@
 \ifE@
  \ifnum\tcount@=\z@\advance\Tcount@\tw@\else\ifnum\tcount@=13
   \advance\Tcount@\tw@\else\advance\Tcount@\tcount@\fi\fi
  \ifnum\scount@=\z@\else\ifnum\scount@=13 \advance\Scount@\thr@@\else
   \advance\Scount@\scount@\fi\fi
 \else
  \ifcase\tcount@\advance\Tcount@\thr@@\or\or\advance\Tcount@\thr@@\or
  \advance\Tcount@\tw@\or\advance\Tcount@6 \or\advance\Tcount@7
  \or\or\or\or\or\advance\Tcount@8 \or\advance\Tcount@9 \or
  \advance\Tcount@12 \or\advance\Tcount@\thr@@\fi
  \ifcase\scount@\or\or\advance\Scount@\thr@@\or\advance\Scount@\tw@\or
  \or\or\advance\Scount@4 \or\advance\Scount@5 \or\advance\Scount@\ten@
  \or\advance\Scount@11 \or\or\or\advance\Scount@12 \or\advance
  \Scount@\tw@\fi
 \fi
 \ifcase\arrcount@\or\or\global\advance\angcount@\@ne\else\fi
 \ifN@\shifted@\firsty@\else\shifted@-\firsty@\fi
 \ifE@\else\advance\shifted@\charht@\fi
 \goal@\secondy@\advance\goal@-\firsty@
 \ifN@\else\multiply\goal@\m@ne\fi
 \setbox\shaft@\hbox{\arrowfont@\char\angcount@}%
 \ifnum\arrcount@=\thr@@
  \getcos@{1.5\p@}%
  \setbox\shaft@\hbox to\wd\shaft@{\arrowfont@
   \rlap{\hskip\dimen@ii
    \smash{\ifNESW@\let\next@\lower\else\let\next@\raise\fi
     \next@\dimen@\hbox{\arrowfont@\char\angcount@}}}%
   \rlap{\hskip-\dimen@ii
    \smash{\ifNESW@\let\next@\raise\else\let\next@\lower\fi
      \next@\dimen@\hbox{\arrowfont@\char\angcount@}}}\hfil}%
 \fi
 \rlap{\smash{\hskip\tocenter@\hskip\firstx@
  \ifnum\arrcount@=\m@ne
  \else
   \ifnum\arrcount@=\thr@@
   \else
    \ifnum\scount@=\m@ne
    \else
     \ifnum\scount@=\z@
     \else
      \setbox\ZER@\hbox{\ifnum\angcount@=117 \arrow@v\else\arrowfont@\fi
       \char\Scount@}%
      \ifNESW@
       \ifnum\scount@=\tw@
        \dimen@\shifted@\advance\dimen@-\charht@
        \ifN@\hskip-\wd\ZER@\fi
        \Nnext@
        \next@\dimen@\copy\ZER@
        \ifN@\else\hskip-\wd\ZER@\fi
       \else
        \Nnext@
        \ifN@\else\hskip-\wd\ZER@\fi
        \next@\shifted@\copy\ZER@
        \ifN@\hskip-\wd\ZER@\fi
       \fi
       \ifnum\scount@=12
        \advance\shifted@\charht@\advance\goal@-\charht@
        \ifN@\hskip\wd\ZER@\else\hskip-\wd\ZER@\fi
       \fi
       \ifnum\scount@=13
        \getcos@{\thr@@\p@}%
        \ifN@\hskip\dimen@\else\hskip-\wd\ZER@\hskip-\dimen@\fi
        \adjust@\shifted@\advance\adjust@\dimen@ii
        \Nnext@
        \next@\adjust@\copy\ZER@
        \ifN@\hskip-\dimen@\hskip-\wd\ZER@\else\hskip\dimen@\fi
       \fi
      \else
       \ifN@\hskip-\wd\ZER@\fi
       \ifnum\scount@=\tw@
        \ifN@\hskip\wd\ZER@\else\hskip-\wd\ZER@\fi
        \dimen@\shifted@\advance\dimen@-\charht@
        \Nnext@
        \next@\dimen@\copy\ZER@
        \ifN@\hskip-\wd\ZER@\fi
       \else
        \Nnext@
        \next@\shifted@\copy\ZER@
        \ifN@\else\hskip-\wd\ZER@\fi
       \fi
       \ifnum\scount@=12
        \advance\shifted@\charht@\advance\goal@-\charht@
        \ifN@\hskip-\wd\ZER@\else\hskip\wd\ZER@\fi
       \fi
       \ifnum\scount@=13
        \getcos@{\thr@@\p@}%
        \ifN@\hskip-\wd\ZER@\hskip-\dimen@\else\hskip\dimen@\fi
        \adjust@\shifted@\advance\adjust@\dimen@ii
        \Nnext@
        \next@\adjust@\copy\ZER@
        \ifN@\hskip\dimen@\else\hskip-\dimen@\hskip-\wd\ZER@\fi
       \fi	
      \fi
  \fi\fi\fi\fi
  \ifnum\arrcount@=\m@ne
  \else
   \loop
    \ifdim\goal@>\charht@
    \ifE@\else\hskip-\charwd@\fi
    \Nnext@
    \next@\shifted@\copy\shaft@
    \ifE@\else\hskip-\charwd@\fi
    \advance\shifted@\charht@\advance\goal@-\charht@
    \repeat
   \ifdim\goal@>\z@
    \dimen@\charht@\advance\dimen@-\goal@
    \divide\dimen@\tan@i\multiply\dimen@\tan@ii
    \ifE@\hskip-\dimen@\else\hskip-\charwd@\hskip\dimen@\fi
    \adjust@\shifted@\advance\adjust@-\charht@\advance\adjust@\goal@
    \Nnext@
    \next@\adjust@\copy\shaft@
    \ifE@\else\hskip-\charwd@\fi
   \else
    \adjust@\shifted@\advance\adjust@-\charht@
   \fi
  \fi
  \ifnum\arrcount@=\m@ne
  \else
   \ifnum\arrcount@=\thr@@
   \else
    \ifnum\tcount@=\m@ne
    \else
     \setbox\ZER@
      \hbox{\ifnum\angcount@=117 \arrow@v\else\arrowfont@\fi\char\Tcount@}%
     \ifnum\tcount@=\thr@@
      \advance\adjust@\charht@
      \ifE@\else\ifN@\hskip-\charwd@\else\hskip-\wd\ZER@\fi\fi
     \else
      \ifnum\tcount@=12
       \advance\adjust@\charht@
       \ifE@\else\ifN@\hskip-\charwd@\else\hskip-\wd\ZER@\fi\fi
      \else
       \ifE@\hskip-\wd\ZER@\fi
     \fi\fi
     \Nnext@
     \next@\adjust@\copy\ZER@
     \ifnum\tcount@=13
      \hskip-\wd\ZER@
      \getcos@{\thr@@\p@}%
      \ifE@\hskip-\dimen@\else\hskip\dimen@\fi
      \advance\adjust@-\dimen@ii
      \Nnext@
      \next@\adjust@\box\ZER@
     \fi
  \fi\fi\fi}}%
 \iflabel@i
  \rlap{\hskip\tocenter@
  \dimen@\firstx@\advance\dimen@\secondx@\divide\dimen@\tw@
  \advance\dimen@\ldimen@i
  \dimen@ii\firsty@\advance\dimen@ii\secondy@\divide\dimen@ii\tw@
  \global\multiply\ldimen@i\tan@i\global\divide\ldimen@i\tan@ii
  \ifNESW@\advance\dimen@ii\ldimen@i\else\advance\dimen@ii-\ldimen@i\fi
  \setbox\ZER@\hbox{\ifNESW@\else\ifnum\arrcount@=\thr@@\hskip4\p@\else
   \hskip\tw@\p@\fi\fi
   $\m@th\tsize@@\label@i$\ifNESW@\ifnum\arrcount@=\thr@@\hskip4\p@\else
   \hskip\tw@\p@\fi\fi}%
  \ifnum\arrcount@=\m@ne
   \ifNESW@\advance\dimen@.5\wd\ZER@\advance\dimen@\p@\else
    \advance\dimen@-.5\wd\ZER@\advance\dimen@-\p@\fi
   \advance\dimen@ii-.5\ht\ZER@
  \else
   \advance\dimen@ii\dp\ZER@
   \ifnum\slcount@<6 \advance\dimen@ii\tw@\p@\fi
  \fi
  \hskip\dimen@
  \ifNESW@\let\next@\llap\else\let\next@\rlap\fi
  \next@{\smash{\raise\dimen@ii\box\ZER@}}}%
 \fi
 \iflabel@ii
  \ifnum\arrcount@=\m@ne
  \else
   \rlap{\hskip\tocenter@
   \dimen@\firstx@\advance\dimen@\secondx@\divide\dimen@\tw@
   \ifNESW@\advance\dimen@\ldimen@ii\else\advance\dimen@-\ldimen@ii\fi
   \dimen@ii\firsty@\advance\dimen@ii\secondy@\divide\dimen@ii\tw@
   \global\multiply\ldimen@ii\tan@i\global\divide\ldimen@ii\tan@ii
   \advance\dimen@ii\ldimen@ii
   \setbox\ZER@\hbox{\ifNESW@\ifnum\arrcount@=\thr@@\hskip4\p@\else
    \hskip\tw@\p@\fi\fi
    $\m@th\tsize@\label@ii$\ifNESW@\else\ifnum\arrcount@=\thr@@\hskip4\p@
    \else\hskip\tw@\p@\fi\fi}%
   \advance\dimen@ii-\ht\ZER@
   \ifnum\slcount@<9 \advance\dimen@ii-\thr@@\p@\fi
   \ifNESW@\let\next@\rlap\else\let\next@\llap\fi
   \hskip\dimen@\next@{\smash{\raise\dimen@ii\box\ZER@}}}%
  \fi
 \fi
}
\def\outCD@#1{\def#1{\Err@{\noexpand#1must not be used within \string\CD}}}
\newskip\preCDskip@
\newskip\postCDskip@
\preCDskip@\z@
\postCDskip@\z@
\def\preCDspace#1{\RIfMIfI@
 \onlydmatherr@\preCDspace\else\advance\preCDskip@#1\relax\fi\else
 \onlydmatherr@\preCDspace\fi}
\def\postCDspace#1{\RIfMIfI@
 \onlydmatherr@\postCDspace\else\advance\postCDskip@#1\relax\fi\else
 \onlydmatherr@\postCDspace\fi}
\def\predisplayspace#1{\RIfMIfI@
 \onlydmatherr@\predisplayspace\else
 \advance\abovedisplayskip#1\relax
 \advance\abovedisplayshortskip#1\relax\fi
 \else\onlydmatherr@\preCDspace\fi}
\def\postdisplayspace#1{\RIfMIfI@
 \onlydmatherr@\postdisplayspace\else
 \advance\belowdisplayskip#1\relax
 \advance\belowdisplayshortskip#1\relax\fi
 \else\onlydmatherr@\postdisplayspace\fi}
\def\PreCDSpace#1{\global\preCDskip@#1\relax}
\def\PostCDSpace#1{\global\postCDskip@#1\relax}
\def\CD#1\endCD{%
 \outCD@\cgaps\outCD@\rgaps\outCD@\Cgaps\outCD@\Rgaps
 \preCD@#1\endCD
 \advance\abovedisplayskip\preCDskip@
 \advance\abovedisplayshortskip\preCDskip@
 \advance\belowdisplayskip\postCDskip@
 \advance\belowdisplayshortskip\postCDskip@
 \vcenter{\offinterlineskip
  \vskip\preCDskip@\Let@\global\colcount@\@ne\global\rowcount@\z@
  \everycr{%
   \noalign{%
    \ifnum\rowcount@=\Rowcount@
    \else
     \getrgap@\rowcount@\vskip\getdim@
     \global\advance\rowcount@\@ne\global\colcount@\@ne
    \fi}}%
  \tabskip\z@
  \halign{&\global\xoff@\z@\global\yoff@\z@
   \getcgap@\colcount@\hskip\getdim@
   \hfil\vrule\height\ten@\p@\width\z@\depth\z@
   $\m@th\displaystyle{##}$\hfil
   \global\advance\colcount@\@ne\cr
   #1\crcr}\vskip\postCDskip@}%
 \preCDskip@\z@\postCDskip@\z@
 \def\getcgap@##1{\ifcase##1\or\getdim@\z@\else\getdim@\standardcgap\fi}%
 \def\getrgap@##1{\ifcase##1\getdim@\z@\else\getdim@\standardrgap\fi}%
 \let\Width@\relax\let\Height@\relax\let\Depth@\relax\let\Rowheight@\relax
 \let\Rowdepth@\relax\let\Colwidth@\relax
}
\let\enddocument\bye
\def\alloc@#1#2#3#4#5{\global\advance\count1#1by\@ne
  \ch@ck#1#4#2%
  \allocationnumber=\count1#1%
  \global#3#5=\allocationnumber
  \wlog{\string#5=\string#2\the\allocationnumber}}
\catcode`\@=\active


\def\frac#1#2{{{\strut {#1}} \over {\strut {#2}}}}

\def\ub#1{{\underbar{#1}}}
\def\id{{1\kern-.235em \text{I}}}



\overfullrule=0pt
\def\ub#1{{\underbar{#1}}}

\RefWarnings
\TagsOnRight

\input ETHpaper.st\relax

\doublespace

\document

\ethpp{93-24}
\bigtitle  SOME ELEMENTS OF CONNES' NON-COMMUTATIVE\\
GEOMETRY, AND SPACE-TIME GEOMETRY \\ \\ \\
\endbigtitle
{\hfill ZU-TH-18/1993}
\vskip.1truecm
{\hfill June 1993}
\vskip 2truecm
\title
SOME ELEMENTS OF CONNES' NON-COMMUTATIVE\\
GEOMETRY, AND SPACE-TIME GEOMETRY \\ \\ \\
\endtitle

\author
A.H. Chamseddine$^1$\footnote"$^*$"{Supported in part by the Swiss
National Foundation (FNS)} \ \ and \ \ J.
Fr\"ohlich$^2$\footnote"$^{\text{\#}}$"{Permanent address: Theoretical
Physics, ETH-H\"onggerberg, CH-8093 Z\"urich}
\endauthor

\maketitle

\vskip 1truecm
\centerline{$^1$  Theoretical Physics, University of Z\"urich, CH-8001
Z\"urich}
\bigskip
\centerline{$^2$  Institut des Hautes Etudes Scientifiques, F-91440
Bures-sur-Yvette}

\vfill\eject

\heading
Introduction  \\
\endheading

Physics -- as we practice it -- rests on two pillars:

\item{(i)} The analysis of (causal sequences of events in) \ub{classical
space-time}, viewed as a four-dimensional, smooth Lorentzian manifold (with
certain good properties).

\item{(ii)} \ub{Quantum theory}.

These two pillars appear to be somewhat incompatible, in the sense that it is
found to be difficult to join them into a unified theoretical framework, or,
in other words, to derive them as two different (limiting) aspects of a
consistent, unified theory. Fortunately, space-time can be taken to be
classical and, for purposes of laboratory physics, Minkowskian, down to
distance scales comparable to the Planck length (corresponding to $\sim$
10$^{19}$ GeV). Thus, the unification of space-time geometry with quantum
theory is not an urgent issue from a pragmatic point of view. However, for the
logical consistency of the building of theoretical physics, joining
space-time geometry with quantum theory would appear to be a fundamental task.

For space-time to directly reveal its ``quantum nature'', it would have to be
explored at distance scales close to the Planck length. Since this is
impossible, our thinking about the problem of unifying space-time geometry
with quantum theory is necessarily speculative and must be guided by
considerations of mathematical consistency, elegance and aesthetic appeal.

One standard idea in the search for a theoretical framework unifying
space-time-  and quantum dynamics is to attempt to formulate a fundamental
quantum theory without any reference to specific classical space-time models
(``background independence'') and to try to view space-time as a derived
(rather than as a fundamental) structure which manifests itself in a certain
limiting regime of the fundamental quantum theory (e.g. as the geometry
connected with an algebra of ``functions on constant loops'' or of ``zero
modes'').

Among a variety of theoretical ideas in this direction the following two
approaches have been pursued most forcefully:

\item{(A)} Superstring theory \Ref{r1},\Ref{r2}.

\item{(B)}  The study of non-commutative spaces and their non-commutative
geometry, as initiated by Connes \Ref{r3}.

The successes and problems of approach (A) are relatively well known among
theoretical physicists. One success of approach (B) consists in a new
perspective in the study of gauge theories \Ref{r4}, in particular of the
standard model \Ref{r5} (see also \Ref{r6}) and of grand unified models
\Ref{r7}. However, since approach (B) has not been worked out in much detail
for examples of infinite-dimensional non-commutative geometries, yet, problems
connected with quantization have remained essentially untouched.

An idea that might be promising is to look for a manifestly
background-independent, or ``invariant'' formulation of superstring theory and
then use the methods of non-commutative geometry of Connes to study its
properties. Since we still do not know a completely precise form of string
field theory, we might settle for a more modest goal: Since string vacua
correspond to superconformal field theories, we might first try to formulate
superconformal field theories in a purely algebraic way, i.e., in a form
independent of a choice of a target space and a target space geometry (see
e.g. \Ref{r8},\Ref{r9}) and then to reconstruct geometrical data from a
superconformal field theory by using methods of non-commutative geometry.

For simplicity, let us consider an $N$=1 unitary superconformal field theory.
[Of course, in the construction of string vacua, one must study $N$=2
superconformal field theories. While $N$=1 theories turn out to generalize
real Riemannian geometry, $N$=2 theories generalize complex K\"ahler geometry
which is more difficult, and the necessary tools have not been fully
developed, yet.] We choose the Ramond sector of an $N$=1 theory. Abstractly,
it can be coded into the following data; (see \Ref{r9} for background
material):

\item{(i)} A \ub{$^*$algebra}, $A$, of operators acting on a separable Hilbert
space $H$; $A$ contains the identity operator.

\item{(ii)} A``\ub{Dirac operator}'' $D$ (the Ramond generator $G_0$ of the
theory) which is a selfadjoint operator on $H$. [Technically, it is useful to
assume that $A$ consists of bounded operators with the property that $[D,a]$
is a bounded operator, for arbitrary $a\in A$.]

\item{} A Laplacian $\triangle$ is defined by setting $-\triangle = D^2 \geq
0$.

\item{(iii)} \ub{$\Bbb Z_2$ grading}: There is a unitary involution, $\Gamma$,
on $H$ such that $\Gamma a = a\Gamma$, for all $a \in A$, ($A$ is ``even''),
but $\Gamma D = - D\Gamma$, on the domain of $D$, ($D$ is ``odd'').
Physically, $\Gamma =(-1)^F$, where $F$ is fermion number.

\item{(iv)} \ub{Conformal invariance}: $H$ carries a projective, unitary
representation of the group PSU(1,1) of M\"obius transformations preserving
the unit circle, with generators $L_0=L_0^*$ and $L_1, L_{-1}$, with
$L_1^*=L_{-1}$. We assume that
$$
D^2\;\equiv\;-\,\triangle\;=\;L_0\;-\,\frac{c}{24}\;\id ,
$$
where $c$ is the central charge of the theory, and that the representation of
PSU(1,1) on $H$ defines a $^*$automorphism group of the algebra $A$.

Formally, one can now inductively define all the generators of the Ramond
algebra \Ref{r9}. It is an intersting problem to isolate the precise
hypotheses needed to prove that the formal Ramond generators obtained from
(ii), (iii), (iv) are well defined and satisfy the Ramond algebra. [One
approach towards solving this problem consists in generalizing the
L\"uscher-Mack theorem \Ref{r10}.]

Our goal in this paper is to show how from data (i) - (iii) one can
reconstruct a generalized (non-commutative version of) Riemannian geometry,
(Sect.~2). Rather than exemplifying non-commutative Riemannian geometry in the
context of superconformal field theory -- which would be a highly desirable
goal that is, however, still somewhat elusive, so far -- we shall, in Sect.~3,
disucss the easier example of Riemannian geometry on finite-dimensional,
generalized commutative and non-commutative spaces, e.g. on a two-sheeted,
four-dimensional manifold, as in \Ref{r5}, and consider analogues of the
Einstein-Hilbert and the Chern-Simons action functionals. Sect.~4 contains
some conclusions and an outlook.

This paper is rather mathematical in its structure, and the applicability of
the mathematical formalism to physical problems remains, at least to a fairly
large extent, a matter of speculation -- really convincing examples are still
missing. Nevertheless, we feel that Prof. C.N. Yang might follow attempts such
as the present one with some benevolence. It is a pleasure to dedicate this
paper to him.

\vfill

\vskip 1truecm

\heading
Some Elements of Non-Commutative Geometry, \Ref{r3}.\\
\endheading

\noindent 2.1. \ \ub{Non-commutative spaces}
\medskip

Let $M$ be a smooth, compact manifold without boundary. All properties of $M$
can be retrieved from the study of the \ub{commutative} $^*$algebra $C^\infty
(M)$ of complex-valued, smooth functions on $M$. Since we assume that $M$ is
compact, $C^\infty (M)$  contains the function identically equal to 1 on $M$.
Connes' idea is to study ``non-commutative spaces'' in terms of
non-commutative $^*$algebras with identity, 1. It remains to be seen what a
good notion of ``non-commutative manifold'' would consist of, \Ref{r3}. Here
we say that any non-commutative, unital $^*$algebra $A$ of bounded operators
(``bounded'' with respect to some $C^*$ norm on $A$) defines a
``non-commutative space'', also denoted by $A$.

\bigskip

\noindent 2.2. \ \ub{Non-commutative differential forms on a non-commutative
space.}
\medskip

According to Connes \Ref{r3}, non-commutative differential forms on a
non-commutative space $A$ are elements of the graded, differential algebra
$\Omega^\cdot (A)$ of ``universal forms'' over $A$:
$$
\Omega^\cdot (A)\;=\;\mathop{\oplus}_{n=0}^\infty\;\Omega^n (A)
\tag\label{2.1}
$$
is a $\Bbb Z$-graded complex vector space such that each $\Omega^n(A)$ is an
$A$ bimodule;
$$
\align
\Omega^n(A)\;&=\;A^{\otimes(n+1)} \big/_{\text{Relations}}\\
&=\;\big\langle a_0 da_1\cdots da_n\,:\,a_0,a_1,\cdots,a_n\enskip \text{in}
\enskip A\big\rangle\big/_{\text{Relations}}\;,
\tag\label{2.2}
\endalign
$$
with $\Omega^0 (A) = A.$

The ``Relations'' are:
$$
da\;=\;da'\quad\text{iff}\enskip a-a'\;=\;\lambda 1, \lambda \in \Bbb C,
\tag\label{2.3}
$$
(in particular, $d1=0$). Clearly, $\Omega^n(A)$ defined by \Ref{2.2}  and
\Ref{2.3} is a left $A$ module. For $\Omega^n(A)$ to be a right $A$ module, we
must impose the \ub{Leibniz rule}:
$$
d(a\cdot b)\;=\;(da)\cdot b\;+\;a\cdot (db).\tag\label{2.4}
$$
Relation \Ref{2.4} could be deformed to read
$$
d(a\cdot b)\;=\;(da)\cdot \theta (b)\;+\;\psi (a)\cdot (db),\tag\label{2.5}
$$
where $\theta$ and $\psi$ are $^*$automorphisms of $A$. We shall, however, not
pursue this generalization here. Rather, we shall view $d$ as an analogue of
\ub{exterior differentiation} and define an $A$-linear map \ $d : \Omega^n (A)
\to \Omega^{n+1} (A)$, $n=0,1,2,\cdots$, by setting
$$
d\bigl( a_0\, da_1\cdots da_n\bigr)\;:=\;1\;da_0\, da_1\cdots da_n.
$$
Since $d 1=0$, see \Ref{2.3}, it follows that
$$
d^2\;=\;0, \tag\label{2.6}
$$
so that $\Omega^\cdot (A)$ is a graded complex of vector spaces.

Thanks to \Ref{2.4} or \Ref{2.5}, $\Omega^\cdot (A)$ is equipped with a
multiplication,
$$
m\;:\;\Omega^n(A)\;\otimes_A\;\Omega^l (A)\;\to\;\Omega^{n+l} (A),
\tag\label{2.7}
$$
$$
\align
\bigl(a_0\,da_1 &\cdots da_n\bigr)\;\cdot\;\bigl( b_0\,db_0 \cdots
db_l\bigr)\\
&=\;a_0\,da_1\cdots d\bigl(a_n \cdot b_0\bigr) \;db_1\cdots db_l\\
&-\;a_0\,da_1\cdots d\bigl(a_{n-1}\,a_n\bigr)\;db_0\,db_1\cdots db_l\\
&+\;a_0\,da_1\cdots d\bigl(a_{n-2}\,a_{n-1}\bigr)\;da_n\,db_0\cdots db_l\\
&-\;\cdots
\endalign
$$
which belongs to $\Omega^{n+l}(A)$.

Thus $\Omega^\cdot (A)$ is an algebra under $m$. Since it contains $\Omega^0
(A) = A$ as a unital subalgebra, it contains an identity $1\in A$.
Furthermore, it becomes a $^*$algebra by defining
$$
(da)^*\;=\;-\,da^*,\quad\text{for all}\quad a\in A,
$$
and hence
$$
\align
(d\alpha)^*\;&=\;(-1)^{deg\,\alpha+1}\;d\alpha^*,\\
deg\,\alpha\;&=\;deg\,\alpha^*\;=\;n,
\tag\label{2.8}
\endalign
$$
for all $\alpha \in \Omega^n(A)$, for all $n$.

The graded, differential algebra $\Omega^\cdot (A)$ plays an important role in
analyzing the ``topology'' of a non-commutative space $A$ using Connes' cyclic
cohomology, \Ref{r3}. The classical theory emerges as a special case.

\bigskip
\noindent 2.3. \ \ub{Vector bundles, connections, hermitian structures}.
\medskip

Classically, the space of sections of a vector bundle over a manifold $M$ can
be described as a finitely generated, projective left module for the ring
$C^\infty (M)$ of smooth functions on $M$, \Ref{r12}. A left module $E$ over
a ring $A$ is \ub{finitely generated} iff there is a finite number of
elements, $s_1, \cdots, s_n$, in $E$ such that every $s\in E$ can be written
as $s = \displaystyle\mathop{\sum}_{j=1}^{n} a_j\,s_j$, for some $a_1, \cdots,
a_n$ in
$A$. The elements $s_1, \cdots, s_n$ form a basis of $E$ iff $0 =
\displaystyle\mathop{\sum}_{j=1}^{n} a_j\,s_j$ implies $a_j=0$, for all
$j=1,\cdots,n$. The left module $E$ is called \ub{free} iff it has a basis; $E$
is called \ub{projective} iff it is a submodule of a free module $F$, i.e.,
there exists a free module $F$ and a submodule $G$ such that $F=E\oplus G$.

The theorem of Swan \Ref{r12} quoted above suggests to interpret the
\ub{space of sections} $E$ \ub{of a vector bundle over a non-commutative
space} described
by a unital $^*$algebra $A$ as a \ub{finitely generated}, \ub{projective left}
$A$ \ub{module}; see \Ref{r3}. [This notion of vector bundles is adequate in
the context of ``real geometry'' but may not be useful in the holomorphic
setting.]

Adapting the classical notion of connection (gauge potential) to the more
general setting of vector bundles over non-commutative spaces, Connes
\Ref{r3} has proposed to define a \ub{connection on} $E$ as a linear map,
$\nabla$, from $E$ to $\Omega^1(A) \otimes_A E$
$$
\nabla\;:\;E\;\longrightarrow\;\Omega^1(A)\;\otimes_A\;E,
\tag\label{2.9}
$$
satisfying the Leibniz rule
$$
\nabla (as)\;=\;da\;\otimes_A\;s\,+\,a\,\nabla s,
\tag\label{2.10}
$$
for all $a\in A$ and all $s\in E$. Defining
$$
\Omega^\cdot (E)\;=\;\mathop{\oplus}_{n=0}^\infty\;\Omega^n\;(E),
\tag\label{2.11}
$$
with $\Omega^n (E) = \Omega^n (A) \otimes_A E$, it is easy to verify that
$\nabla$ extends to $\Omega^\cdot (E)$, with
$$
\nabla\;:\;\Omega^n(E)\;\longrightarrow\;\Omega^{n+1} (E)
\tag\label{2.12}
$$
satisfying
$$
\nabla (\alpha\phi)\;=\;d\alpha\phi\;+\;(-1)^{deg\,\alpha}\,\alpha\nabla\phi,
\tag\label{2.13}
$$
for all homogeneous forms $\alpha\in\Omega^\cdot(A)$ and all
$\phi\in\Omega^\cdot(E)$.

This permits us to define the \ub{curvature}, $R(\nabla)$, of a connection
$\nabla$ on $E$ by
$$
R(\nabla)\;:=\;-\,\nabla^2\;:\;E\;\longrightarrow\;\Omega^2 (A) \otimes_A E;
\tag\label{2.14}
$$
$R(\nabla)$ is an $A$-linear map satisfying
$$
R(\nabla)(as)\;=\;a\; R(\nabla)\;s,
\tag\label{2.15}
$$
for all $a \in A$ and all $s\in E$. By \Ref{2.13} and \Ref{2.14}, the
definition of $R(\nabla)$ extends to $\Omega^\cdot (E)$, and one has (using
\Ref{2.13} and $d^2=0$) that
$$
R(\nabla)(\alpha\phi)\;=\;\alpha\;R(\nabla)\;\phi,
\tag\label{2.16}
$$
for arbitrary $a\in\Omega^\cdot(A)$ and arbitrary $\phi\in\Omega^\cdot(E)$.

The following result from module theory \Ref{r13} permits us to rewrite
$R(\nabla)$ in a more concrete form: Let $E$ and $F$ be two left modules over
a ring $A$. Define $\theta_{EF} : E^* \otimes_A F \to Hom_A (E,F)$ by setting
$\theta_{EF} (\sigma \otimes_A t) (s) = \sigma (s) t$, for arbitrary $\sigma
\in E^*$ (the space of $A$-linear functionals on $E$) and arbitrary
$s\in E$ and $t\in F$. Then $\theta_{EF}$ is an \ub{isomorphism} from $E^*
\otimes_A F$ to $Hom_A(E,F)$ iff $E$ is finitely generated and projective.
Applying this result to $R(\nabla)$, we set $E: = E$, $F := \Omega^2
(A)\otimes_A E$ and note that $R(\nabla)\in Hom_A(E,F)$. Since $E$ is finitely
generated and projective, it follows that $R(\nabla)$ can be written as
$$
R(\nabla)\;=\;\sum_{\alpha,\beta}\;\varepsilon_\alpha \otimes_A
F^\alpha\;_\beta \otimes_A e^\beta.
\tag\label{2.17}
$$
for some elements $\varepsilon_\alpha \in E^*$, $R^\alpha\,_\beta \in
\Omega^2(A)$ and $e^\beta \in E$, with
$$
R(\nabla)s\;=\;\sum_{\alpha,\beta}\;\varepsilon_\alpha (s) F^\alpha\,_\beta
\otimes_A e^\beta.
\tag\label{2.18}
$$
In spite of the fact that representation \Ref{2.17} is not unique,
it turns out to be useful.
[For example, it permits one to define traces: \ \
trace$_E$ $R(\nabla) =
\displaystyle\mathop{\sum}_{\alpha,\beta}\; \varepsilon_\alpha (e^\beta)
F^\alpha\,_\beta \in \Omega^2 (A)$, \break
trace$_E$ $R(\nabla)^2 =
\displaystyle\mathop{\sum}_{\alpha,\beta,\gamma,\delta} \varepsilon_\gamma
(e^\beta) F^\gamma\,_\delta \;\varepsilon_\alpha (e^\delta) F^\alpha\,_\beta
\in \Omega^4 (A), \cdots $]

Next, we recall the notion of \ub{hermitian vector bundles}. An element $a\in
A$ is said to be positive iff $a = \displaystyle\mathop{\sum}_i b_i^* b_i$,
$b_i\in A$, (where the series is assumed to converge in the $C^*$ norm on $A$
to an element of $A$). We say that a vector bundle $E$ over $A$ is
\ub{hermitian} iff there is a map $\langle\cdot,\cdot\rangle : E\times E \to
A$, called a \ub{hermitian inner product} on $E$, with the properties:

\itemitem{(i)} $\langle as, bt\rangle = a\langle s,t\rangle\,b^*$, for all
$a,b$ in $A$ and all $s,t$ in $E$; $\langle\cdot,\cdot\rangle$ is linear in
the first and anti-linear in the second argument.

\itemitem{(ii)} $\langle s,s\rangle \geq 0$, for all $s\in E (= 0
\Leftrightarrow s=0)$.

\itemitem{(iii)} The anti-linear map $s \mapsto \langle\cdot,s\rangle$ defines
an \ub{isomorphism} from $E$ to $E^*$.

One now shows without difficulty that $\langle\cdot,\cdot\rangle$ extends
uniquely to a hermitian inner product on $\Omega^\cdot (E)$ with values in the
algebra $\Omega^\cdot (A)$, with
$$
\langle \alpha\phi,\beta\psi\rangle\;=\;\alpha\langle\phi,\psi\rangle\beta^*,
\tag\label{2.19}
$$
for all $\alpha,\beta$ in $\Omega^\cdot(A)$ and all $\phi,\psi$ in
$\Omega^\cdot(E)$.

Note that it follows easily from (ii) that
$$
\langle\phi,\psi\rangle^*\;=\;\langle\psi,\phi\rangle.
\tag\label{2.20}
$$
One says that a connection $\nabla$ on a hermitian vector bundle $E$ over $A$
is unitary iff
$$
d\langle s,t\rangle\;=\;\langle\nabla s,t\rangle\;-\;\langle s,\nabla
t\rangle. \tag\label{2.21}
$$
[The minus sign on the R.S. of \Ref{2.21} is forced upon us by the convention
that $(da)^* = -da^*$, for $a\in A$.] One then effortlessly shows that
$$
d\langle\phi,\psi\rangle\;=\;\langle\nabla\phi,\psi\rangle\;-\;
(-1)^{\text{deg}\,\phi\,+\,\text{deg}\,\psi}\;\langle\phi,\nabla\psi\rangle,
\tag\label{2.22}
$$
for arbitrary homogeneous $\phi,\psi$ in $\Omega^\cdot (E)$.

Using Connes' cyclic cohomology \Ref{r3} one can now go on to define Chern
characters of vector bundles $E$ over a non-commutative space $A$ which are
pairings between the $K$-theory of $A$ and even cyclic cocycles. We shall not
pursue this theme here but refer the interested reader to the literature, in
particular to Connes' book \Ref{r3} and to \Ref{r14}.

What is more important for our theme is to introduce a notion of
differentiable structure on a non-commutative space.

\bigskip

\noindent 2.4. \ \ub{Differentiable structure on a non-commutative space}
\medskip
We recall that, among the basic data specifying a conformal field theory was
not only a non-commutative space described by a unital $^*$algebra $A$, but
also a Dirac operator $D$ such that $[D,(\cdot)]$ acts as a derivation on $A$.
This is what is required to define a differentiable structure on $A$.

Thus, consider a non-commutative space corresponding to a unital $^*$algebra
$A$. We define an \ub{even $K$-cycle} for $A$ to consist of the following
data:
\item{(i)} A $^*$representation, $\pi$, of $A$ on a separable Hilbert space
$H$. [Usually, we may assume that $\pi$ is a faithful representation, and we
shall therefore write $a$ for both, the element $a\in A$ and the bounded
operator $\pi (a)$ on $H$.]

\item{(ii)} A (possibly unbounded) selfadjoint operator $D$ on $H$ such that
$$
[D,a]\quad\text{is bounded, for all}\quad a\in A;
\tag\label{2.23}
$$
$$
(D^2+\id)^{-1}\quad\text{is a compact operator}.
\tag\label{2.24}
$$
This condition expresses the idea that the non-commutative space under
consideration is compact.

\item{(iii)} A unitary involution $\Gamma$ on $H$ (i.e.,
$\Gamma=\Gamma^*=\Gamma^{-1}$) such that
$$
A\quad\text{is \ub{even} under $\Gamma$ and $D$ is \ub{odd} under $\Gamma$},
\tag\label{2.25}
$$
(i.e., $\Gamma a = a\Gamma$, for all $a \in A$, and $\Gamma D+D\Gamma=0$, on
the domain of $D$).

If (iii) is omitted one speaks of \ub{odd} $K$-cycles. An odd $K$-cycle
$(\pi,H,D)$ determines an even $K$-cycle, ($\widetilde\pi,\widetilde H,
\widetilde D,
\Gamma$), by setting $\widetilde\pi = \pi\oplus\pi$, $\widetilde H = H \oplus
H$,
$$
\widetilde D\;=\; {{0\quad D}\choose {D\quad 0}} \quad
\text{and}\enskip\Gamma\;=\;
{{\id\quad 0}\choose {0-\id}}.
$$
Given a $K$-cycle $(\pi,H,D)$ for $A$, we can replace the somewhat monstrous
graded, differential algebra $\Omega^\cdot (A)$ of universal forms by a more
manageable one, the differential algebra $\Omega_D^\cdot (A)$ defined as
follows: We define a $^*$representation $\pi^\cdot$ of $\Omega^\cdot(A)$ on
$H$ by setting
$$
\pi^\cdot (a_0\,da_1\cdots da_n)\;=\;a_0\,[D,a_1]\cdots[D,a_n],
\tag\label{2.26}
$$
for arbitrary $a_0,a_1,\cdots,a_n$ in $A$. As shown by Connes and Lott in
\Ref{r5},\Ref{r3}, the subalgebra $ker\,\pi^\cdot+ d\,ker\,\pi^\cdot$ of
$\Omega^\cdot (A)$ is a two-sided ideal in $\Omega^\cdot(A)$; (the proof is an
easy application of the Leibniz rule for $d$). This enables them to define a
graded, differential algebra $\Omega_D^\cdot(A)$ by setting
$$
\Omega_D^\cdot (A)\;=\;\Omega^\cdot(A)
\big/_{(ker\,\pi^\cdot\,+\,d\,ker\,\pi^\cdot)}. \tag\label{2.27}
$$
Note that
$$
\Omega_D^0(A)\;=\;\Omega^0(A)
\big/_{ker\,\pi^\cdot}\;=\;A\big/_{ker\,\pi}\;=\;A,
\tag\label{2.28}
$$
since $ker\,\pi^\cdot \big/_{\Omega^0(A)} = ker\,\pi = \{ 0\} - \pi$ has been
assumed to be faithful -- and hence
$$
\align
\Omega_D^1(A)\;&=\;\Omega^1 (A)\big/_{ker\,\pi^\cdot \big/ \Omega^1(A)}\\
&\simeq\;\pi^\cdot \bigl(\Omega^1 (A)\bigr)\;=\;\biggl\{ \sum_i\;
a_0^i\,\bigl[ D,a_1^i\bigr]\;:\;a_0^i, a_1^i\;\in\;A\biggr\}
\tag\label{2.29}\endalign
$$
and
$$
\align
\Omega_D^2(A)\;&=\;\Omega^2(A)\big/_{(ker\,\pi^\cdot \big/_{\Omega^2(A)}
\,+\,d\,
ker\,\pi^\cdot\big/_{\Omega^1(A)})}\\
&\simeq\;\pi^\cdot\bigl(\Omega^2(A)\bigr)
\big/_{\pi^\cdot(d\,ker\,\pi^\cdot\big/_{\Omega^1(A)})}.
\tag\label{2.30}
\endalign
$$
The space $Aux := \pi^\cdot \bigl( d\,ker\,\pi^\cdot \big/_{\Omega^1(A)}\bigr)$
is
called the ``space of auxiliary fields'' \Ref{r5}. We note that
$$
\align
Aux\;&=\;\biggl\{ \sum_i\bigl[D,a_0^i\bigr]\bigl[D,a_1^i\,\bigr]\;:\; \sum_i
a_0^i \bigl[D, a_1^i\bigr]\;=\;0\biggr\}\\
&=\;\biggl\{ -\sum_i a_0^i \bigl[D,[D,a_1^i]\bigr]\;:\; \sum_i a_0^i\,
\bigl[D, a_1^i\bigr]\;=\;0\biggr\}.
\tag\label{2.31}
\endalign
$$
In the classical situation, where $A=C^\infty (M)$, for some even-dimensional,
smooth Riemannian spin manifold $M$, and $D=/\!\!\!\partial_M$, the Dirac
operator on $M$,
$$
Aux\;=\;A.
\tag\label{2.32}
$$
Since $[D,(\cdot)]$ satisfies the Leibniz rule, $\Omega_D^\cdot (A)$ is a left
and right $A$ module. \ub{Assuming} that, for a given $K$-cycle $(\pi,H,D)$,
$\Omega_D^1(A)$ is \ub{finitely generated} and \ub{projective}, the following
definition becomes meaningful.
\medskip
\ub{Definition}. \ $\Omega_D^1(A)$, viewed as a finitely generated, projective
left $A$ module, is called the (space of sections of the) \ub{cotangent
bundle} associated with the non-commutative space $A$ with differentiable
structure given by $(\pi,H,D)$.

Thus if $\Omega_D^1(A)$ is finitely generated and projective it is, according
to Sect. 2.3, a vector bundle over $A$, and we can study connections on
$\Omega_D^1(A)$. Let $\nabla$ be a connection on $\Omega_D^1(A)$, i.e.,
$$
\nabla\;:\;\Omega_D^1(A)\;\longrightarrow\;\Omega_D^1(A) \otimes_A
\Omega_D^1(A)
$$
is a linear map satisfying
$$
\nabla (a\omega)\;=\;da \otimes_A \omega\;+\;a\,\nabla\omega,
\tag\label{2.33}
$$
for all $a\in A \simeq \pi(A)$ and all $\omega\in\Omega_D^1(A)\simeq \pi^\cdot
\bigl(\Omega^1(A)\bigr)$. One defines the torsion, $T(\nabla)$, of a
connection $\nabla$ on $\Omega_D^1(A)$ by the formula
$$
T(\nabla)\;=\;d\;-\;m \circ \nabla,
\tag\label{2.34}
$$
see \Ref{r11}. One easily verifies that $T(\nabla)$ is an $A$-linear map
from $\Omega_D^1(A)$ to $\Omega_D^2(A)$; in particular,
$$
T(\nabla)(a\omega)\;=\;a\,T(\nabla)\omega,
\tag\label{2.35}
$$
for all $a\in A$ and all $\omega\in\Omega_D^1(A)$. Since $\Omega_D^2(A)$ is
generated by products of pairs of elements in $\Omega_D^1(A)$, one can always
construct, from a given connection $\nabla$ on $\Omega_D^1(A)$, a new
connection $\nabla'$ whose \ub{torsion}, $T(\nabla')$, \ub{vanishes},
\Ref{r15}. [This follows from arguments similar to those leading to
eq.~\Ref{2.17}.]

\bigskip
\noindent 2.5. \ \ub{Integration theory and Hilbert spaces of forms}.
\medskip
Following Connes \Ref{r3}, one says that a $K$-cycle $(\pi,H,D)$ for a
non-commutative space described by a unital $^*$algebra $A$ is
$(d,\infty$)-summable iff
$$
\text{trace}_H\;\bigl(D^2+\id\bigr)^{-p}\;<\;\infty\;,\quad
\text{for all}\enskip p\;> \frac d 2\;.
\tag\label{2.36}
$$
Let $Tr_\omega$ denote the Dixmier trace \Ref{r3}. The Dixmier trace is a
positive, cyclic trace on the algebra $B(H)$ of all bounded operators on $H$
which vanishes on trace-class operators.

We define the integral of a form $\alpha \in \Omega^\cdot (A)$ over a
non-commutative space $A$ by setting
$$
\align
\int \alpha\;&:=\;\mathop{\lim}_{\varepsilon\searrow 0}\;Tr_\omega\;
\bigl(\pi^\cdot(\alpha)(D^2+\varepsilon\id)^{-d/2}\bigr)\\
&\equiv\;Tr_\omega\;\bigl(\pi^\cdot(\alpha)\mid D\mid^{-d}\bigr).
\tag\label{2.37}
\endalign
$$
[The limit $\varepsilon\searrow 0$ exists trivially, since $Tr_\omega \bigl(
\pi^\cdot(\alpha)(D^2+\varepsilon\id)^{-d/2}\bigr)$ is actually independent of
$\varepsilon$.]

Unfortunately, the $K$-cycles encountered in supersymmetric quantum field
theory are not $(d,\infty)$-summable, for any finite $d$, but there are plenty
of so-called $\theta$-summable $K$-cycles \Ref{r3},\Ref{r14}, meaning that
$$
\text{trace}_H\;e^{-\beta\,D^2}\;<\;\infty,
\tag\label{2.38}
$$
for any $\beta > 0$. In this case, one may attempt to define the integral of
an element $\alpha \in \Omega^\cdot(A)$ by the formula
$$
\int \alpha\;:=\;\mathop{\text{Lim}}_{\beta\searrow 0}\,_\omega\;
\frac{\text{trace}_H\;\bigl(\pi^\cdot(\alpha)\;e^{-\beta\,D^2}
\bigr)}{\text{trace}_H\;\bigl(e^{-\beta\,D^2}\bigr)}\;,\tag\label{2.39}
$$
where the notation \ Lim$_\omega$ indicates that the ``limit'' is defined in
terms of a suitable mean on the space of uniformly bounded functions of
$\beta\in (0,1]$; see \Ref{r3}. [The definition \Ref{2.39} is useful, e.g., if
\  trace$_H \,e^{-\beta D^2}$ is bounded by \ exp const $\beta^{-s}$, as
$\beta\searrow 0$, for some $s<1$.] In the examples of Sect.~3 (which are
$(d,\infty)$-summable, for some $d<\infty$), the two definitions \Ref{2.37}
and \Ref{2.39} agree, but \Ref{2.39} has the advantage that it may still be
meaningful for $\theta$-summable $K$-cycles with $d=\infty$.

Connes has shown that if \ $(D^2+\varepsilon\id)^{-\frac{ds}{2}}$ is trace
class
for $s>1$ and $\displaystyle\mathop{\lim}_{s\searrow 1}\;(s-1)$ trace$_H(D^2 +
\varepsilon \id)^{-\frac{ds}{2}}$  exists then
$$
\int\alpha\;=\;\text{const}\;\mathop{\lim}_{s\searrow
1}\;(s-1)\;\text{trace}_H\;
\bigl(\pi^\cdot (\alpha) (D^2+\varepsilon\id)^{-\frac{ds}{2}}\bigr),
$$
and the result is independent of the choice of the mean $\omega$.

When $d<\infty$ and $\int(\cdot)$ is defined by \Ref{2.37}, or $\int(\cdot)$
is defined by \Ref{2.39} and the behaviour of \ trace$_H\;e^{-\beta D^2}$ is
suitably constrained, as $\beta\searrow 0$, then
$$
\int a \alpha\;=\;\int\alpha a, \quad\text{for all}\enskip a\in A\enskip
\text{and all}\enskip\alpha\in\Omega^\cdot(A).
\tag\label{2.40}
$$
The integral permits us to define a scalar product on the space
$\Omega^\cdot(A)$: For $\alpha$ and $\beta$ in $\Omega^\cdot(A)$, we set
$$
(\alpha,\beta)\;:=\;\int\alpha\beta^*\;.
\tag\label{2.41}
$$
This is linear in the first argument and anti-linear in the second argument
and is positive-semidefinite. Let $\widetilde H$ denote the completion of
$\pi^\cdot \bigl(\Omega^\cdot(A)\bigr)$, modulo the kernel of $(\cdot,\cdot)$,
in the norm determined by $(\cdot,\cdot)$. Clearly $\widetilde H$ is a Hilbert
space. It carries a $^*$representation of $A$ by bounded operators on
$\widetilde H$, determined by the equation
$$
\align
\bigl(\widetilde\pi (a)\widehat\alpha,\widehat\beta\bigr)\;
:&=\;\int a \alpha\beta^*\;=\;\int\alpha\beta^* a\\
&=\;\int\alpha(a^*\beta)^*\;=\;\bigl(\widehat\alpha,\widetilde\pi(a^*)
\widehat\beta\bigr),
\tag\label{2.42}
\endalign
$$
where $\widehat\alpha$ and $\widehat\beta$ are the vectors in $\widetilde H$
corresponding
to the elements $\alpha,\beta$ in $\Omega^\cdot(A)$. We let $\bar A$ denote
the von Neumann algebra obtained from $\widetilde\pi(A)$ by taking the weak
closure in $B(\widetilde H)$.

The Hilbert space $\widetilde H$ has a filtration into subspaces
$$
\widetilde H^{(0)} \subset \widetilde H^{(1)}\subseteq\cdots\subseteq
\widetilde H^{(n-1)} \subseteq \widetilde H^{(n)}
\subseteq\cdots\subseteq\widetilde H,
\tag\label{2.43}
$$
where $\widetilde H^{(n)}$ is defined to be the closed subspace of $\widetilde
H$
obtained by taking the closure of \ $\displaystyle\mathop{\sum}_{k=0}^n
\pi^\cdot\bigl(\Omega^k(A)\bigr)$, modulo the kernel of $(\cdot,\cdot)$, in
the norm determined by $(\cdot,\cdot)$. Let $P_D^{(n)}$ denote the orthogonal
projection onto $\widetilde H^{(n)}$. It is reasonable to define the space
$\widehat\Omega^n(A)$ of ``square-integrable $n$-forms'' by setting
$$
\widehat\Omega^n(A)\;:=\;\bigl(\id - P_D^{(n-1)}\bigr)\;\widetilde H^{(n)}
\;\equiv\; \widetilde H^{(n)} \ominus \widetilde H^{(n-1)}.
\tag\label{2.44}
$$
In the classical case $\bigl(A=C^\infty(M),\cdots\bigr)$ $\widehat\Omega^n(A)$
is
precisely the \ub{space of square-integrable} \ub{de Rham $n$-forms}.

The scalar product $(\cdot,\cdot)$ permits us to choose canonical
representatives in the equivalence classes in
$$
\Omega^\cdot(A)\big/_{(ker\,\pi^\cdot + d\,ker\,\pi^\cdot)}\;\simeq\;
\pi^\cdot\bigl(\Omega^\cdot(A)\bigr)\big/_{\pi^\cdot(d\,ker\,\pi^\cdot)}
$$
which are identified with the elements of $\Omega_D^\cdot(A)$: With an
equivalence class $[\alpha]$, $\alpha\in\Omega^n(A)$, defining an element of
$\Omega_D^n(A)$ we associate the vector $\alpha^\perp$ in $\widetilde H$
defined
by
$$
\alpha^\perp\;:=\;\bigl( \id - P_{d\,ker^{n-1}}\bigr)\;\widehat\alpha\;,
\tag\label{2.45}
$$
where \ $P_{d\,ker^{n-1}}$ \ is the orthogonal projection onto the subspace of
\
$\widetilde H$ \  spanned by \break
$d\,\widehat{ker}\,\pi^\cdot\big/_{\Omega^{n-1}(A)}$. We
define $\Omega_D^\perp (A)$ to be the linear space spanned by the forms
$\{\alpha^\perp:\alpha$ a homogeneous element of $\Omega^\cdot(A)\}$. Since
\ $\int a\alpha=\int\alpha a$, for all $a\in A$, $\alpha\in\Omega^\cdot(A)$,
and since $d\,ker\,\pi^\cdot\big/_{\Omega^n(A)}$ \ is closed under left and
right
multiplication by elements of $A$, for all $n$, $\Omega_D^\perp(A)$ is a left
and right $A$ module.

We define $H^\perp$ to be the Hilbert space of differential forms obtained by
taking the closure of $\Omega_D^\perp(A)$ in the norm determined by the scalar
product $(\cdot,\cdot)$ introduced in \Ref{2.41}. Clearly $H^\perp \subseteq
\widetilde H$, with equality in the
classical case \ $\bigl(A=C^\infty(M), D=/\!\!\!\partial_M,\cdots\bigr)$.
Since $\Omega_D^{\perp n}(A)$ is a left and right $A$ module, for all $n$,
$H^\perp$ carries a $^*$representation, $\pi^\perp$, of $A$, and it has a
filtration into subspaces $H^{\perp (0)} \subset H^{\perp (1)} \subseteq
\cdots\subseteq H^{\perp (n)} \subseteq\cdots\subseteq H^\perp$ which are
invariant subspaces for $\pi^\perp (A)$. By \Ref{2.28} and \Ref{2.29},
$$
H^{\perp (0)}\;=\;H^{(0)}\quad,\quad H^{\perp (1)}\;=\;H^{(1)},
$$
and
$$
\alpha^\perp\;=\;\widehat\alpha,
\tag\label{2.46}
$$
for all $\alpha\in\Omega^0(A) = \Omega_D^0(A)=A$ and all
$\alpha\in\Omega_D^1(A)=\Omega_D^{\perp 1}(A)=\pi^\cdot\bigl(\Omega^1
(A)\bigr)$.

In the classical case, we have that $H^\perp = \widetilde H$, as
$$
\alpha^\perp\;=\;\widehat\alpha,\quad\text{for all}\enskip
\alpha\in\widehat\Omega^n(A)\enskip\text{and all}\enskip n.
\tag\label{2.47}
$$
Moreover, for $\alpha\in\widehat\Omega^n(A)$, the operator
$$
\widehat d\widehat\alpha\;:=\;P_D^{(n+1)}\;\widehat{\pi^\cdot(d\alpha}),
\tag\label{2.48}
$$
$n=0,1,2,\cdots$, satisfies
$$
\widehat d\,^2\;=\;0.\tag\label{2.49}
$$
It defines standard \ub{exterior differentiation}.

Another interesting special case is the following one: Suppose that $F$ is an
operator on $H$ with $F^2=\lambda\id$, for some $\lambda\geq 0$, and such that
$[F,a]$ is a compact operator on $H$ with the property that $\mid[ F,a]\mid^l$
is trace class, for $l>d$. Suppose, moreover, that $F\Gamma+\Gamma F=0$, where
$\Gamma$ defines the $\Bbb Z_2$-grading on $H$. [An example is $F$ = sign $D$,
under suitable assumptions on $A$ and $(\pi,H,D,\Gamma)$; see \Ref{r3}.] For
$\alpha = a_0\,da_1\cdots da_n \in\Omega^n(A)$, $n=0,1,2,\cdots$, we define
$$
\pi^\cdot(\alpha)\;=\;a_0[F,a_1]\cdots[F,a_n].
$$
For $x\in B(H)$ we define
$$
[F,x]_\Gamma\;=\;
\cases
Fx\,-\,xF &\quad\text{if  $x\Gamma\;=\;\Gamma x\qquad(x$  even})\\
Fx\,+\,xF &\quad\text{if  $x\Gamma\;=\;-\,\Gamma x\quad(x$  odd}).
\endcases
$$
Then
$$
\bigl[ F,\pi^\cdot (\alpha)\bigr]_\Gamma\;=\;\pi^\cdot (d\alpha)
\tag\label{2.50}
$$
and hence
$$
\Omega_F^\cdot (A)\;\simeq\;\pi^\cdot\bigl(\Omega^\cdot(A)\bigr),
\tag\label{2.51}
$$
since, for $\alpha\in\;ker\,\pi^\cdot$, $0=\bigl[F,\pi^\cdot
(\alpha)\bigr]_\Gamma = \pi^\cdot (d\alpha)$, i.e.,
$d\alpha\in\;ker\,\pi^\cdot$ .

We define integration by setting
$$
\int (\cdot)\;=\;Tr_\omega(\cdot).
$$
This enables us to define an analogue of $\widetilde H$, the Hilbert space of
differential forms, and, in this case, $H^\perp = \widetilde H$, because of
\Ref{2.51}. We may now define an operator $\widehat d$ on $\widetilde H$ by
setting
$$
\widehat d\widehat\alpha\;:=\;\widehat{d\alpha},\quad\text{with}\enskip
\widehat d\,^2
\;=\;0. \tag\label{2.52}
$$
By \Ref{2.50}, $\widehat d$ is well defined, and $\widehat d\,^2 =0$ follows
from
$d^2=0$. [The situation described here may be important in the study of ``BRST
geometry'', conformal geometry \Ref{r3} and complex geometry.]

In the situation just described and in the classical case considered in
equs.~\Ref{2.47} through \Ref{2.49}, the Hilbert space of ``differential
forms'' is $\widetilde H$ which is a $\Bbb Z_2$-graded \break
($\Bbb Z$-graded, resp.)
complex for the operator $\widehat d$ defined in \Ref{2.52} (\Ref{2.48},
resp.).
On the Hilbert space $\widetilde H$, one defines a ``Dirac operator on
differential forms'', $\widetilde D$, by setting
$$
\widetilde D\;=\;\widehat d\;+\;\widehat d\,^*.
\tag\label{2.53}
$$

For topics like the definition of $C^n$-differentable structures on
non-commutative spaces and cyclic homology and cohomology, we refer the reader
to the literature, in particular \Ref{r3},\Ref{r14},\Ref{r16}. [One key idea
is to define integration of ``top-dimensional forms'' by $\int\Gamma\alpha$,
where $\Gamma$ is the $\Bbb Z_2$-grading on $H$; see \Ref{r3}.]

\bigskip

\noindent 2.6. \ \ub{A hermitian structure on differential forms}.
\medskip

The purpose of this section is to equip $\Omega_D^1(A)$ with a canonical
hermitian structure. This will permit us to introduce a natural notion of
unitary connections on $\Omega_D^1(A)$.

We start with some general considerations. Suppose that $\widehat v$ is a
vector
in $\widetilde H^{(0)}$. Then $\widehat v$ defines an operator $v^{op}$ on
$\widetilde H$
affiliated with the von Neumann algebra $\bar A$ (defined as the weak closure
of $\widetilde\pi (A)$ in $B(\widetilde H))$. Since $\widehat{\widetilde\pi
(A)}$  is dense
in $\widetilde H^{(0)}$, there exists a sequence $\{ b_\kappa\} \subset
\widetilde\pi (A)$ such that
$$
\widehat v\;=\;\mathop{s-\lim}_\kappa\;\widehat b_\kappa.
$$
We define $v^{op}$ by setting
$$
\align
(v^{op}\,\widehat a',\widehat a'')\;&=\;\lim_\kappa\;\int
b_\kappa\,a'\,(a'')^*\\
&=\;\lim_\kappa\;\bigl( \widehat b_\kappa,\,\widehat{a''(a'})^*\bigr)\\
&=\;\bigl(\widehat v,\,\widehat{a''(a'})^*\bigr).
\tag\label{2.54}
\endalign
$$
The domain of $v^{op}$ contains $\widehat{\widetilde\pi (A})$. For, if
$a\in\widetilde\pi(A)$ then
$$
\align
0\;\leq\;(v^{op}\,\widehat a,&v^{op}\,\widehat a)\;=\;\lim_\kappa \int
b_\kappa\,a\,a^* b_\kappa^*\\
&=\;\lim_\kappa\int a\, a^*\, b_\kappa^* b_\kappa\;=\;
\lim_\kappa\; (a\,a^*\,\widehat{b_\kappa^*},\widehat{b_\kappa^*})\\
&\leq\;\Vert a\,a^*\Vert \lim_\kappa \;(\widehat{b_\kappa^*},
\widehat{b_\kappa^*})\;=\;\Vert a\,a^*\Vert \lim_\kappa \int
b_\kappa^*\,b_\kappa\\
&=\;\Vert a\,a^*\Vert \lim_\kappa \int b_\kappa\,b_\kappa^*\\
&=\;\Vert a\,a^*\Vert (\widehat v,\widehat v).
\endalign
$$
If $v^{op}$ is a bounded operator then $\{ b_\kappa\}$ can be chosen
such that $\Vert b_\kappa\Vert$ is uniformly bounded, and it follows from
\Ref{2.54} that
$$
v^{op}\;=\;\mathop{w-\lim}_\kappa\;b_\kappa\;\in\;\bar A.
\tag\label{2.55}
$$
Next, let $\alpha$ and $\beta$ be in $\pi^\cdot\bigl(\Omega^\cdot(A)\bigr)$.
We define
$$
\langle\alpha,\beta\rangle_D\;:=\;P_D^{(0)}(\alpha\beta^*)\;\equiv\;
P_D^{(0)}(\alpha\beta^*)^{op},
\tag\label{2.56}
$$
where $P_D^{(0)}$ is the orthogonal projection onto the subspace $\widetilde
H^{(0)}$ of $\widetilde H$. By \Ref{2.56},
$$
\align
\bigl(\langle\alpha,\beta\rangle_D,\,\widehat a\bigr)\;
&=\;\int \langle\alpha,\beta\rangle_D\,a^*\\
&=\;\int\alpha\beta^*\;a^*\\
&=\;\int a^*\,\alpha\beta^*\;=\;\bigl(\widehat a^*,\langle\beta,\alpha\rangle_D
\bigr). \tag\label{2.57}
\endalign
$$
 From what we have shown above and definition \Ref{2.56} it follows that
$\langle\alpha,\beta\rangle_D$ defines an operator affiliated with $\bar A$.
As shown in \Ref{r11}, it is actually a bounded operator and hence belongs to
$\bar A$. By using \Ref{2.57}, it has been shown in \Ref{r11} that:
\item{(i)}
$$
\langle a\alpha, b\beta\rangle_D\;=\;a\langle\alpha,\beta\rangle_D\; b^*,
\tag\label{2.58}
$$
for arbitrary $\alpha,\beta$ in $\pi^\cdot\bigl(\Omega^\cdot(A)\bigr)$ and
arbitrary $a$ and $b$ in $\bar A$.
\item{(ii)}
$$
\langle\alpha,\alpha\rangle_D\;\geq\;0,\quad\text{for arbitrary}\enskip
\alpha\in\pi^\cdot \bigl(\Omega^\cdot(A)\bigr);
\tag\label{2.59}
$$
\item{(iii)} the anti-linear map $\alpha\mapsto\langle\cdot,\alpha\rangle_D$
defines an isomorphism from $\pi^\cdot\bigl(\Omega^\cdot(A)\bigr)$ to the
space of linear functionals on $\pi^\cdot\bigl(\Omega^\cdot(A)\bigr)$
extending continuously to linear functionals on $\widetilde H$ with values in
$\bar A$.

We conclude from (i) -- (iii) that,\  since \ $\Omega_D^1(A)\simeq\pi^\cdot
\bigl(\Omega^1(A)\bigr)$,\  $\langle\cdot,\cdot\rangle_D$\  defines a\break
\ub{generalized hermitian structure on} $\Omega_D^1(A)$ \ub{with values in}
$\bar A$.

Since we interpret $\Omega_D^1(A)$ as the (space of sections of the)
cotangent bundle over the non-commutative space described by $A$ we can view
$\langle\cdot,\cdot\rangle_D$ as the non-commutative analogue of a
\ub{Riemannian metric}. Apparently, it is uniquely determined by the $K$-cycle
$(\pi, H, D)$ on $A$ and the choice of integration, $\int (\cdot)$. Since
$\Omega_D^1(A)$ is a left \ub{and} right $A$ module, the metric
$\langle\cdot,\cdot\rangle_D$ on $\Omega_D^1(A)$ is \ub{unitary invariant}:
If $U(\bar A)$ denotes the group of unitary elements of $\bar A$ then
$$
\langle \alpha\,u,\,\beta\,u\rangle_D\;=\;\langle\alpha,\beta\rangle_D,
\tag\label{2.60}
$$
for arbitrary $\alpha,\beta \in \Omega_D^1(A)$ and arbitrary $u\in U(\bar A)$.

\bigskip
\noindent 2.7. \  \ub{Riemann-, Ricci- and scalar curvature}\,;
\ub{``Levi-Civita'' connections  on} $\Omega_D^1(A)$.
\medskip
In this section, we shall assume that $\Omega_D^1(A)$ is a finitely generated,
projective  left $A$ module. Thus, by the results of the last subsection, \
$\Omega_D^1(A)$ \ is then a \ub{hermitian vector} \ub{bundle} over $A$, the
cotangent bundle over $A$. Let $\nabla$ be a connection on $\Omega_D^1(A)$.
Thus
$$
\nabla\;:\;\Omega_D^1 (A)\;\longrightarrow\;\Omega_D^1(A)
\otimes_A\;\Omega_D^1(A)
$$
is a linear map satisfying the Leibniz rule \Ref{2.33}. By \Ref{2.14},
\Ref{2.15} the \ub{Riemann} \break
\ub{curvature} of $\nabla$ is defined by
$$
R(\nabla)\;:=\;-\,\nabla^2
\tag\label{2.61}
$$
and is an $A$-linear map from $\Omega_D^1(A)$ to $\Omega_D^2(A)\otimes_A
\Omega_D^1(A)$. Since $\Omega_D^1(A)$ is finitely generated and projective and
$\Omega_D^2(A)\otimes_A\Omega_D^1(A)$ is a left $A$ module, we may apply
eq.~\Ref{2.17} and write
$$
R(\nabla)\;=\;\sum_{\alpha,\beta} \varepsilon_\alpha \otimes_A
R^\alpha\,_\beta \;\otimes_A\;e^\beta ,
\tag\label{2.62}
$$
where $\varepsilon_\alpha\in\Omega_D^1(A)^*$ (which, thanks to the hermitian
structure defined on $\Omega_D^1(A)$, is actually isomorphic to
$\overline{\Omega_D^1(A})$), \ $R^\alpha\,_\beta\in\Omega_D^2(A)$, and
$e^\beta\in\Omega_D^1(A)$, for all $\alpha,\beta = 1,2,3,\cdots$ . Now, by
\Ref{2.30}, $\Omega_D^2(A)$ is defined as a space of equivalence classes:
$$
\Omega_D^2(A)\;\simeq\;\pi^\cdot\bigl(\Omega^2(A)\bigr)\big/_{\pi^\cdot (d\,
ker\,\pi^\cdot \big/_{\Omega^1(A)})}.
\tag\label{2.63}
$$
If we want to identify $R^\alpha\,_\beta$ with an element of $\pi^\cdot\bigl(
\Omega^2(A)\bigr)$, (i.e., with a well defined operator on the Hilbert space
$H$), we shall choose the representative $R^{\alpha,\perp}_{\enskip\;\beta}
\;\in
\Omega_D^{\perp 2} (A)$ defined by \Ref{2.45}, for $n=2$. No such choices have
to be made for $\varepsilon^\alpha$ and $e^\beta$, since
$\Omega_D^1(A)\simeq\pi^\cdot\bigl( \Omega^1(A)\bigr)$. In the classical case,
the choice $R^{\alpha,\perp}_{\enskip\;\beta}$ for $R^\alpha\,_\beta$
identifies
$R^{\alpha,\perp}_{\enskip\;\beta}$ with an ordinary (de Rham) differential
2-form,
by \Ref{2.47}, and \Ref{2.61}, \Ref{2.62} reduce to the standard definition of
the Riemann
curvature tensor.

Representation \Ref{2.62} enables us to define the \ub{Ricci}- and \ub{scalar
curvature} of $\nabla$ as follows:
$$
Ric (\nabla)\;:=\;\sum_\alpha \varepsilon_\alpha\otimes_A P_1\;
\biggl( \sum_\beta R^{\alpha,\perp}_{\enskip\;\beta}\;\cdot\;e^\beta\biggr),
\tag\label{2.64}
$$
where $P_1 := P_D^{(1)} - P_D^{(0)}$ is the orthogonal projection onto the
closed subspace $\widetilde H^{(1)} \ominus \widetilde H^{(0)}$ of ``1-forms''
in $\widetilde H$. Furthermore, we define the scalar curvature, $r(\nabla)$, of
$\nabla$ by setting
$$
r(\nabla)\;:=\;\sum_\alpha \varepsilon_\alpha \biggl( P_1\,\biggl( \sum_\beta
R^{\alpha,\perp}_{\enskip\;\beta}\;\cdot\;e^\beta\biggr)\biggr).
\tag\label{2.65}
$$
Since $\otimes_A$ and \ $\cdot$ \ are $A$-distributive and associative,
$Ric(\nabla)$ and $r(\nabla)$ are defined \ub{invariantly} by eqs.~\Ref{2.64}
and \Ref{2.65}. These equations show that
$$
Ric(\nabla)\in\Omega_D^1(A)^* \otimes_A \overline{\Omega_D^1(A)},\enskip
r(\nabla)\in\bar A,
\tag\label{2.66}
$$
where $\overline{\Omega_D^1(A)}$ denotes the closure of
$\Omega_D^1(A)\simeq\pi^\cdot\bigl(\Omega^1(A)\bigr)$ in the norm determined
by the scalar product $(\cdot,\cdot)$ defined in \Ref{2.41}.

The \ub{Einstein-Hilbert action} in non-commutative geometry is now defined by
$$
I(\nabla)\;:=\;\kappa \int r (\nabla)\;+\;\Lambda \int\id,
\tag\label{2.67}
$$
where $\kappa$ is related to Newton's constant and $\Lambda$ is the
cosmological constant; see \Ref{r11}.

A connection $\nabla$ on $\Omega_D^1(A)$ is said to be unitary if, for all
$\alpha$ and $\beta$ in $\Omega_D^1(A)$,
$$
d\langle\alpha,\beta\rangle\;=\;\langle\nabla\alpha,\beta\rangle\;-\;
\langle\alpha,\nabla\beta\rangle,
\tag\label{2.68}
$$
see eq.~\Ref{2.21}, Sect. 2.3. \ As in eq.~\Ref{2.34}, Sect. 2..4., the
torsion of $\nabla$ is defined by
$$
T(\nabla)\;=\;d - m \circ \nabla.
$$

It is tempting to define a \ub{Levi-Civita connection} to be a \ub{unitary}
connection, $\nabla_{LC}$, on $\Omega_D^1(A)$, whose torsion,
$T(\nabla_{LC})$, vanishes.

It is straightforward to show that, in the classical case, $I(\nabla_{LC})$,
as given by \Ref{2.67}, reduces to the usual Einstein-Hilbert action (with
cosmological constant $\Lambda)$, \Ref{r11}.

\bigskip

\noindent \ub{Remarks}.
\medskip
\bf (1)\rm \  \ In general, $\Omega_D^1(A)$ is not a free left $A$ module,
i.e., the cotangent bundle over a non-commutative space is usually not a
trivial bundle; as one would expect.
\medskip
\bf (2)\rm \ \ The cotangent bundle of a non-commutative space need not admit
any Levi-Civita connection, and -- if it admits such connections -- the
Levi-Civita connection may not be unique.
\medskip
\bf (3)\rm \ \ If $\Omega_D^1(A)$ is a free left $A$ module then, by
definition of free modules, it has a basis. It is natural to choose a basis
$\{ e^\beta\}_{\beta=1}^N$ which is orthonormal with respect to the canonical
hermitian structure $\langle\cdot,\cdot\rangle_D$ on $\Omega_D^1(A)$, i.e.,
$$
\big\langle e^\alpha, e^\beta\big\rangle_D\;=\;\delta^{\alpha\beta}\;1.
\tag\label{2.69}
$$
The basis elements $e^\alpha$ are analogues of the ``vielbein'' used in
Cartan's formalism of Riemannian geometry. The automorphisms of
$\Omega_D^1(A)$ are then generated by unitary $N\times N$ matrices,
$M=(M^\alpha\,_\beta)$, with matrix elements in $A$.

One may now define the Cartan structure equations in non-commutative geometry,
(see \Ref{r11}): \ Let \ $\omega^\alpha\,_\beta \in \Omega_D^1(A)$ be defined
by
$$
\nabla\,e^\alpha\;=\;-\,\omega^\alpha\,_\beta\;\otimes_A\;e^\beta
\tag\label{2.70}
$$
and let $T^\alpha \in \Omega_D^2(A)$ be given by
$$
T^\alpha\;=\;T (\nabla)\;e^\alpha\; .
\tag\label{2.71}
$$
Finally, we define \ $R^\alpha\,_\beta \in \Omega_D^2(A)$ \ by setting
$$
R(\nabla)\;e^\alpha\;=\;R^\alpha\,_\beta\;\otimes_A\;e^\beta \;.
\tag\label{2.72}
$$
Then the Cartan equations are
$$
T^\alpha\;=\;d e^\alpha\;+\;\omega^\alpha\,_\beta\;e^\beta,
\tag\label{2.73}
$$
$$
R^\alpha\,_\beta\;=\;d\omega^\alpha\,_\beta\;+\;\omega^\alpha\,_\gamma\;
\omega^\gamma\,_\beta.
\tag\label{2.74}
$$
If $\{\varepsilon_\alpha\}$ is the basis of $\Omega_D^1(A)^*$ dual to the
basis $\{ e^\alpha\}$ of $\Omega_D^1(A)$ then formula \Ref{2.62} gives
$$
R(\nabla)\;=\;\varepsilon_\alpha\;\otimes_A\;R^\alpha\,_\beta\;
\otimes_A\;e^\beta,
\tag\label{2.75}
$$
and if $\varepsilon_\alpha$ is identified with the element $e_\alpha \equiv
e^\alpha$ of $\Omega_D^1(A)$, using the hermitian structure
$\langle\cdot,\cdot\rangle_D$ on $\Omega_D^1(A)$, then
$$
\align
I(\nabla)\;&=\;\kappa \int R^{\alpha,\perp}_{\enskip\;\beta}\;e^\beta\;
e_\alpha^*\;+\;\Lambda \int 1 \\
&=\;\kappa\bigl( R^{\alpha,\perp}_{\enskip\;\beta}\; e^\beta,\;e_\alpha\bigr)\;
+\;
\Lambda \int 1\;.
\tag\label{2.76}
\endalign
$$
See \Ref{r11}.
\medskip
\bf (4)\rm \ \  An alternative approach to defining a generalized
Einstein-Hilbert action goes as follows: In the classical case, it is easy to
show that $\int r (\nabla)$ is proportional to the constant term in the
Laurent series expansion of
$$
-\;\frac{\frac{d}{d\beta}\;\text{trace}_H\,\bigl(e^{-\beta D^2}
\bigr)}{\text{trace}_H\,\bigl(e^{-\beta D^2}\bigr)}\;=\;
\frac{\text{trace}_H\,\bigl(D^2\,e^{-\beta D^2}\bigr)}{\text{trace}_H \,
\bigl(e^{-\beta D^2}\bigr)}
$$
around $\beta=0$ which we denote by $\int D^2$.

Hence if $(\pi,H,D)$ is a $K$-cycle for a unital $^*$algebra $A$ one may
define a generalized Einstein-Hilbert action by setting
$$
I(D)\;=\;\kappa \int D^2\;+\;\Lambda\int 1\;.
\tag\label{2.77}
$$
However, in general, this definition of $I$ is \ub{not} equivalent to the one
given in eq.~\Ref{2.67}, as we have checked for the examples discussed in
Sect. 3.

The approach sketched here suffers from an ambiguity: In general the algebra,
$\pi(A)'$, of all bounded operators on $H$ commuting with the operators of
$\pi(A)\simeq A$ contains a non-trivial subspace, $B_{odd}$, of \ub{odd}
operators,
$$
B_{odd}\;=\;\bigl\{ b\in\pi (A)'\;:\;b\Gamma\;=\;-\,\Gamma b\bigr\}.
$$
Then two ``Dirac operators'', $D$ and $D'$, on $H$ for which $[D,a] = [D',a]$,
for all $a\in A$, may differ by an operator $b$ affiliated with $B_{odd}$,
i.e., $D'=D+b$. If $b$ is a compact perturbation of $D$ then our definition of
integration in eq.~\Ref{2.37} is independent of $b$. Thus, perturbations $b$
affiliated with $B_{odd}$ and relatively compact with respect to $D$ describe
the ambiguities in the definition of the ``Dirac operator'' on $H$ which
propagate into the definition of $I(D)$, as given in \Ref{2.77}. In attempting
to eliminate them one must presumably return to the tools developed in Sects.
2.5 -- 2.7 and derive expressions for a Levi-Civita spin connection.
\medskip
\bf (5)\rm \ \ To incorporate gauge fields in this formalism, one is led,
according to Connes \Ref{r3}, to study hermitian vector bundles $E$ over $A$,
as in Sect. 2.3, but with $\Omega^\cdot (A)$ replaced by $\Omega_D^\cdot(A)$,
with connections $\nabla : E\to \Omega_D^1(A) \otimes_A E$, whose curvature,
$R(\nabla)$, is given by the formula in eq.~\Ref{2.17}, i.e.,
$$
R(\nabla)\;=\;-\,\nabla^2\;=\;\sum_{\alpha,\beta} \varepsilon_\alpha\;
\otimes_A\;F^{\alpha,\perp}_{\enskip\;\beta}\;\otimes\;e^\beta,
$$
with $\varepsilon_\alpha\in E^*$, $e^\beta \in E$ and
$F^{\alpha,\perp}_{\enskip\;\beta}\;\in\;\Omega_D^{\perp^2} (A)$. \ The
Yang-Mills
action functional \Ref{r3} is then defined by
$$
YM(\nabla)\;=\;\sum_{\alpha,\beta} \int F^{\alpha,\perp}_{\enskip\;\beta}\;
F^{\beta,\perp}_{\enskip\;\alpha}\;.
$$

\vskip 1truecm

\heading
Examples: \ Einstein-Hilbert and Chern-Simons\\
action for two-sheeted space-times.\\ \\
\endheading
In this section we illustrate Connes' formalism sketched in Sect.~2 in the
context of some simple examples. As in \Ref{r5}, we choose (Euclidean)
space-time, $X$, to consist of two copies of a four-dimensional spin manifold
$M$:
$$
X\;=\;M \times \Bbb Z_2.
\tag\label{3.1}
$$
We consider a non-commutative space described by an algebra $A$ given by
$$
A\;=\;{\Cal A}_1 \otimes C^\infty (M) \oplus {\Cal A}_2 \otimes C^\infty (M),
\tag\label{3.2}
$$
where ${\Cal A}_1$ and ${\Cal A}_2$ are finite-dimensional, unital
$^*$algebras over the real or complex numbers. It is convenient to think of
elements of $A$ as operators of the form
$$
\pmatrix
\id \otimes a_1 & 0 \\
\; &\;\\
0               & \id \otimes a_2
\endpmatrix
\tag\label{3.3}
$$
where $a_i$ is a smooth function on $M$ with values in
${\Cal A}_i$, $i=1,2$, and $\id$ is the identity in the Clifford algebra,
$Cliff (T^*M)$, of Dirac matrices over $M$. To define a differentiable
structure on $A$, we consider even $K$-cycles $(\pi,H,D,\Gamma)$ for $A$,
with:
\medskip
\itemitem{(a)} $\pi \;=\;\pi_1 \oplus \pi_2$.

\itemitem{(b)} $\pi_i$ \ is a representation of ${\Cal A}_i \otimes C^\infty
(M)$ on a Hilbert space $L^2 (S_i, \tau_i, dv)$, where $S_i$ is a bundle of
spinors on $M$ with values in a finitely generated, projective, hermitian left
${\Cal A}_i$ module $E_i$; the scalar product on $L^2(S_i,\tau_i,dv)$ is given
by
$$
(\psi_1,\psi_2)\;=\;\int_M dv_x\;\tau_i\;\bigl\langle \psi_1(x),
\psi_2(x)\big\rangle_i,
\tag\label{3.4}
$$
where $\tau_i$ is a normalized trace on ${\Cal A}_i$,
$\langle\cdot,\cdot\rangle_i$ denotes the hermitian structure on $E_i, i=1,2$,
and $dv_x$ is the volume element on $M$. Then we define $H$ by
$$
H\;=\;L^2\,(S_1,\tau_1,dv) \oplus L^2\,(S_2,\tau_2,dv).
\tag\label{3.5}
$$

\itemitem{(c)} The Dirac operator is given, for example, by
$$
D\;=\;
\pmatrix
\big/\!\!\!\!\nabla_M \otimes \id_1 & \gamma^5 \otimes \phi \\
\;&\;\\
\gamma^5 \otimes \phi^* &\big/\!\!\!\!\nabla_M \otimes \id_2
\endpmatrix,
\tag\label{3.6}
$$
where $\big/\!\!\!\!\nabla_M$ is the standard covariant Dirac operator on $M$,
and $\id_i$ is the identity operator in ${\Cal A}_i$, $i=1,2$; $\phi$ is a
homomorphism from $E_2$ to $E_1$, and $\phi^*$ is the adjoint homomorphism
from $E_1$ to $E_2$. Finally, \ $\gamma^5=\gamma^1\gamma^2\gamma^3\gamma^4$,
$\{ \gamma^a, \gamma^b\} = - 2\delta^{ab}$, $(\gamma^a)^* = - \gamma^a$,
$\gamma^\mu = e^\mu\,_a$ $\gamma^a$, where $e^\mu\,_a\, (x)\, \partial_\mu$
is a basis of the tangent space, $T_xM$, of $M$ at $x$, with
$e^\mu\,_a\;e^\nu\,_b\;\delta^{ab} = g^{\mu\nu}$, and $g_{\mu\nu}$ is a
Riemannian metric on $M$. We choose $dv_x$ to be the volume element
corresponding to the metric $g_{\mu\nu}(x)$.

\itemitem{(d)} The $\Bbb Z_2$ grading on $M$ is given by
$$
\Gamma\;=\;
\pmatrix
\gamma^5 & 0 \\
\; & \;\\
0 & -\gamma^5
\endpmatrix
. \tag\label{3.7}
$$

A second interesting example is obtained as follows: \ We choose $A$ to have
the form
$$
A\;=\;{\Cal A} \otimes C^\infty (M),
$$
where ${\Cal A}$ is a finite-dimensional, unital $^*$algebra.

An odd $K$-cycle \ $(\pi_0,H_0,D)$ is given by
\medskip
\itemitem{(a')} a representation $\pi_0$ of $A$ on a
\itemitem{(b')} a Hilbert space $H_0 = L^2 (S, \tau, dv)$, where $S,\tau$ and
$dv$ are as above; and
\itemitem{(c')} $D\;\overset{e.g.}\to{=}\;\big/\!\!\!\!\nabla_M
\;\otimes\;\id$.
\medskip
 From this $K$-cycle we obtain an even $K$-cycle by setting
\medskip
\itemitem{(a'')} $\pi = \pi_0 \oplus \pi_0$;
\itemitem{(b'')} $H = H_0 \oplus H_0$;
\itemitem{(c'')}
$$
D\;=\;\pmatrix
\big/\!\!\!\!\nabla_M \otimes\id & i\,\gamma^5\otimes \phi\\
\; &\;\\
-\,i\,\gamma^5\phi & -\,\big/\!\!\!\!\nabla_M\otimes\id
\endpmatrix
, \tag\label{3.8}
$$
where $\phi = \phi^* \in \;End (E)$; and
\itemitem{(d'')}
$$
\Gamma\;=\;\pmatrix
0 & \id\otimes\id\\
\; &\;\\
\id\otimes\id & 0
\endpmatrix
. \tag\label{3.9}
$$

Now
$$
\pi(a)\;=\;\pmatrix
\id\otimes \pi_0 (a) & 0 \\
\; &\; \\
0 & \id \otimes \pi_0 (a)
\endpmatrix
, \enskip a \in A,
$$
clearly commutes with $\Gamma$, and one easily checks that $D$, as given in
\Ref{3.8}, anticommutes with $\Gamma$.

To these examples we shall now apply the methods developed in Sect.~2.

\bigskip

\noindent 3.1. \ \ub{Generalized Einstein-Hilbert actions for two-sheeted
space-time geometries}.
\medskip

In this section, we briefly review the example of general relativity on a
two-sheeted space-time proposed in \Ref{r11}. Let Euclidean space-time, $X$,
be as in eq.~\Ref{3.1} and the algebra $A$ as in \Ref{3.2}, with ${\Cal A}_1 =
{\Cal A}_2 = \Bbb C$. We choose an even $K$-cycle $(\pi,H,D,\Gamma)$ for $A$
as in (a) - (d); see eqs.~\Ref{3.4} through \Ref{3.7}. Then the ``cotangent
bundle'' $\Omega_D^1 (A)$ is a free left and right $A$ module, with a basis
$\{ e^N\}_{N=1}^5$ given by
$$
e^a\;=\;\pmatrix
\gamma^a & 0\\
\;&\;\\
0 & \gamma^a\endpmatrix
\enskip =\enskip
\pmatrix
\gamma^\mu\,e_\mu^a & 0 \\
\; &\;\\
0 & \gamma^\mu\,e_\mu^a\endpmatrix
,\enskip a = 1,2,3,4,
\tag\label{3.10}
$$
$$
e^5\;=\;\pmatrix
0 &\gamma^5\\
\;&\;\\
-\,\gamma^5 & 0\endpmatrix
. \tag\label{3.11}
$$
The hermitian structure on $\Omega_D^1(A)$ is given by the trace on 8$\times$8
matrices, normalized such that $tr \id = 1$. Hence
$$
\big\langle e^N, e^M\big\rangle_M\;=\;tr\;\bigl( e^N\,(e^M)^*\bigr)\;=\;
\delta^{NM}.
\tag\label{3.12}
$$
Using the generalized Cartan formalism, eqs.~\Ref{2.69} - \Ref{2.74}, we find
that, in this example, the components of a connection $\nabla$ on
$\Omega_D^1(A)$ in the basis $\{ e^N\}_{N=1}^5$ are given by
$$
\omega^N\,_M\enskip=\enskip \pmatrix
\gamma^\mu\,\omega_{1\mu M}^N & \gamma^5\,\phi\,l^N\,_M\\
\;&\;\\
-\gamma^5\,\phi\,\widetilde l^N\,_M & \gamma^\mu\,\omega_{2\mu M}^N
\endpmatrix
. \tag\label{3.13}
$$
The unitarity of $\nabla$ then implies that
$$
\omega_{i\mu M}^N\;=\;-\,\omega_{i\mu N}^M,\quad i=1,2,\quad\text{and}\quad
\widetilde l^N\,_M\;=\;-\,l^M\,_N .
\tag\label{3.14}
$$
The expressions for the components $R^{N,\perp}_{\enskip\;_M}$ of the
curvature,
$R(\nabla) = - \nabla^2$, of $\nabla$ are found to be given by
$$
R^{N,\perp}_{\enskip\;M}\;=\;\pmatrix
\gamma^{\mu\nu}\,R_{1\mu\nu M}^N & \gamma^5\,\phi\,Q_{\mu M}^N \\
\;&\;\\
-\gamma^5\,\phi\,\widetilde Q_{\mu M}^N & \gamma^{\mu\nu}\,R_{2\mu\nu M}^N
\endpmatrix
, \tag\label{3.15}
$$
with
$$
\align
R_{i\mu\nu M}^N\;&=\;\partial_\mu \omega_{i\nu M}^N - \partial_\nu
\omega_{i\mu M}^N + \omega_{i\mu L}^N \omega_{i\nu M}^L - \omega_{i\nu L}^N
\omega_{i\mu M}^L, i=1,2, \\
\gamma^{\mu\nu}\;:&=\;\frac 1 2\;\bigl( \gamma^\mu \gamma^\nu - \gamma^\nu
\gamma^\mu\bigr), \\
Q_{\mu M}^N\;&=\;\partial_\mu \, l^N\,_M\,+\,\omega_{1\mu M}^N - \omega_{2\mu
M}^N + \omega_{1\mu L}^N \, l^L\,_M\,-\,l^N\,_L\, \omega_{2\mu M}^L, \\
\widetilde Q_{\mu M}^N\;&=\;- \partial_\mu \,l^M\,_N\,+\,\omega_{1\mu M}^N -
\omega_{2\mu M}^N\,+\,l_L\,^N\,\omega_{1\mu M}^L - \omega_{2\mu L}^N\, l_M\,^L
{}.
\endalign
$$
Imposing the condition that the torsion, $T(\nabla)$, of $\nabla$ vanishes one
deduces that
$$
\omega_{1\mu b}^a\;=\;\omega_{2\mu b}^a\;\equiv\;\omega_{\mu b}^a
$$
is the classical Levi-Civita connection derived from the metric \ $g_{\mu\nu}
= e_\mu^a \delta_{ab} e_\nu^b$ on $M$, for $a,b = 1, \cdots, 4$;
$$
\align
l^a\,_b\;&=\;l^b\,_a\;,\;a,b\;=\;1,\cdots,4,\quad\text{and}\enskip
l^5\,_a\;=\;-\,l^a\,_5, \\
\omega_{1\mu 5}^a\;&=\;-\,\omega_{2\mu 5}^a\;=\;\phi\,l^a\,_b\;e^b\,_\mu\;; \\
l^5\,_a\;e^a\,_\mu\;&=\;-\,\partial_\mu\;\phi^{-1}.
\tag\label{3.16}
\endalign
$$
The Einstein-Hilbert action defined in \Ref{2.76} is then calculated to be
$$
\align
I(\nabla)\;=\;\kappa\int_M&\biggr[ 2r-4\phi\nabla_\mu \partial^\mu \phi^{-1} +
4\phi^2\,l^a\,_a\;l^5\,_5 \\
&+\;\phi^2\bigl((l^a\,_a)^2\,-\,l^a\,_b\;l^b\,_a\bigr)\biggr] \sqrt{g}\;
d^4x\,+\,\Lambda \;vol. (M),
\tag\label{3.17}
\endalign
$$
where $r$ is the scalar curvature of the classical Levi-Civita connection. The
fields $l^a\,_b$ and $l^5\,_5$ turn out to decouple. Setting $\phi =
e^{-\sigma}$ and eliminating \ $l^a\,_b$, $l^5\,_5$, one finds \Ref{r11}.
$$
I(\nabla)\;=\;2\kappa\int_M \bigl[ r - 2\partial_\mu \sigma \partial^\mu
\sigma\bigr] \;\sqrt{g}\; d^4x + \Lambda\;vol. (M).
\tag\label{3.18}
$$
Thus, in this approach, the theory of gravity on $X=M\times \Bbb Z_2$ is
equivalent to general relativity on $M$, with an additional, massless scalar
field $\sigma$ that couples to gravity via the metric on $M$. [Some results
concerning this model have been reported in \Ref{r18}.] Geometrically,
$e^{-\sigma(x)}$  \ is a measure for the distance between the two copies of
$M$ at a point $x\in M$. [For generalizations see \Ref{r19}.]

It is worthwhile to compare expression \Ref{3.18} to the one obtained from
definition \Ref{2.17}
of the Einstein-Hilbert action. The total ``Dirac
operator'' $D'$, with a spin connection
determined from \ $\omega^N\,_M$, \ see \
\Ref{3.13}, \ and with \ $l^a\,_b = l^5\,_5 = 0$, \ is given by \break
$D' = D - \frac 1 2
 \enskip /\!\!\!\partial \sigma \otimes$ $1\enskip 0 \choose 0\enskip 1$. \ \
The action \ $I(D')$,\  defined as in \Ref{2.77}, comes out to be
$$
I(D')\;=\;\kappa\int_M\bigl[ ar\,+\,b\,e^{-2\sigma}\bigr]
\sqrt{g}\;d^4x\,+\,\Lambda\;vol. (M),
$$
for some constants $a$ and $b$. Apparently, it does \ub{not} coincide with
\Ref{3.18}. However, had we chosen the $\Bbb Z_2$-grading \ $\Gamma$ =
$\gamma^5\enskip 0 \choose 0\enskip \gamma^5$ \ and set
$$
D'\;:=\;\pmatrix
\big/\!\!\!\!\nabla_M  & \delta\,\gamma^5 /\!\!\!\partial\;\sigma\\
\; &\;\\
\delta\,\gamma^5 /\!\!\!\partial\;\sigma   & \big/\!\!\!\!\nabla_M
\endpmatrix
, \quad \delta\enskip\text{some constant},
$$
we would have obtained
$$
I(D')\;=\;\kappa\int\bigl[
ar-\delta^2\partial_\mu\,\sigma\,\partial^\mu\,\sigma\bigr]\;+\;\Lambda\;vol.
(M),
$$
which essentially agrees with \Ref{3.18}. [Details of these calculations have
been carried out in \Ref{r17}.]


\bigskip
\noindent 3.2. \ \ub{The standard model coupled to gravity}.
\medskip

Connes and Lott \Ref{r5}, \Ref{r3} have shown how to construct the tree-level
Lagrangian of the standard model from the formalism of non-commutative
geometry. In order to construct the electroweak sector, they use the formalism
sketched in Sect.~2.3 and remark (5) of Sect. 2.7. \ They use an algebra $A$
as in \Ref{3.2}, with
$$
{\Cal A}_1\;=\;\Bbb H\quad\text{(the real algebra of quaternions)},\quad {\Cal
A}_2\;=\;\Bbb C.
$$
The Dirac operator is chosen as in \Ref{3.6}, with
$$
\phi\;=\;e^{-\sigma}\;\phi_0,
$$
for some constant homomorphism $\phi_0$; see \Ref{r3}, \Ref{r5}. They set
$\sigma=0$. However, since we have identified $\sigma$ as a dynamical field
coupled to gravity, we choose $\sigma$ to be an arbitrary function of $x\in
M$.

In order to couple the quarks to colour - SU(3), Connes and Lott choose the
corresponding spinors to take values in an $A$-$B$ bimodule, where $A$ is as
above, and $B = \bigl(\Bbb C \oplus \Bbb M_3 (\Bbb C)\bigr) \otimes C^\infty
(M)$. For three generations of fermions and a suitable choice of the
Kobayashi-Maskawa mixing matrix, they obtain precisely the Lagrangian of the
standard model \ub{including the Higgs field}. Their construction has the
following interesting features:

\itemitem{--} The Higgs field is identified with a component of the
generalized electro-weak gauge connection and thus aquires a geometrical
significance.

\itemitem{--} For the Higgs potential \ub{not} to vanish one must require
\ub{more than one generation of} \ub{fermions}.

\itemitem{--} At the tree level, the cosmological constant of the Connes-Lott
Lagrangian \ub{vanishes} \ub{naturally}.

One may now proceed to calculate the one-loop effective potential of the
theory, as in \Ref{r20}. Making the \ub{heuristic} ansatz that, to order
$\hbar$, the cosmological constant of the theory retains the form imposed by
the formalism of non-commutative geometry at the tree level, one finds an
explicit expression for the effective potential, $V^{(1)} (H, \sigma; g)$, to
order $\hbar$, where $H$ denotes the vacuum expectation value of the Higgs
field, and $g_{\mu\nu}$ is the metric on $M$. Choosing $g_{\mu\nu}
\equiv \eta_{\mu\nu}$
to be the flat metric (a delicate point in the argument\bf
!\rm), one may proceed to minimize $V^{(1)} (H,\sigma;\eta)$ in $H$ \ub{and}
in $\sigma$. The result of the calculation \Ref{r21} is quite surprizing: The
minimization in $\sigma$ yields one new relation between the parameters of the
standard model Lagrangian which, together with the requirement that $V^{(1)}
(H,\sigma;\eta)$ be stable in $H$ and $\sigma$, fixes the mass, $m_t$, of the
top quark to satisfy the bounds \ 146 GeV $< m_t <$ 147.5 GeV and yields a
relation between $m_t$ and the mass, $m_H$, of the Higgs that constrains $m_H$
to lie between $\sim$ 110 GeV \ and $\sim$ 150 GeV; \Ref{r21}.

Of course, these predictions have, \ub{at best}, heuristic value, since the
problem of fixing the form of the cosmological constant to order $\hbar$ and
higher by imposing natural, geometrical constraints is not understood.
However, they \ub{do} suggest that gravitational effects may play a role in
understanding masses of fermions and Higgses and that methods of
non-commutative geometry may be useful in understanding these problems.

\bigskip
\noindent 3.3. \ \ub{Chern-Simons actions and gauge theories of gravitation}.
\medskip

The purpose of this section is to briefly review some recent results \Ref{r22}
on the Chern-Simons action in non-commutative geometry. We consider a
non-commutative space described by an algebra
$A={\Cal A}\otimes C^\infty (M)$, where ${\Cal A}$ is a finite-dimensional,
unital $^*$algebra. The differentiable structure of $A$ is given by an odd
$K$-cycle $(\pi_0, H_0, D)$ for $A$ with properties $(a'),(b')$ and $(c')$
specified at the beginning of Sect. 3. Let us first consider the case where
the dimension, $d$, of $M$ is \ub{odd}, and $M$ is a homology sphere. We
consider a vector bundle $E$ over $A$ given by $A$ itself. The components of a
connection on $E$ are then given by one-forms $\pi_0(\alpha)\in\Omega_D^1(A)$,
and the corresponding curvature is obtained from $\pi_0(d\alpha
+\alpha^2)\in\Omega_D^{\perp 2}(A)$. The Chern-Simons forms are given by
$$
\align
\vartheta^{(3)}\;&:=\;\pi_0(\alpha)\,\pi_0(d\alpha)\;+\;\frac 2 3\;
\pi_0(\alpha^3)\,\in\,\pi_0\bigl(\Omega^3(A)\bigr), \\
\vartheta^{(5)}\;&:=\;\pi_0(\alpha)\,\pi_0\bigl((d\alpha)^2\bigr)\,+\,\frac 3
2 \;\pi_0(\alpha^3)\,\pi_0(d\alpha)\;+\;\frac 3 5 \;
\pi_0(\alpha^5)\,\in\,\pi_0 \bigl(\Omega^5(A)\bigr),
\endalign
$$
where $\pi_0(d\alpha)$ is chosen to belong to $\Omega_D^{\perp 2} (A)$; etc..
Chern-Simons actions are now defined by
$$
I_{CS}^{(d,d)}(\alpha)\;:=\;i\int\vartheta^{(d)}\;\equiv\;i\;Tr_\omega
\bigl(\vartheta^{(d)}\mid D\mid^{-d}\bigr);
\tag\label{3.19}
$$
see eq.~\Ref{2.37}, Sect.~2.5. They turn out to agree with the classical
Chern-Simons actions.

The formalism of non-commutative geometry allows us to consider Chern-Simons
actions in the case where $M$ is \ub{even-dimensional}, $(d=2,4,\cdots)$: We
choose the algebra $A$ as above, but consider an \ub{even} $K$-cycle
$(\pi,H,D,\Gamma)$ defined as in $(a'') - (d'')$; see eqs.~\Ref{3.8},
\Ref{3.9}. Then we define the Chern-Simons action by
$$
I_{CS}^{(d+1,d)}\;:=\;\int\Gamma\vartheta^{(d+1)}\;\equiv\;Tr_\omega\;
\bigl(\Gamma\vartheta^{(d+1)}\mid D\mid^{-d}\bigr)\;.
\tag\label{3.20}
$$
What kind of actions do we obtain? For $d=2$ and ${\Cal A} = \Bbb M_n(\Bbb
C)$, for example, we obtain a two-dimensional topological gauge theory with
action
$$
I_{CS}^{(3,2)}\;=\;i\,c \int_{M^2} tr\;(\phi F),
\tag\label{3.21}
$$
where \ $F=F_{ij}\; dx^i \wedge dx^j$, $F_{ij} = \partial_i A_j - \partial_j
A_i
+ \bigl[ A_i, A_j\bigr]$, with $A\in u(n)$, and $c$ is a constant. This is the
theory first considered in \Ref{r23}. We could also have considered the action
$$
I_{CS}^{(5,2)}\;:=\;Tr_\omega\;\bigl( \Gamma\,\vartheta^{(5)}\mid
D\mid^{-2}\bigr)
$$
and would have obtained
$$
I_{CS}^{(5,2)}\;=\;i\,c\int_{M^2}tr\;\bigl[-\,\phi\,\nabla\phi\wedge\nabla\phi
\;+\;\phi^3\,F\bigr],
\tag\label{3.22}
$$
where $\nabla$ denotes covariant differentiation in the adjoint representation
of ${\Cal A}$. Similarly,
$$
I_{CS}^{(5,4)}\;=\;i\,c\int_{M^4} tr\;\bigl(\phi\,F\wedge F\bigr),
$$
an action of interest in connection with Donaldson theory, \Ref{r25}.

A particularly interesting example is obtained when one chooses ${\Cal A} =
Cliff_{\Bbb R}(SO(4))$. As usual one requires that the connection $\alpha$ is
hermitian. After somewhat lengthy calculations \Ref{r22}\  one finds that \
$I_{CS}^{(5,4)}$ \ and \ $I_{CS}^{(7,4)}$ \ determine Lagrangians for
\ub{topological} \ub{gravity theories} formulated as metric-independent gauge
theories of a vierbein and a spin connection coupled to a $Cliff_{\Bbb
R} (SO(4))$-valued scalar field $\phi$. Details concerning these theories go
beyond the scope of this review; but see \Ref{r22}.

\vskip 1truecm

\heading
Conclusions and outlook. \\
\endheading

In this survey we have discussed some elements of Connes' non-commutative
geometry and indicated some applications of the formalism; mostly in the
context of classical field theory and for spaces which are ``close'' to
classical commutative spaces but which are not manifolds in the classical
sense. We have found that when general relativity is formulated on generalized
spaces, fields such as $\sigma$ and $B_{\mu\nu}$ appear in the theory which
also appear in supergravity and superstring theory and receive a geometrical
interpretation: They describe the geometrical structure of \ub{discrete}
internal spaces. It is tempting to imagine that what we have found is the
``classical regime'', the geometry of a ``space of zero modes'', of a putative
quantum theory of space-time structure  which one may hope  can be
formulated within the formalism of \ub{infinite-dimensional} non-commutative
geometry.

We have also seen that many familiar topological field theories can be derived
from Chern-Simons theories on generalized commutative and non-commutative
spaces, typically products of a classical manifold with a discrete commutative
or non-commutative ``internal space''. We have studied finite-dimensional
examples. According to the program described in the introduction, one should
extend these attempts to infinite-dimensional examples. This might shed new
light on string field theory which, at least for open, bosonic strings, has
the form of a Chern-Simons theory \Ref{r25}. An attempt to fit Witten's open
string field theory into Connes' formalism of non-commutative geometry has
been described in \Ref{r26}, but further work in this direction appears to be
necessary before these problems will be understood more fully. As suggested by
the work in \Ref{r26}, it is tempting to think that Connes' theory of
foliations (see \Ref{r3}) will be useful in understanding gauge fixing and
BRST cohomology in a deeper way which play a vital role in the quantization of
all theories with infinite-dimensional symmetries.

\vskip 1truecm

\noindent \ub{Acknowledgements}. \ We thank B. Bleile, O. Grandjean and
especially A. Connes, G. Felder and K. Gaw\c edzki for very useful discussions.
We are
grateful to A. Connes for providing us with advance copies of refs.~\Ref{r3}
and \Ref{r5}.


\vfill\eject


\heading
References\\
\endheading
\newstyle\list1#1{[#1]}
\list
\item\label{r1} M. Green, J. Schwarz and E. Witten, ``Superstring theory'',
Cambridge University Press, (1987).

\item\label{r2} A.M. Polyakov, ``Gauge fields and strings'', Harwood Academic
Publ., Chur (1987).

\item\label{r3} A. Connes, ``Non commutative differential geometry'', Publ.
I.H.E.S. \ub{62}, 41-144 (1985).\hfill\break
A. Connes, ``G\'eom\'etrie non commutative'', Inter\'editions, Paris
1990.\hfill\break
A. Connes, ``Non commutative geometry'', book to be published by Academic
Press.

\item\label{r4} M. Dubois-Violette, R. Kerner and J. Madore, ``Non-commutative
geometry and new models of gauge theory'', Preprint, Orsay 1988, J. Math.
Phys. \ub{31}, 316 (1990).

\item\label{r5} A. Connes and J. Lott, Nucl. Phys. B\ub{18} (Proc. Suppl.),
29-47 (1990); in ``New symmetry principles in quantum field
theory'', p. 53 (1992); editors: J. Fr\"ohlich et al., Plenum Pub. \hfill\break
D. Kastler, Marseille preprints, CPT-91/P2610, P2611, and
CPT-92/P2814;\hfill\break
D. Kastler and M. Mebkhout, ``Lectures on non-commutative geometry and
applications to elementary particles'', book to be published.

\item\label{r6} R. Coquereaux, G. Esposito-Far\`ese and G. Vaillant, Nucl.
Phys. B \ub{353}, 689 (1991);\hfill\break
R. Coquereaux, G. Esposito-Far\`ese and F. Scheck, Int. J. Mod. Phys. \ub{A7}
(1992) 6555.

\item\label{r7} A.H. Chamseddine, G. Felder and J. Fr\"ohlich, Phys. Lett.
\ub{296} B, 109 (1992); Nucl. Phys. B\ub{395}, 672 (1993);\hfill\break
A.H. Chamseddine and J. Fr\"ohlich, ``SO(10) unification in non-commutative
geometry'', preprint 1993.

\item\label{r8} D. Gepner, ``Exactly solvable string compactification on
manifolds of SU(N) holonomy'', Phys. Lett. \ub{199}B (1987) 380;\hfill\break
T. Eguchi,  A. Taormina and S.K. Yang, ``Superconformal algebras and string
compactification on manifolds with SU(N) holonomy'', Nucl. Phys. B\ub{315},
193 (1989).

\item\label{r9} F. Gabbiani and J. Fr\"ohlich, ``Operator algebras and
conformal field theory'', Commun. Math. Phys., to appear.

\item\label{r10} M. L\"uscher and G. Mack; unpubl. manuscript, 1976.

\item\label{r11} A.H. Chamseddine, G. Felder and J. Fr\"ohlich, ``Gravity in
non-commutative geometry'', Commun. Math. Phys., to appear.

\item\label{r12} R.G. Swan, Trans. Amer. Math. Soc. \ub{105}, 264-277 (1962).

\item\label{r13} P. Hilton and Y.-C. Wu, ``A course in modern algebra'', John
Wiley, New York 1974.

\item\label{r14} A. Jaffe, A. Lesniewski and K. Osterwalder, ``Quantum
$K$-theory,~I. The Chern character'', Commun. Math. Phys. \ub{118}, 1-14
(1988) .\hfill\break
K. Ernst, P. Feng, A. Jaffe and A. Lesniewski,
``Quantum $K$-theory,~II. Homotopy invariance of the Chern character'', J.
Funct. Anal. \ub{90}, 355 (1990).

\item\label{r15} B. Bleile, ``Some aspects of non-commutative geometry'',
diploma thesis 1993;\hfill\break
O. Grandjean, private communication.

\item\label{r16} J.-L. Loday, ``Cyclic homology'', Springer-Verlag,
Berlin-Heidelberg-New York, 1992.

\item\label{r17} A. Connes, private communication;\hfill\break
A.H. Chamseddine and J. Fr\"ohlich (unpublished).

\item\label{r18} B.C. Xanthopoulos and T. Zannias, Phys. Rev. D \ub{40}, 2564
(1989).

\item\label{r19} A.H. Chamseddine and J. Fr\"ohlich, in preparation.

\item\label{r20} C. Ford, D.R.T. Jones, P.W. Stephenson and M.B. Einhorn,
``The effective potential and the renormalisation group'', Preprint
UM-TH-92-21, 1992; the original work is in: E. Weinberg and S. Coleman, Phys.
Rev. \ub{D7}, 1888 (1973).

\item\label{r21} A.H. Chamseddine and J. Fr\"ohlich, ``Constraints on the
Higgs and top quark masses from effective potential and non-commutative
geometry'', Z\"urich University preprint ZU-TH-16/1993.

\item\label{r22} A.H. Chamseddine and J. Fr\"ohlich, in preparation.

\item\label{r23} A.H. Chamseddine and D. Wyler, ``Topological gravity in
1+1 dimensions'', Nucl. Phys. B \ub{340}, 595 (1990);\hfill\break
E. Witten, ``Surprises with topological field theories'', published in College
Station String Workshop, p. 50 (1990).

\item\label{r24} E. Witten, ``Topological quantum field theory'', Commun.
Math. Phys. \ub{117}, 353 (1988).

\item\label{r25} E. Witten, ``String field theory and non-commutative
geometry'' Nucl. Phys. B \ub{268}, 253 (1986).

\item\label{r26} H.-W. Wiesbrock, ``The $C^*$-algebra of bosonic strings'',
Commun. Math. Phys. \ub{136}, 369 (1991).

\endlist

\enddocument